\documentclass[aps,prb,twocolumn,footinbib,superscriptaddress,longbibliography]{revtex4-2}
\usepackage{graphicx}
\usepackage{color}
\usepackage{bm}
\usepackage{amsmath,amssymb}
\usepackage{bbold}
\usepackage[cal=boondoxo]{mathalfa}
\allowdisplaybreaks
\usepackage[bookmarksnumbered,bookmarksopen]{hyperref}

\usepackage{url}
\usepackage{braket}
\let\av\braket

\makeatletter
\DeclareFontEncoding{LS2}{}{\noaccents@}
\makeatother
\DeclareFontSubstitution{LS2}{stix}{m}{n}
\DeclareSymbolFont{largesymbolsstix}{LS2}{stixex}{m}{n}
\DeclareMathDelimiter{\lAngle}{\mathopen}{largesymbolsstix}{"EC}{largesymbolsstix}{"12}
\DeclareMathDelimiter{\rAngle}{\mathclose}{largesymbolsstix}{"ED}{largesymbolsstix}{"13}
\newcommand* {\Av}[1]{\lAngle\,#1\,\rAngle}

\usepackage{array}
\newcolumntype {L}{>{$}l<{$}}             
\newcolumntype {C}{>{$}c<{$}}             
\newcolumntype {R}{>{$}r<{$}}             
\newcolumntype {s}[1]{@{\hspace*{#1}}}    
\newcolumntype {S}[1]{@{\extracolsep{#1}}} 

\hypersetup{
   colorlinks=true,       
   citecolor=blue,        
}
\newcommand* {\vek}[1]{{\bm{\mathrm{#1}}}}
\newcommand* {\vekc}[1]{{\bm{\mathcal{#1}}}}
\newcommand* {\kc}{\mathcal{k}}
\newcommand* {\kk}{\vek{k}}
\newcommand* {\pp}{\vek{p}}
\newcommand* {\rr}{\vek{r}}
\newcommand* {\ee}{\mathrm{e}}
\newcommand* {\frack}[2]{{\Ts\frac{#1}{#2}}}
\newcommand* {\Ds}{\displaystyle}
\newcommand* {\Ts}{\textstyle}
\newcommand* {\Ec}{\mathcal{E}}
\newcommand* {\Bc}{\mathcal{B}}
\newcommand* {\Nc}{\mathcal{N}}
\newcommand* {\Xc}{\mathcal{X}}
\newcommand* {\Zc}{\mathcal{Z}}
\newcommand* {\SX}{Y}
\renewcommand* {\cp}{\mbox{cp}}
\newcommand{\mm}[3][\mathcal{\SX}]{\HK_{#2 \, #3}^{\,#1}}
\newcommand* {\Ns}{N_\mathrm{s}}

\newcommand* {\HK}{\mathcal{H}} 
\newcommand* {\Hcb}{\mathsf{H}} 
\newcommand* {\BtoP}{{\vekc{P} \leftarrow \vekc{B}}}
\newcommand* {\EtoM}{{\vekc{M} \leftarrow \vekc{E}}}
\newcommand* {\bohrmag}{\mu_\mathrm{B}}

\let\myRe\Re
\let\myIm\Im
\renewcommand{\Re}{\myRe\mathrm{e}\,}
\renewcommand{\Im}{\myIm\mathrm{m}\,}

\begin{document}

\title{Collinear Orbital Antiferromagnetic Order and
Magnetoelectricity in Quasi-2D Itinerant-Electron
Paramagnets, Ferromagnets and Antiferromagnets}

\thanks{Dedicated to Ulrich R\"ossler on the occasion
of his 80th birthday.}

\author{R. Winkler}
\affiliation{Department of Physics, Northern Illinois University, DeKalb, IL
60115, USA}
\affiliation{Materials Science Division, Argonne National Laboratory,
Argonne, Illinois 60439, USA}
\affiliation{Materials Research Laboratory, University of Illinois at
Urbana-Champaign, Urbana, Illinois, 61801, USA}
\affiliation{Materials Science and Engineering, University of Illinois at
Urbana-Champaign, Illinois, 61801, USA}
\affiliation{Institut f\"ur Theoretische Physik, Universit\"at Regensburg, 93040 Regensburg, Germany}

\author{U. Z\"ulicke}
\affiliation{School of Chemical and Physical Sciences and MacDiarmid
Institute for Advanced Materials and Nanotechnology, Victoria University
of Wellington, PO Box 600, Wellington 6140, New Zealand}
\affiliation{Kavli Institute for Theoretical Physics, University of California,
Santa Barbara, CA 93106, USA}
\affiliation{Materials Science Division, Argonne National Laboratory,
Argonne, Illinois 60439, USA}
\affiliation{Department of Physics, Northern Illinois University, DeKalb, IL
60115, USA}

\date{June 10, 2020}

\begin{abstract}
We develop a comprehensive quantitative theory for
magnetoelectricity in magnetically ordered quasi-2D systems whereby
in thermal equilibrium an electric field can induce a magnetization
and a magnetic field can induce an electric polarization.  This
effect requires that both space-inversion and time-reversal symmetry
are broken.  Antiferromagnetic order plays a central role in this
theory.  We define a N\'eel operator $\vek{\tau}$ such that a
nonzero expectation value $\av{\vek{\tau}}$ signals collinear
antiferromagnetic order, in the same way a magnetization signals
ferromagnetic order.  While a magnetization is even under space
inversion and odd under time reversal, the operator $\vek{\tau}$
describes a toroidal moment that is odd both under space inversion
and under time reversal.  Thus the magnetization and the
toroidal moment $\av{\vek{\tau}}$ quantify complementary aspects of
collinear magnetic order in solids.
Focusing on quasi-2D systems, itinerant-electron ferromagnetic order
can be attributed to dipolar equilibrium currents that give rise to
a magnetization.  In the same way, antiferromagnetic order arises
from quadrupolar equilibrium currents that generate the toroidal
moment $\av{\vek{\tau}}$.
In the magnetoelectric effect, the electric-field-induced
magnetization can then be attributed to the electric manipulation of
the quadrupolar equilibrium currents.
We develop a $\kk \cdot \pp$ envelope-function theory for the
antiferromagnetic diamond structure that allows us to derive
explicit expressions for the N\'eel operator $\vek{\tau}$.
Considering ferromagnetic zincblende structures and
antiferromagnetic diamond structures, we derive quantitative
expressions for the magnetoelectric responses due to electric and
magnetic fields that reveal explicitly the inherent duality of these
responses required by thermodynamics.  Magnetoelectricity is found
to be small in realistic calculations for quasi-2D electron systems.
The magnetoelectric response of quasi-2D hole systems turns out to
be sizable, however, with moderate electric fields being able to
induce a magnetic moment of one Bohr magneton per charge carrier.
Our theory provides a broad framework for the manipulation of
magnetic order by means of external fields.

\end{abstract}

\maketitle

\section{Introduction}
\label{sec:intro}

The technological viability of alternative spin-based electronics
prototypes \cite{bad10, sin12, hof15} hinges on the ability to
efficiently manipulate magnetizations using electric currents or
voltages.  Various basic device architectures are currently being
explored that could offer the crucially needed electric
magnetization control. One promising approach utilizes
antiferromagnetic materials \cite{jun16, bal18}, while another
employs current-induced spin torques \cite{che09, mir10, mel14,
jun16a}. A third interesting avenue has been opened by harnessing
the magnetoelectric effect \cite{dzy59, lan84, ode70, sir94, geh94,
fie05} in multiferroic materials \cite{eer06, tok14, don15, fie16,
spa17} for switching the magnetization of an adjacent ferromagnetic
contact \cite{her14, tra15}.  Results obtained in our work point to
an appealing alternative possibility, whereby intrinsic
magnetoelectric couplings in ferromagnetic and antiferromagnetic
quasi-twodimensional (quasi-2D) itinerant electron systems provide a
nondissipative mechanism for electric control of magnetizations.  We
present a comprehensive theoretical study of magnetoelectricity in
these paradigmatic nanoelectronic structures that have the potential
to become blueprints for future spintronic devices.

\begin{table*}
  \caption{\label{tab:compareME} Magnitude of the magnetoelectric
  effect in the quasi-2D electron and hole systems considered in
  this work, compared with the values that have been demonstrated or
  that can be reasonably expected in selected known magnetoelectric
  materials.  Among bulk materials, we consider the paradigmatic
  Cr$_2$O$_3$ \cite{heh08a, hal14} as well as TbPO$_4$ that has the
  largest value of the magnetoelectric-tensor components
  $|\alpha_{ij}|$ recorded for a single-phase material \cite{rad84}.
  Also, we include heterostructures made of \mbox{GaMnAs}
  \cite{saw10} and FeRh/BTO \cite{che14}.  The latter has the
  current record value for $|\alpha_{ij}|$.  We list values for
  components $|\alpha_{ij}|$, as well as estimates for the
  achievable magnetizations $\mathsf{M}$ per charge carrier (in the
  quasi-2D electron and hole systems) or per magnetic atom in the
  unit cell (for heterostructures and bulk
  magnetoelectrics). $\epsilon_0$ and $\mu_0$ denote the electric
  permittivity and magnetic permeability of vacuum, respectively.}
  \renewcommand{\arraystretch}{1.4}
\begin{tabular}{@{}C*{6}{s{1.0em}C}@{}}
\hline\hline \rule{0pt}{2.5ex}
\mbox{material} &
{\renewcommand{\arraystretch}{0.95}\begin{array}[t]{@{}C@{}}
  2D electrons \\ (FM InSb)
\end{array}} & 
{\renewcommand{\arraystretch}{0.95}\begin{array}[t]{@{}C@{}}
  2D holes \\ (FM InSb)
\end{array}} & 
\mbox{GaMnAs} & \mbox{FeRh/BTO} & \mbox{Cr$_2$O$_3$} & \mbox{TbPO$_4$} \\ \hline
|\alpha_{ij}| \,\, ( \sqrt{\epsilon_0/\mu_0} ) &
1.9 {\times} 10^{-6} \footnotemark[1]
\footnotetext{This work [Fig.~\ref{fig:e-mom-150}(a)].} &
1.3 {\times} 10^{-4} \footnotemark[2]
\footnotetext{This work [Fig.~\ref{fig:h-mom}(a)].} &
4.0 {\times} 10^{-3} \footnote{Derived from data given in Fig.~2 of Ref.~\cite{saw10}.} &
4.8 {\times} 10^3 \footnote{Derived from measured value of
$\mu_0\, \alpha_{ij}$~\cite{che14}.} &
3.1 {\times} 10^{-4} \footnote{Ref.~\cite{heh08a}.} &
9.0 {\times} 10^{-2} \footnote{Derived from measured value of $\mu_0\,
\alpha_{ij}$~\cite{rad84}, using SI-unit value quoted in
Ref.~\cite{che14}.} \\
\mathsf{M} \,\, (\bohrmag) &
2 {\times} 10^{-2} \footnotemark[1] &
0.6 \footnotemark[2] &
2 \footnote{Value per Mn acceptor atom derived from data given in Fig.~2 of
Ref.~\cite{saw10}.} &
2 \footnote{Value per Fe atom estimated from $\Delta \mathcal{M} \sim
550\,$emu/cm$^3$ \cite{che14}.} &
1 {\times} 10^{-3} \footnote{Value per Cr atom estimated for
$\Ec \sim 10\,$MV/cm in Ref.~\cite{hal14}.} &
2 \footnote{Value per Tb atom estimated for $\Ec \sim 10\,$MV/cm.}
\\ \hline \hline
\end{tabular}
\end{table*}

Ordinarily, when matter is exposed to an electric field $\vekc{E}$,
the field generates a polarization $\vekc{P}$, while a magnetic
field $\vekc{B}$ generates a magnetization $\vekc{M}$.  Counter to
this familiar behavior, magnetoelectric media also develop an
equilibrium magnetic response $\vekc{M}$ to an electric stimulus
$\vekc{E}$, and an electric response $\vekc{P}$ to a magnetic
stimulus $\vekc{B}$ \cite{dzy59, ode70, lan84, sir94, geh94, fie05}.
A systematic understanding of magnetoelectricity can be based on an
expansion of the free-energy density $F$ as a function of the
externally applied electric field $\vekc{E}$ and magnetic field
$\vekc{B}$ \cite{lan84, fie05},
\begin{align}\label{eq:Fdens}
  F (\vekc{E}, \vekc{B})
  & = F(\vek{0}, \vek{0})
  - \mathcal{P}_i^\mathrm{s} \Ec_i
  - \mathcal{M}_i^\mathrm{s} \Bc_i
  \nonumber \\* & \hspace{1em} {}
  - \frack{1}{2} \, \chi^\Ec_{i j}\, \Ec_i \Ec_j
  - \frack{1}{2} \, \chi^\Bc_{i j}\, \Bc_i \Bc_j
  \nonumber \\* & \hspace{1em} {}
  - \alpha_{i j}\, \Ec_i\, \Bc_j
  - \frack{1}{2} \beta_{i j k}\, \Ec_i \Bc_j \Bc_k
  - \frack{1}{2} \gamma_{i j k}\, \Bc_i \Ec_j \Ec_k
  \nonumber \\* & \hspace{1em} {}
  - {} \ldots \quad .
\end{align}
The first two lines in Eq.\ (\ref{eq:Fdens}) pertain to ordinary
electromagnetic phenomena
\footnote{In Eq.\ (\ref{eq:Fdens}), $\vekc{P}^\mathrm{s}$
($\vekc{M}^\mathrm{s}$) describes a spontaneous polarization
(magnetization). Such a contribution can arise intrinsically or due
to proximity to a polarized (magnetized) medium.  The spontaneous
polarization $\vekc{P}^\mathrm{s}$ can be nonzero only if space
inversion symmetry is broken, whereas a nonzero magnetization
$\vekc{M}^\mathrm{s}$ requires broken time-reversal symmetry. The
quantities $\chi^\Ec_{i j}$ ($\chi^\Bc_{i j}$) are the elements of
the material's electric (magnetic) susceptibility tensor.},
whereas terms in the third line are associated with
magnetoelectricity.  In particular, the magnetoelectric tensor
$\alpha_{i j}$ characterizes the generation of an electric
polarization by a magnetic field and of a magnetization by an
electric field, as is clear from the explicit expressions for the
polarization $\vekc{P} = - \partial F / \partial \vekc{E}$,
\begin{subequations}
  \label{eq:moments:gen-def}
  \begin{align}
    \mathcal{P}_i & = \mathcal{P}_i^\mathrm{s}
    + \chi^\Ec_{i j}\, \Ec_j
    \nonumber \\* & \hspace{1em} {}
    + \alpha_{i j} \Bc_j
    + \frack{1}{2} \beta_{i j k}\, \Bc_j \Bc_k
    + \gamma_{j k i}\, \Bc_j \Ec_k
    + \ldots \,\, ,
    \label{eq:moments:gen-def:pol}
  \end{align}
  and the magnetization $\vekc{M} = - \partial F / \partial \vekc{B}$,
  \begin{align}
    \mathcal{M}_i & = \mathcal{M}_i^\mathrm{s}
    + \chi^\Bc_{i j}\, \Bc_j
    \nonumber \\* & \hspace{1em} {}
    + \alpha_{j i} \Ec_j
    + \beta_{j k i}\, \Ec_j \Bc_k
    + \frack{1}{2} \gamma_{i j k}\, \Ec_j \Ec_k
    + \ldots \,\, .
    \label{eq:moments:gen-def:mag}
  \end{align}
\end{subequations}
Here and in the following, we have denoted by $\partial/\partial
\vek{a}$ the gradient vector
$(\partial_{a_x}, \partial_{a_y}, \partial_{a_z})$ of derivatives
w.r.t.\ the Cartesian components of a vector $\vek{a} \equiv (a_x,
a_y, a_z)$.  In both Eqs.\ (\ref{eq:moments:gen-def:pol}) and
(\ref{eq:moments:gen-def:mag}), the first line embodies conventional
electromagnetism in the solid state \cite{lan84}, whereas terms in
the second line of these equations are ramifications of the
magnetoelectric effect \cite{lan84, ode70}.  The appearance of the
same set of coefficients $\alpha_{i j}$, $\beta_{i j k}$, and
$\gamma_{i j k}$ in these equations indicates a deep connection
between the microscopic mechanisms causing a magnetically induced
polarization and the microscopic mechanisms causing an electrically
induced magnetization. As shown in the present work, quasi-2D
systems facilitate the detailed discussion and thorough elucidation of
the underlying mechanisms for such dual magnetoelectric responses.
They also present a promising platform for exploiting
magnetoelectricity in device applications.

As the product of $\vekc{E}$ and $\vekc{B}$ is odd under space
inversion and time reversal, a nonzero tensor $\alpha_{i j}$ is
permitted only for systems with space-inversion symmetry and
time-reversal symmetry both broken \cite{lan84}.  Terms proportional to the
tensors $\beta_{i j k}$ and $\gamma_{i j k}$ embody higher-order
magnetoelectric effects \cite{asc68, gri94, fie05}. Systems in which
only space-inversion (time-reversal) symmetry is broken can have
nonzero tensors $\beta_{i j k}$ ($\gamma_{i j k}$), while $\alpha_{i j}
= 0$.  As an example for the latter in the context of the present work,
we show that paramagnetic quantum wells in zincblende-structure
materials exhibit the higher-order magnetoelectric effect associated
with the tensor $\beta_{i j k}$.

The magnetoelectric effect has been studied experimentally for a
range of materials including ferromagnetic, antiferromagnetic and
multiferroic systems \cite{ode70, fie05, eer06, fie09, riv09}.
Existing theoretical studies of the magnetoelectric effect have
either focused on elucidating general properties of the tensors
$\alpha_{i j}$, $\beta_{i j k}$ and $\gamma_{i j k}$ based on
symmetry \cite{sch73, riv94, wat18a} or developed first-principles
methods for their numerical calculation \cite{ess10, mal10, mal12,
sca12, spa13, tho16} and semiclassical approaches~\cite{gao14}.
These works considered insulators where magnetoelectric effects are
well-defined as a bulk property.  Typically, these works have also
limited their scope to investigating only one of the two dual
magnetoelectric responses.  As a result, the microscopic basis for
the intrinsic symmetry of electric and magnetic responses has been
rarely discussed \cite{rad62}.  In contrast, the conceptually
transparent and practically important quantum-well system considered
in the present work provides a versatile, unified theoretical
framework for describing magnetoelectricity in paramagnets,
ferromagnets and antiferromagnets, covering both the electrically
induced magnetization and the magnetically induced polarization and
demonstrating explicitly how these two effects are intrinsically
related. Furthermore, the quasi-2D systems studied here are unusual
examples of \emph{metals} exhibiting magnetoelectricity in
equilibrium, i.e., in the absence of transport currents.
Specifically, the in-plane magnetic field generates an electric
polarization perpendicular to the 2D plane and a perpendicular
electric field induces an in-plane magnetization
\cite{misc:2Dmetal}. The reduced dimensionality of the quantum-well
systems guarantees that these manifestations of magnetoelectricity
are well-defined and also accessible experimentally.  The
magnetoelectric coupling per volume is proportional to the width $w$
of the quasi-2D system, and in antiferromagnetic and
halfmetallic-ferromagnetic quasi-2D systems, it is also proportional
to the sheet density $\Ns$.  Thus, unlike magnetoelectricity in bulk
materials, it is easily tunable in quasi-2D systems.  While the
magnitude of magnetoelectric-tensor components are similar to the
moderate values in the classic magnetoelectric Cr$_2$O$_3$, the
electric-field-induced magnetization per particle is comparable to
the values found in current record-breaking multiferroics.  See the
comparison of relevant magnitudes provided in
Table~\ref{tab:compareME}. The unusual situation where an electric
field can generate a large magnetization per particle in a system with
small magnitude of magnetoelectric-tensor components arises
because the magnetoelectric response in our metallic quasi-2D
systems is associated with the itinerant charge carriers whose
density per unit cell is small.

\begin{figure}[t]
  \includegraphics[width=\linewidth]{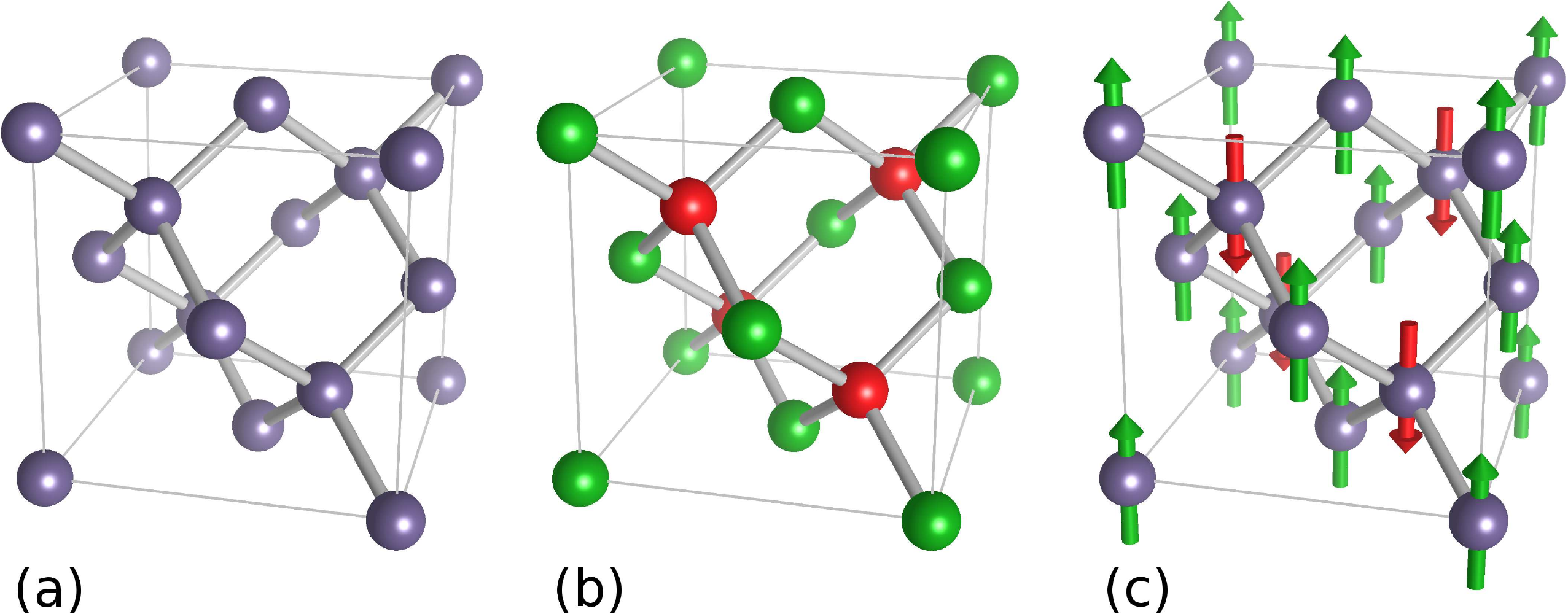}
  \caption{Variations of the diamond structure considered in this
  work.  (a) Inversion-symmetric diamond structure.  (b) Zincblende
  structure that breaks inversion symmetry.  (c) Antiferromagnetic
  diamond structure that breaks time-reversal symmetry $\Theta$ and
  inversion symmetry $I$ (though the joint operation $\Theta I$
  remains a good symmetry).  Materials with structure (a) are not
  magnetoelectric. Those with structure (b) become magnetoelectric
  when they are magnetized, whereas materials with structure (c) are
  intrinsically magnetoelectric.}
  \label{fig:diamond}
\end{figure}

Our realistic theoretical study focuses on the technologically
important class of materials realizing variants of the diamond
structure; see Fig.~\ref{fig:diamond}.  As discussed earlier,
magnetoelectricity only occurs in situations where both
space-inversion and time-reversal symmetry are broken. Hence, the
magnetoelectric effect is absent in paramagnetic materials having
the inversion-symmetric \cite{misc:diamond} diamond structure
[Fig.~\ref{fig:diamond}(a)].  In contrast, the zincblende structure
[Fig.~\ref{fig:diamond}(b)] breaks inversion symmetry.  In addition,
time-reversal symmetry is broken in magnetized samples with ordered
spin magnetic moments or with an orbital magnetization due to
dissipationless equilibrium currents.  Such a magnetization can be
caused by a Zeeman coupling of the charge carriers to an applied
magnetic field, or by a ferromagnetic exchange field \cite{jun06,
die14} that is present in the material itself or induced by
proximity to a ferromagnet.  The origin of the magnetization is
largely irrelevant for the microscopic mechanism of
magnetoelectricity so that we denote all these scenarios jointly as
ferromagnetically ordered.  We demonstrate in this work the
emergence of finite magnetoelectric couplings in ferromagnetically
ordered quantum wells made from materials having a zincblende
structure.  We find that already in the absence of external fields,
the interplay of broken space-inversion and time-reversal symmetry
generates a collinear orbital antiferromagnetic order of the charge
carriers that renders these systems to be actually ferrimagnetic.
The magnetoelectric effect can then be viewed as arising from the
manipulation of the equilibrium current distributions underlying the
orbital antiferromagnetic order.  Specifically, an electric field
affects these currents in a way reminiscent of the Lorentz force
such that the modified currents give rise to a magnetization
component in addition to, and oriented at an angle to the
ferromagnetic order in the system. In contrast, an external magnetic
field $\vekc{B}$ applied perpendicularly to the ferromagnetic order
can induce an electric dipole moment via a mechanism resembling the
Coulomb force, where the scalar potential is replaced by the vector
potential for $\vekc{B}$. This mechanisms for magnetoelectricity in
quantum wells made from ferromagnetic zincblende semiconductors
differs fundamentally from the electric-field control of the spontaneous
magnetization $\vekc{M}^\mathrm{s}$ in these systems~\cite{mat15, misc:gamnas}. 

Magnetoelectricity occurs most prominently in antiferromagnetically
ordered materials, where an electrically induced magnetization is
not masked by an intrinsic magnetization in the system.  Similar to
ferromagnetic order, antiferromagnetic order can have a spin
component and an orbital component, and we can have spontaneous
antiferromagnetic order due to a staggered exchange field in the
material, but the order can also be induced in both paramagnets and
ferromagnets.  Here we consider the antiferromagnetic diamond
structure shown in Fig.~\ref{fig:diamond}(c). To study the
magnetoelectricity exhibited in quantum wells made from such a
material, we develop a $\kk\cdot \pp$ envelope-function theory for
itinerant-electron diamond antiferromagnets, which is in itself an
important result presented in this work. On the basis of this
theory, we are able to define an operator $\vek{\tau}$ in terms of
itinerant-electron degrees of freedom such that a nonzero
expectation value $\av{\vek{\tau}}$ signals collinear
antiferromagnetic order in the same way that a nonzero expectation
value $\av{\vek{\sigma}}$ of the charge carriers' spin operator
$\vek{\sigma}$ signals ferromagnetic order of spins.  Applying our
theoretical framework to antiferromagnetically ordered quantum wells
placed into external magnetic and electric fields, we reveal them to
exhibit magnetoelectric couplings remarkably similar to those found
for the ferromagnetically ordered zincblende quantum wells described
above. The magnetoelectric response of the antiferromagnetic system
can be related to the modification of the quadrupolar
equilibrium-current distribution associated with antiferromagnetic
order by external electric and magnetic fields.  This is in line
with the fact that the magnetoelectric tensor $\alpha_{ij}$ behaves
under symmetry transformations like a magnetic quadrupole moment
\cite{spa08}, i.e., both of these second-rank material tensors
require broken space-inversion symmetry and broken time-reversal
symmetry and these tensors share the same pattern of nonzero
components, though microscopically they are generally not simply
related with each other.

Analytical results obtained from effective two-band models of confined
charge carriers elucidate the basic physical phenomena associated with
magnetoelectricity in para-, ferro- and antiferromagnetic quantum wells.
Accurate numerical calculations utilizing realistic $8 \times 8$ and
$14 \times 14$ $\kk \cdot \pp$ Hamiltonians establish a typically large,
practically relevant magnitude of the electric-field-induced magnetization
in hole-doped quantum wells made from zincblende ferromagnets or
diamond-structure antiferromagnets.
The ability to illustrate the full complementarity of
magnetoelectric responses within the same microscopic theory
distinguishes our approach from most previous ones \cite{geh94}.  We
show that our explicit results for the magnetic responses provide an
important benchmark for general theories of magnetoelectricity
\cite{spa08, gao18, shi18}.  Our findings provide a platform for
further systematic studies aimed at manipulating charges, currents,
and magnetic order in solids.

The remainder of this Article is organized as follows.  In
Sec.~\ref{sec:responses}, we define the relevant quantities of
interest for our study, establishing the relation between the
thermodynamic definitions of polarization
(\ref{eq:moments:gen-def:pol}) and magnetization
(\ref{eq:moments:gen-def:mag}) and the electromagnetic definitions
of these quantities.  We then proceed, in Sec.~\ref{sec:para-ferro},
to calculate magnetoelectric responses of quasi-2D electron and hole
systems realized in zincblende heterostructures having a Zeeman spin
splitting due to an external magnetic field or due to the coupling
to ferromagnetic exchange fields.  In Sec.~\ref{sec:AFM}, we develop
a general framework for the $\kk \cdot \pp$ envelope-function
description of antiferromagnetic order.  We use this framework to
perform a comprehensive analysis of magnetoelectric phenomena in
quantum wells made from diamond-structure antiferromagnets.
Section~\ref{sec:bound} is devoted to deriving an upper bound on the
magnitude of magnetoelectric-tensor components in quasi-2D
systems~\cite{bro68a}.  We summarize our conclusions and provide a
brief outlook in Sec.~\ref{sec:concl}.
Appendix~\ref{app:current-induced-magnetization} reviews
current-induced magnetization, a phenomenon that shares some
apparent similarities with the magnetoelectric effect.  Ancillary
results are presented in Appendices~\ref{app:mag-mom-op}
and~\ref{app:orb-magnet}.

\section{Electric and magnetic responses in quasi-2D
systems}
\label{sec:responses}

We consider a quasi-2D system in the $(x,y)$ plane with open
boundary conditions in the $z$ direction in the presence of a
perpendicular electric field $\Ec_z$ and an in-plane magnetic field
$\vekc{B}_\| = (\Bc_x, \Bc_y)$~\cite{misc:2Dmetal}. Throughout
this work, vectors like $\vekc{B}_\|$ that have only in-plane
components will be indicated by a subscript `$\|$', and their vanishing
$z$ component will be suppressed. Very generally, the polarization
and magnetization can be obtained from the free-energy density $F$
via the relations \cite{lan84}
\begin{subequations}
  \label{eq:PM:free-en:def}
  \begin{align}
    \mathcal{P}_z & = - \frac{\partial F}{\partial \Ec_z} \; ,
    \label{eq:PM:free-en:pol:def} \\
    \vekc{M}_\| & = - \frac{\partial F}{\partial \vekc{B}_\|} \; .
    \label{eq:PM:free-en:total-mag:def}
  \end{align}
\end{subequations}
More accurately, the polarization and magnetization only depend on
the change of the free energy $\delta F \equiv F(\Ec_z, \vekc{B}_\|)
- F(0, \vek{0})$ due to the fields $\Ec_z$ and $\vekc{B}_\|$.

To simplify the analysis, we assume that only the itinerant charge
carriers in the quasi-2D system contribute to the electric and
magnetic response.  We assume that the confining potential $V(z)$ of
the quasi-2D system includes the electrostatic potential due to
compensating charges and external gates that ensure overall charge
neutrality and that are assumed to be fixed in space.  Also, we
assume that the potential $V(z)$ defining a quantum well for the
quasi-2D system is symmetric, i.e., $V(-z) = V(z)$.  We denote the
Hamiltonian for the charge carriers by $H$.  The electric field
$\Ec_z$ enters $H$ via the additional potential $e \Ec_z z$, where
$e\equiv |e|$ is the elementary charge.  The magnetic field
$\vekc{B}_\|$ enters $H$ via the vector potential $\vekc{A}$ that is
related to the magnetic field via $\vekc{B}_\| = \vek{\nabla}
\times \vekc{A}$, with $\vek{\nabla}$ denoting the gradient
w.r.t.\ the position vector $\rr \equiv (x, y, z)$.  In addition,
$\vekc{B}_\|$ may enter $H$ via a Zeeman term \mbox{$(g/2) \bohrmag
\, \vek{\sigma}\cdot \vekc{B}_\|$}, where $g$ denotes the $g$ factor,
$\bohrmag \equiv e\hbar / (2m_0)$ is the Bohr magneton, with $m_0$
being the mass of free electrons, and $\vek{\sigma}$ is a
dimensionless spin operator~\footnote{Within generic $2\times 2$
models, $\vek{\sigma}$ is typically represented by the vector of
Pauli matrices.  Representations of the spin operator in more
general multi-band models are discussed, e.g., in
Ref.~\cite{win03}.}.  The eigenstates of $H$ associated with
eigenvalues $E_{n \kk_\|}$ have the general form
\begin{equation}
\Psi_{n \kk_\|}(\rr) = \frac{\ee^{i \kk_\| \cdot \rr}}{2\pi} \,
\Phi_{n \kk_\|}(z) \; .
\end{equation}
Here $n$ labels the quasi-2D subbands, and $\kk_\| \equiv (k_x, k_y)$
is the in-plane wave vector. The free-energy density can then be written
in the form
\begin{equation}
  \label{eq:free-en:def}
  F = \frac{1}{w} \sum_{n} \int \frac{d^2 k_\|}{(2\pi)^2}
  \; f (E_{n \kk_\|}) \; E_{n\kk_\|} \quad ,
\end{equation}
where $w$ is the width of the quantum well, and $f(E)$ denotes the Fermi
distribution function. We will later assume zero temperature so that
$f(E)$ becomes a step function $f(E) = \Theta(E_\mathrm{F} - E)$,
with the Fermi energy $E_\mathrm{F}$.

Using the expression (\ref{eq:free-en:def}) for the free-energy density,
the polarization becomes
\begin{equation}
  \label{eq:pol-tot:def}
  \mathcal{P}_z = \mathcal{P}_z^\mathrm{e} + \mathcal{P}_z^\mathrm{q} \,\, ,
\end{equation}
with
\begin{subequations}
\begin{align}
  \mathcal{P}_z^\mathrm{e} & = -\frac{1}{w} \sum_{n} \int
  \frac{d^2 k_\|}{(2\pi)^2} \; f (E_{n \kk_\|}) \, \frac{\partial
  E_{n\kk_\|}}{\partial \Ec_z} \; , \\
  \mathcal{P}_z^\mathrm{q} & = -\frac{1}{w} \sum_{n} \int
  \frac{d^2 k_\|}{(2\pi)^2} \; E_{n \kk_\|} \, \frac{\partial
  f(E_{n\kk_\|})}{\partial \Ec_z} \; .
\end{align}
\end{subequations}
The first term $\mathcal{P}_z^\mathrm{e}$ arises from the
$\Ec_z$-dependence of the energies $E_{n \kk_\|}$ of occupied
states. The second term $\mathcal{P}_z^\mathrm{q}$ represents a
quantum-kinetic \cite{hua87} contribution to $\mathcal{P}_z$ that
accounts for changes in the equilibrium occupation-number
distribution arising from a change of $\Ec_z$. Hence, in the
low-temperature limit, $\mathcal{P}_z^\mathrm{q}$ reflects
$\Ec_z$-induced changes in the shape or topology of the Fermi
surface.

Using the Hellmann-Feynman theorem and assuming the only
explicit $\Ec_z$-dependence in the Hamiltonian $H$ to be the
potential $e\Ec_z z$~\footnote{In general, $H$ may also
contain spin-orbit terms that depend explicitly on $\Ec_z$.
We ignore such terms in the present discussion of the electric
polarization.}, we find
\begin{equation}\label{eq:pol:def}
\mathcal{P}_z^\mathrm{e} = - \frac{e}{w} \sum_{n} \int
\frac{d^2 k_\|}{(2\pi)^2} \; f (E_{n \kk_\|}) \, \av{z}_{n\kk_\|} \; ,
\end{equation}
where
\begin{equation}
  \av{z}_{n\kk_\|} = \int dz\, \left| \Phi_{n\kk_\|} (z) \right|^2\, z
\end{equation}
denotes the displacement of an electron in the state
$\Phi_{n\kk_\|}(z)$.  Thus the term $\mathcal{P}_z^\mathrm{e}$
coincides with the electrostatic definition of polarization as the
volume average of microscopic electric dipole moments \cite{jac99,
res10}. In a quasi-2D system with open boundary conditions in the
$z$ direction (and overall charge neutrality as assumed above), the
electrostatic polarization $\mathcal{P}_z^\mathrm{e}$ is
unambiguously defined independently of the origin of the coordinate
system.  It avoids the technical problems inherent in studies of the
bulk (3D) polarization \cite{res10}.  The average displacement of
the occupied states is
\begin{subequations}
  \begin{align}
    \label{eq:z:av:def}
    \av{z} & = \frac{1}{\Ns} \sum_{n} \int
    \frac{d^2 k_\|}{(2\pi)^2} \, f (E_{n \kk_\|}) \,
    \av{z}_{n\kk_\|}
    \; , \\ &
    = \frac{1}{\Ns} \int dz \: \rho(z) \, z \; ,
  \end{align}
\end{subequations}
where
\begin{equation}
  \rho (z) = \sum_{n} \int \frac{d^2 k_\|}{(2\pi)^2} \,\, f (E_{n \kk_\|}) \,\,
  \left| \Phi_{n\kk_\|} (z) \right|^2
\end{equation}
is the 3D number density, and $\Ns = \int dz \, \rho (z)$ is
the 2D (sheet) density of charge carriers in the quantum well.  Thus
we can rewrite the polarization (\ref{eq:pol:def}) as
\begin{equation}
  \label{eq:polToAvZ}
  \mathcal{P}_z^\mathrm{e} = \mathcal{P}_0 \, P
  \equiv \mathcal{P}_0 \frac{\av{z}}{w} \quad ,
\end{equation}
where $\mathcal{P}_0 \equiv -e \Ns$, and the dimensionless
number $P = \av{z}/w$ describes the average polarization per
particle.

Similar to the polarization $\mathcal{P}_z$, the magnetization
$\vekc{M}_\|$ is also the sum of two contributions,
\begin{equation}
  \vekc{M}_\| = \vekc{M}_\|^\mathrm{e} + \vekc{M}_\|^\mathrm{q} \,\, ,
\end{equation}
with
\begin{subequations}
\begin{align}
  \vekc{M}_\|^\mathrm{e} & = -\frac{1}{w} \sum_{n} \int
  \frac{d^2 k_\|}{(2\pi)^2} \; f (E_{n \kk_\|}) \, \frac{\partial
  E_{n\kk_\|}}{\partial \vekc{B}_\|} \; , \\
  \vekc{M}_\|^\mathrm{q} & = -\frac{1}{w} \sum_{n} \int
  \frac{d^2 k_\|}{(2\pi)^2} \; E_{n \kk_\|} \, \frac{\partial
  f(E_{n\kk_\|})}{\partial \vekc{B}_\|} \; ,
\end{align}
\end{subequations}
which again represent the electromagnetic and the quantum-kinetic
effects of $\vekc{B}_\|$, respectively. Given that $\vekc{B}_\|$
generally enters the Hamiltonian $H$ via both the vector potential
$\vekc{A}$ and also via the Zeeman term, the contribution
$\vekc{M}_\|^\mathrm{e}$ can be split further into orbital and spin
contributions,
\begin{equation}
  \label{eq:mag-tot:def}
  \vekc{M}_\|^\mathrm{e} = \vekc{M}_\|^\mathrm{o} + \vekc{S}_\| \; ,
\end{equation}
where
\begin{subequations}
\label{eq:mag-parts:def}
\begin{align}
  \vekc{M}_\|^\mathrm{o} & = -\frac{1}{w} \sum_{n} \int
  \frac{d^2 k_\|}{(2\pi)^2} \; f (E_{n \kk_\|}) \, \Big\langle \sum_{j=x,y}
  \frac{\partial H}{\partial \mathcal{A}_j}
  \frac{\partial \mathcal{A}_j}{\partial \vekc{B}_\|}
  \Big\rangle_{n \kk_\|}\, , \nonumber \\*
  & = - \frac{e}{w} \sum_n \int
  \frac{d^2 k_\|}{(2\pi)^2} \,\, f (E_{n \kk_\|}) \,\,
  \hat{\vek{z}} \times \av{\{ z\, , \vek{v}_\| \}}_{n \kk_\|} \,\, ,
  \label{eq:orb-mag:def} \\
  \vekc{S}_\| & = -\frac{1}{w} \sum_{n} \int \frac{d^2 k_\|}{(2\pi)^2}
  \; f (E_{n \kk_\|}) \, \Big\langle \frac{\partial H}{\partial \vekc{B}_\|}
  \Big\rangle_{n \kk_\|} \; , \nonumber \\*
  & = -\frac{g \bohrmag}{2 w} \sum_{n}
  \int \frac{d^2 k_\|}{(2\pi)^2} \, f (E_{n\kk_\|}) \,
  \av{\vek{\sigma}}_{n\kk_\|} \; .  \label{eq:spin-mag:def}
\end{align}
\end{subequations}
To obtain Eqs.\ (\ref{eq:orb-mag:def}) and (\ref{eq:spin-mag:def}), we
used once again the Hellmann-Feynman theorem.  The first term
$\vekc{M}_\|^\mathrm{o}$ represents the in-plane orbital magnetization
\cite{jac99, res10}.  In Eq.\ (\ref{eq:orb-mag:def}), the symbol
$\vek{v}_\| \equiv \partial H / (\partial \hbar \kk_\|)$ denotes the
in-plane component of the velocity operator, and
\begin{equation}
  \label{eq:z-v:av:def}
  \av{\{ z\, , \vek{v}_\| \}}_{n\kk_\|} = \int dz \;\;
  \Phi^\dagger_{n\kk_\|} (z)\,\, \{ z\, , \vek{v}_\| \} \,\, \Phi_{n\kk_\|} (z)
  \,\, ,
\end{equation}
with $\{A,B\} \equiv \frac{1}{2} (AB + BA)$.  The expression
(\ref{eq:orb-mag:def}) is associated with the vector potential
$\vekc{A} = z \, \vekc{B}_\| \times \hat{\vek{z}}$ that is adopted
throughout our work as the appropriate gauge for quasi-2D
systems. This is the reason why Eq.\ (\ref{eq:orb-mag:def}) differs
from the conventional formula for the orbital
magnetization \cite{res10} that is obtained for the symmetric
gauge \cite{whi07} $\vekc{A}^\mathrm{sym} \equiv \frac{1}{2} \,
\vekc{B} \times \rr$, see Appendix~\ref{app:mag-mom-op}\@. Similar
to $\mathcal{P}_z^\mathrm{e}$, the magnetization
$\vekc{M}_\|^\mathrm{o}$ of a quasi-2D system avoids the technical
problems inherent in studies of the bulk (3D) orbital magnetization
\cite{res10}; it is unambiguously defined independently of the origin of
the coordinate system.

An orbital magnetization $\vekc{M}_\|^\mathrm{o}$ is generally
accompanied by a nonvanishing in-plane current distribution
\begin{subequations}
  \label{eq:2D-current:def}
  \begin{equation}
    \label{eq:2D-current:def:av}
    \vekc{j}_\| (z) = -e \sum_n \! \int \! \frac{d^2 k_\|}{(2\pi)^2}
    \, f (E_{n\vek{k}_\|}) \,  \vek{j}_\| (z, n \kk_\|) \,\, ,
  \end{equation}
  with
  \begin{equation}
    \vek{j}_\| (z, n \kk_\|) =
    \Re \! \Big[ \Phi_{n\kk_\|}^\dagger (z) \,
    \vek{v}_\| \, \Phi_{n\kk_\|} (z) \Bigr] \,\, ,
  \end{equation}
\end{subequations}
though in thermal equilibrium, the total current $\vekc{J}_\|
= \int dz \, \vekc{j}_\| (z)$ is always zero. These currents
$\vekc{j}_\| (z)$ are nondissipative because they are not driven by
an electric field.  (Throughout this work, we assume $\vekc{E}_\| =
\vek{0}$ for the in-plane electric field.) Direct experimental observation
of the currents $\vekc{j}_\| (z)$ seems impossible, as their nature
appears to preclude any ability to make contact to them. However, their
ramification in terms of the magnetization $\vekc{M}_\|^\mathrm{o}$ is detectable.

The second term $\vekc{S}_\|$ in Eq.\ (\ref{eq:mag-tot:def})
represents the spin magnetization, given in Eq.\
(\ref{eq:spin-mag:def}) in terms of the dimensionless spin
polarization $\av{\vek{\sigma}}_{n\kk_\|}$ of individual states.  We
rewrite this as
\begin{equation}
  \vekc{S}_\| = -\frac{g \bohrmag \Ns}{2 w} \, \vek{S}_\| \; ,
\end{equation}
where $\vek{S}_\|$ is the dimensionless average spin polarization of
the entire system.  Similarly, it is convenient to define
$\vekc{M}_\|^\mathrm{o} = \mathcal{M}_0 \, \vek{M}_\|^\mathrm{o}$
with $\mathcal{M}_0 \equiv -\bohrmag \Ns / w$ and
dimensionless $\vek{M}_\|^\mathrm{o}$ so that we get
\begin{subequations}
 \label{eq:total-mag:def}
\begin{align}
  \vekc{M}_\|^\mathrm{e}
  & = \mathcal{M}_0 \left( \vek{M}_\|^\mathrm{o} + \frac{g}{2}\, \vek{S}_\|
         \right) \,\, , \\
  & = \frac{\mathcal{M}_0}{\Ns}
  \int \frac{d^2 k_\|}{(2\pi)^2} \, f (E_{n\kk_\|})
  \nonumber \\ & \hspace{1.5em} {} \times
  \left[ \frac{2 m_0}{\hbar} \,
  \hat{\vek{z}} \times \av{\{ z\, , \vek{v}_\| \}}_{n \kk_\|}
  + \frac{g}{2} \, \av{\vek{\sigma}}_{n\kk_\|} \right] \,\, .
 \label{eq:total-mag:def:split}
\end{align}
\end{subequations}

A polarization $\mathcal{P}_z^\mathrm{e}$ represents the dipole term
($l=1$) in a multipole expansion of a charge distribution $\rho(z)$
\cite{jac99}.  Similarly, an orbital magnetization
$\vekc{M}_\|^\mathrm{o}$ represents the dipole term ($l=1$) in a
multipole expansion of a current distribution $\vekc{j}_\| (z)$.
Charge neutrality of a localized charge distribution $\rho(z)$
generally requires a vanishing monopole ($l=0$) for the multipole
expansion of $\rho(z)$.  Similarly, a localized current distribution
$\vekc{j}_\| (z)$ requires a vanishing monopole for the multipole
expansion of $\vekc{j}_\| (z)$.  An equilibrium current distribution
$\vekc{j}_\| (z)$ that breaks time-reversal symmetry is permitted in
ferromagnets and in antiferromagnets \cite{misc:currents}.  The
finite magnetization in ferromagnets implies that the equilibrium
current distribution $\vekc{j}_\|(z)$ includes a dipolar component
($l=1$), whereas the vanishing magnetization in antiferromagnets
requires equilibrium currents to be (at least) of quadrupolar type
($l=2$).

For finite systems, the lowest nonvanishing multipole in a multipole
expansion is generally independent of the origin of the coordinate
system and in that sense well-defined, whereas higher multipoles
depend on the choice for the origin \cite{jac99}. We therefore
limit our discussion below to the lowest nonvanishing multipole.
As mentioned above, in infinite periodic crystals, even the lowest
nonvanishing multipole moment requires a more careful treatment
\cite{res10}.

As integrals can be more easily and more reliably calculated
numerically than derivatives \cite{pre92}, it is more
straightforward to evaluate numerically the integrals defining the
electromagnetic parts $\mathcal{P}_z^\mathrm{e}$,
$\vekc{M}_\|^\mathrm{o}$, and $\vekc{S}_\|$ of the response
functions.  On the other hand, it is more difficult to evaluate
accurately the full response functions $P_z$ and $\vekc{M}_\|$ that
require a numerical differentiation of the free energy $F$ as a
function of the applied external fields \cite{ger68, amo87}.  A
detailed account of these technical issues is beyond the scope of
the present work.  In the following, we thus focus on
$P_z^\mathrm{e}$, $\vekc{M}_\|^\mathrm{o}$, and $\vekc{S}_\|$ alone.
This is adequate for scenarios where the quantum-kinetic parts
$P_z^\mathrm{q}$ and $\vekc{M}_\|^\mathrm{q}$ of the response
functions are less important, which we have found to be generally
the case for a strong confinement $V(z)$.
Within the framework of the analytical perturbative calculations of
the magnetoelectric effect discussed below, the external fields
$\Ec_z$ and $\vekc{B}_\|$ do not change the occupation of individual
states.  Quantum-kinetic contributions $\mathcal{P}_z^\mathrm{q}$
and $\vekc{M}_\|^\mathrm{q}$ thus do not arise in the analytical
calculations.

\section{Magnetoelectricity in zincblende paramagnets and ferromagnets}
\label{sec:para-ferro}

\subsection{The model}
\label{sec:para-ferro:model}

The diamond crystal structure is shown in Fig.~\ref{fig:diamond}(a).
Space inversion is a good symmetry in diamond so that electronic
states are at least twofold degenerate throughout the Brillouin zone
\cite{misc:diamond}.  The diamond structure is realized in group-IV
semiconductors including C, Si, and Ge.  In a zincblende structure,
the atomic sites in a diamond structure are alternatingly occupied
by two different atoms such as Ga and As or In and Sb
[Fig.~\ref{fig:diamond}(b)].  Thus spin degeneracy of the electronic
states is lifted in paramagnetic zincblende structures except for
$\kk = \vek{0}$.  

Spontaneous ferromagnetic order is realized in semiconductors with
zincblende structure such as GaMnAs \cite{jun14} and InMnSb
\cite{woj03}, where the ferromagnetic coupling between local Mn
moments is mediated by itinerant holes \cite{jun06, die14}.  In
ferromagnetic GaMnAs, the magnetization resides mostly in the
Mn-impurity spins (with magnetic moment $5\,\bohrmag$) \cite{die14,
jun06}.  We want to focus here on ferromagnetic InSb, where the
effective spin magnetic moment of holes $\sim 2\kappa \bohrmag$
(with Luttinger parameter $\kappa = 15.6$) is more than an order of
magnitude larger than in GaAs ($\kappa = 1.2$), so that the
magnetization density residing in spin-polarized itinerant InSb
holes can easily exceed the magnetization density due to Mn spins
(assuming hole densities comparable to the densities of Mn
acceptors, although the hole densities can also be controlled
independently by means of external gates).  In the present work, we
thus focus on the itinerant carriers, assuming for conceptual
clarity that the spontaneous magnetization $\vekc{M}^\mathrm{s}$ is
fixed~\cite{misc:gamnas}.
The more complicated band structure of holes can only be
satisfactorily approached in less transparent numerical
calculations.  Therefore, we complement the calculations for holes
with more transparent calculations for electron systems.

For common semiconductors with a zincblende structure, such as GaAs,
InAs, and InSb, the electronic states in a quantum well can be
described by a multiband Hamiltonian~\cite{win03}
\begin{equation}
  \label{eq:Kane-ham}
  \HK = \HK_k + V(z) + \HK_\mathrm{D} + \HK_\Zc + e\Ec_z z
  \quad .
\end{equation}
Here $\HK_k$ is the inversion-symmetric part of $\HK$, and
$\HK_\mathrm{D}$ subsumes Dresselhaus terms due to bulk inversion
asymmetry (BIA).  $V(z)$ is the quantum-well confinement, so that
the wave vector $\kk_\| = (k_x, k_y)$ is a good quantum number,
whereas $k_z$ becomes the operator $-i\partial_z$.  An external
electric field $\Ec_z$ can be included in $\HK$ by adding the
potential $e\Ec_z z$.  Similarly, an external
in-plane magnetic field $\vekc{B}_\|$ can be included in $\HK$ via
the vector potential $\vekc{A} = z \, \vekc{B}_\| \times
\hat{\vek{z}}$.  In $\HK_k + \HK_D$ we then replace $\kk$ by the
kinetic wave vector $\vekc{k} = \kk + \frac{e}{\hbar} \vekc{A}$.
The Zeeman term $\HK_\Zc$ includes contributions from both the
external field $\vekc{B}_\|$ and possibly a ferromagnetic exchange
interaction represented by an internal exchange field $\vekc{X}_\|$
that is likewise assumed to be in-plane.  A finite exchange field
$\vekc{X}_\|$ corresponds to a finite spontaneous magnetization
$\vekc{M}^\mathrm{s}$ in the expansion (\ref{eq:Fdens}).  For $\Xc =
0$, the system is a paramagnet, where the lowest-order term in the
expansion (\ref{eq:Fdens}) that depends only on $\vekc{B}$ is $-
\frack{1}{2} \mu_0^{-1}\, \chi^\Bc_{i j}\, \Bc_i \Bc_j$, signifying
the fact that the system's magnetization scales with the applied
field $\vekc{B}$ until the system is fully spin-polarized.  For the
magnetoelectric effect studied here, a finite Zeeman term $\HK_\Zc$
indicates, first of all, a breaking of time-reversal symmetry so
that the origin of $\HK_\Zc$ is largely irrelevant for the
microscopic mechanism yielding the magnetoelectric response.
Nonetheless, as to be expected, we will see below that only for $\Xc
\ne 0$ or a fully spin-polarized paramagnet, the final result for
the lowest-order magnetoelectric contribution to the free energy
(\ref{eq:Fdens}) can be expressed via a tensor $\alpha_{ij}$,
whereas in partially spin-polarized paramagnets the linear
dependence of $\HK_\Zc$ on $\vekc{B}_\|$ is the reason why in lowest
order we get terms in Eq.\ (\ref{eq:Fdens}) that are weighted by a
third-rank tensor $\beta_{i j k}$.

The diagonalization of the Hamiltonian (\ref{eq:Kane-ham}) yields
the eigenenergies $E_{n \kk_\|}$ with associated bound states
$\Phi_{n \kk_\|} (z) \equiv \braket{z| n\kk_\|}$, where $n$ is the
subband index.
In the numerical calculations presented below, we use for $\HK$ the
$8 \times 8$ Kane model and the $14 \times 14$ extended Kane model
as defined in Table C.5 of Ref.\ \cite{win03}.  Confinement in the
quasi-2D system is due to a finite potential well $V(z) = V_0 \,
\Theta (|z| - w/2)$ with barrier height $V_0$.  The numerical
solution of $\HK$ is based on a quadrature method \cite{win93a}.  We
evaluate $k$-space integrals such as Eq.\ (\ref{eq:pol:def}) by
means of analytic quadratic Brillouin-zone integration~\cite{win93}.

Before presenting numerical results for multi-band models, we
illustrate the physical origin and ramifications of
magnetoelectricity in zincblende-semiconductor quantum wells by
analytical calculations.  Specifically, we consider a $2 \times 2$
model for the $\Gamma_6$ conduction band
\begin{subequations}
  \label{eq:ham-cb}
  \begin{equation}
    \Hcb = \Hcb_k + V(z) + \Hcb_\mathrm{D} + \Hcb_\Zc + e\Ec_z z \,\, ,
  \end{equation}
  with
  \begin{align}
    \label{eq:ham-cb:kin}
    \Hcb_k & = \frac{\hbar^2  \kc^2}{2 m} \,\, , \\ \rule{0pt}{4.0ex}
    \Hcb_\mathrm{D} & = d \left(\left\{ \kc_x, \kc_y^2 - \kc_z^2 \right\}
      \sigma_x + \cp \right) \,\, ,
    \label{eq:ham-cb:dressel}\\
    \Hcb_\Zc & = \vekc{Z} \cdot \vek{\sigma} \,\, ,
    \label{eq:ham-cb:zeeman}
  \end{align}
\end{subequations}
where $m$ denotes the effective mass, $\Hcb_\mathrm{D}$ is the
Dresselhaus term with prefactor $d$, $\cp$ denotes cyclic
permutation of the preceding term, $\vek{\sigma} \equiv (\sigma_x,
\sigma_y, \sigma_z)$ is the vector of Pauli matrices, and $\Hcb_\Zc$
is the Zeeman term that depends on the total field $\vekc{Z} \equiv
(g/2) \, \bohrmag \, \vekc{B}_\| + \vekc{X}_\|$. Considering the
transparent $2 \times 2$ model $\Hcb$ turns out to be useful because
it captures the important physical trends, even though it does not
include certain details [such as nonparabolicity and corrections to
the Dresselhaus spin splitting (\ref{eq:ham-cb:dressel})] that are
included in $\HK$ and that would be required for a quantitatively
reliable account of specific experiments.  The relation between the
simplified Hamiltonian $\Hcb$ and the more complete Hamiltonian
$\HK$ is discussed in more detail, e.g., in Ref.~\cite{win03}.

From now on, the direction of $\vekc{Z}$ is chosen as the
spin-quantization axis for convenience.  We will be interested in
terms at most quadratic in $\vekc{k}_\|$ and linear in $\Bc_\|$,
where the latter is justified for weak fields $\Bc_\|$, i.e., when
the well width $w$ is smaller than the magnetic length
$\sqrt{\hbar/|e\Bc_\||}$.  Then the Hamiltonian $\Hcb$ becomes
\cite{misc:gfak}
\begin{widetext}
  \begin{subequations}
    \label{eq:ham-cb:rot}
    \begin{align}
      \Hcb & = \frac{\hbar^2  \kc^2}{2 m} + V(z)
      + d \, k_z^2 \left[ \left( \kc_x \sin \varphi_\Zc
          + \kc_y \cos \varphi_\Zc \right) \sigma_x
        - \left( \kc_x \cos \varphi_\Zc
          - \kc_y \sin \varphi_\Zc \right) \sigma_z \right]
      + \Zc \, \sigma_z
      + e \Ec_z z \,\, , \\
      & = \frac{\hbar^2  \kc_z^2}{2 m} + V(z)
      + \frac{\hbar^2}{2 m} \left( \vekc{k}_\| - \kk_0 \right)^2
      - \frac{\hbar^2 k_0^2}{2 m}
      + \Zc \, \sigma_z
      + e \Ec_z z \,\, ,
      \label{eq:ham-cb:rot:k0}
    \end{align}
  \end{subequations}
\end{widetext}
with
\begin{equation}\label{eq:FM:k0}
  \kk_0  = \frac{m}{\hbar^2} \, d \, k_z^2 \left[
  \begin{pmatrix}
    \cos \varphi_\Zc \\ - \sin \varphi_\Zc
  \end{pmatrix}
  \sigma_z -
  \begin{pmatrix}
    \sin \varphi_\Zc \\ \cos \varphi_\Zc
  \end{pmatrix}
  \sigma_x \right] \,\, ,
\end{equation}
and $\varphi_\Zc$ is the angle between the total Zeeman field
$\vekc{Z}$ and the crystallographic direction $[100]$.  The
usefulness of writing $\Hcb$ as in Eq.\ (\ref{eq:ham-cb:rot:k0})
will become clear later on.

For $\Ec_z = 0$ and $\Bc_\| = 0$, the Hamiltonian is
\begin{equation}
  \Hcb = \Hcb^{(0)} + \Hcb_\Zc^{(0)}
   + \Hcb_\mathrm{D}^{(1)} \,\, ,
\end{equation}
with
\begin{subequations}
  \begin{eqnarray}
    \label{eq:ham-cb:0}
    \Hcb^{(0)} &=& \frac{\hbar^2 k_z^2}{2 m} + V(z) \,\, , \\*
    \label{eq:ham-cb:0Z}
    \Hcb_\Zc^{(0)} & = &
    \Zc\, \sigma_z \,\, , \\
    \label{eq:ham-cb:1:D}
    \Hcb_\mathrm{D}^{(1)} &=&
     \frac{\hbar^2}{2 m} \left( \kk_\| - \kk_0 \right)^2
     - \frac{\hbar^2k_0^2}{2 m} \,\,  .
  \end{eqnarray}
\end{subequations}
The eigenstates of $\Hcb^{(0)} + \Hcb_\Zc^{(0)}$ are
$\ket{\nu\sigma^{(0)}} \equiv \ket{\nu} \otimes \ket{\sigma}$, with
associated eigenvalues $E_{\nu\sigma,\vek{0}} \equiv E_\nu^{(0)} +
\sigma \,\Zc$, where $\sigma = \pm 1$.  Treating
$\Hcb_\mathrm{D}^{(1)}$ in first order, the subband dispersions
are
\begin{widetext}
\begin{equation}
  \label{eq:disp-cb}
  E_{\nu \sigma, \kk_\|}
  = E_\nu^{(0)} + \frac{\hbar^2 k_\|^2}{2m}
   + \sigma \sqrt{d \, \av{k_z^2} \, k_\|^2
   + \Zc \left[\Zc - 2 d \, \av{k_z^2} \,
     (k_x \cos \varphi_\Zc - k_y \sin \varphi_\Zc) \right]} \quad ,
\end{equation}
\end{widetext}
with $\av{k_z^2} = \braket{\nu | k_z^2 | \nu}$.  For $\Zc = 0$,
the spectrum $E_{\nu\sigma,\kk_\|}$ satisfies time-reversal symmetry,
$E_{\nu \sigma,- \kk_\|} = E_{\nu \sigma,\kk_\|}$.  For $\Zc \ne
0$, the relation $E_{\nu \sigma, - \kk_\|} \ne E_{\nu \sigma,\kk_\|}$
reflects broken time-reversal symmetry.  The latter is a prerequisite
for the magnetoelectric effect, as discussed above.

\begin{figure}[b]
  \includegraphics[width=\linewidth]{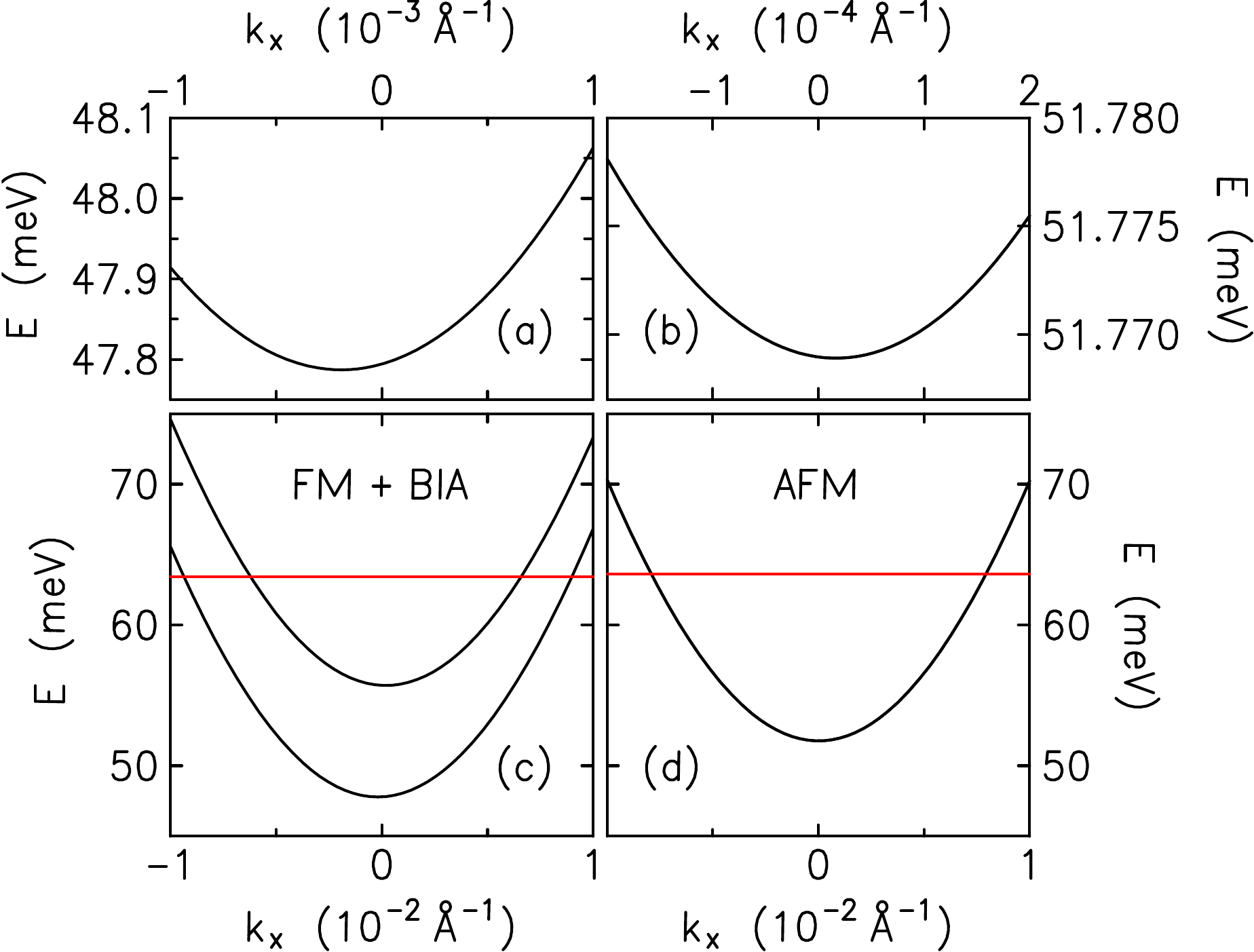}
  \caption{Dispersion of the lowest electron subbands in a quantum
  well with width $w = 150\,$\AA\ and barrier height $V_0 = 1.2\,$eV
  for $\Ec_z = \Bc_\| = 0$. The red lines indicate the Fermi
  energy for an electron density $N_s = 1.0 \times
  10^{11}\,$cm$^{-2}$. Left column [panels (a), (c)]: Dispersion
  $E_{0\pm, k_x}$ for ferromagnetic InSb with $\Xc_x =
  8\,$meV and BIA.  Right column [panels (b), (d)]: Dispersion $E_{0
  k_x}$ for a diamond antiferromagnet with InSb band-structure
  parameters (without BIA) and $\mathcal{\SX}_x = 50\,$meV. The
  upper panels (a), (b) show a zoom-in of the same dispersion as in
  panels (c), (d) near $k_\| = 0$.}
  \label{fig:e-disp}
\end{figure}

Figures~\ref{fig:e-disp}(a) and~\ref{fig:e-disp}(c) illustrate the
dispersion (\ref{eq:disp-cb}) for a quasi-2D electron system in a
ferromagnetic InSb quantum well with $\Xc_x = 8\,$meV, width
$w = 150\,$\AA, and with an electron density $N_s = 1.0 \times
10^{11}\,$cm$^{-2}$. The numerical calculations in
Fig.~\ref{fig:e-disp} are based on the more accurate multiband
Hamiltonian $\HK$ introduced above. Band parameters for InSb are
taken from Ref.~\cite{win03}.

\subsection{$\Ec$-induced magnetization}
\label{sec:EtoM}

In this section, we evaluate the equilibrium magnetization induced
by an electric field $\Ec_z$ using perturbation theory, which is
justified by the fact that the $\Ec$-induced magnetization is
commonly a small fraction of the spontaneous magnetization
$\mathcal{M}^\mathrm{s}_\|$ in the system. (See
Table~\ref{tab:susceptibilities} below for typical numbers.)  We
start from the Hamiltonian (\ref{eq:ham-cb:rot}).  Specializing to
$\Bc_\| = 0$ yields
\begin{equation}
  \label{eq:ham-cb:EtoM:tot}
  \Hcb_\EtoM = \Hcb^{(0)} + \Hcb_\Zc^{(0)}
   + \Hcb_\mathrm{D}^{(1)} + \Hcb_\Ec^{(1)} \,\, ,
\end{equation}
with $\Hcb^{(0)}$, $\Hcb_\Zc^{(0)}$, and $\Hcb_\mathrm{D}^{(1)}$
given by Eqs.\ (\ref{eq:ham-cb:0}), (\ref{eq:ham-cb:0Z}), and
(\ref{eq:ham-cb:1:D}), respectively, and
\begin{equation}
  \Hcb_\Ec^{(1)} = e \Ec_z \, z \,\, .
  \label{eq:ham-cb:E}
\end{equation}
Treating the electric field $\Ec_z$ in first-order perturbation
theory, the eigenstates become
\begin{equation}
  \label{eq:eigen:z}
  \ket{\nu \sigma^{(1)}} = \Bigl( \ket{\nu}
  + e\Ec_z \, \sum_{\nu' \ne \nu} c_{\nu'\nu} \ket{\nu'}
  \Bigr) \otimes \ket{\sigma} \; ,
\end{equation}
with expansion coefficients
\begin{equation}
  c_{\nu'\nu} = \frac{\braket{\nu' | z | \nu }}
  {E_\nu^{(0)} - E_{\nu'}^{(0)}} \; .
\end{equation}
It will be seen below that, for the calculation of the
electric-field-induced magnetization, we can ignore the modification
of the states $\ket{\nu\sigma^{(1)}}$ due to $\Hcb_\mathrm{D}^{(1)}$
that yields an effect of higher order in the Dresselhaus coefficient
$d$.  In the following, $\av{\dots}$ denotes the average in the
unperturbed state $\ket{\nu}$, whereas $\Av{\dots}$ denotes the
average in the perturbed state $\ket{\nu\sigma^{(1)}}$ in the
presence of the external field inducing the magnetoelectric
response.

For the equilibrium magnetization (\ref{eq:orb-mag:def}), we need to
evaluate expectation values $\Av{\{ z\, , \vek{v}_\| (\kk_\|) \}}$
using the velocity operator associated with the Hamiltonian
(\ref{eq:ham-cb:EtoM:tot})
\begin{equation}
  \label{eq:vel:k}
  \vek{v}_\| (\kk_\|)
  = \frac{\partial \Hcb_\EtoM}{\partial \, \hbar \kk_\|}
  = \frac{\hbar}{m} \left( \kk_\| - \kk_0 \right) \,\, .
\end{equation}
We get
\begin{widetext}
  \begin{subequations}
    \label{eq:z-v:av}
    \begin{align}
      \Av{\{ z\, , \vek{v}_\| (\kk_\|) \}}
      & = \Av{z} \Av{\vek{v}_\| (\kk_\|)}
          + \left( \Av{\{ z, \vek{v}_\| (\kk_\|) \}}
                   - \Av{z} \Av{\vek{v}_\|(\kk_\|)} \right) \\
      & = \Av{z} \Av{\vek{v}_\| (\kk_\|)}
      - \frac{\hbar}{m} \bigl( \Av{\{ z, \kk_0 \}}
                   - \Av{z} \Av{\kk_0} \bigr) \\
      & = \Av{z} \Av{\vek{v}_\| (\kk_\|)}
      - \frac{d}{\hbar} \left[
      \begin{pmatrix}
        \cos \varphi_\Zc \\ - \sin \varphi_\Zc
      \end{pmatrix}
      \Av{\{ z, k_z^2 - \Av{k_z^2} \} \, \sigma_z}
      - \begin{pmatrix}
        \sin \varphi_\Zc \\ \cos \varphi_\Zc
      \end{pmatrix}
      \Av{\{ z, k_z^2 - \Av{k_z^2} \} \, \sigma_x}
      \right] \label{eq:z-v:av:exact} \\
      & = \Av{z} \Av{\vek{v}_\| (\kk_\|)}
      - \sigma \, \frac{d}{\hbar}
      \begin{pmatrix}
        \cos \varphi_\Zc \\ - \sin \varphi_\Zc
      \end{pmatrix}
      \biggl[ \av{ \{ z, k_z^2 - \av{k_z^2} \}}
        + 2 e\Ec_z \sum_{\nu' \ne \nu} c_{\nu'\nu} \,
      \braket{\nu | \{ z, k_z^2 - \av{k_z^2} \} - \av{z} k_z^2 | \nu'}
      \biggr] \label{eq:z-v:av:perturb} \\[-1ex]
      & = \Av{z} \Av{\vek{v}_\| (\kk_\|)}
      - \sigma \, \frac{2 d \, e\Ec_z}{\hbar}
      \begin{pmatrix}
        \cos \varphi_\Zc \\ - \sin \varphi_\Zc
      \end{pmatrix}
        \sum_{\nu' \ne \nu} c_{\nu'\nu} \, \braket{\nu | \{ z, k_z^2 - \av{k_z^2} \} | \nu'}
      \; . \label{eq:z-v:av:final}
    \end{align}
  \end{subequations}
\end{widetext}
Up to Eq.\ (\ref{eq:z-v:av:exact}), the steps are exact in the sense
that they do not assume a perturbative treatment of $\Hcb_\EtoM$.
To obtain Eq.\ (\ref{eq:z-v:av:perturb}), we exploited the fact that the
eigenstates $\ket{\nu}$ of the unperturbed problem can be chosen such
that all matrix elements in Eq.\ (\ref{eq:z-v:av}) become real.  For the
last line of Eq.\ (\ref{eq:z-v:av}), we assumed that the
potential $V(z)$ is symmetric.  The first term in Eq.\
(\ref{eq:z-v:av:final}) yields a vanishing contribution when summed
over the equilibrium Fermi sea, as it is proportional to the
system's total equilibrium current.  Therefore, a nonzero
magnetization is due to the second term in Eq.\
(\ref{eq:z-v:av:final}), which yields a contribution independent of
the wave vector $\kk_\|$.
Summing over the Fermi sea and assuming a small density
$\Ns$ such that only the lowest subband $\nu = 0$ is
occupied, we obtain for the magnetization~(\ref{eq:orb-mag:def})
\cite{misc:gfak}
\begin{equation}
  \vekc{M}_\|^\mathrm{o} = \mathcal{M}_0 \,\, e\Ec_z w \,\, \lambda_d \,\, \xi (\Zc)
  \begin{pmatrix} \sin\varphi_\Zc \\ \cos \varphi_\Zc
  \end{pmatrix} \,\, ,
  \label{eq:EtoM:final}
\end{equation}
with
\begin{equation}
  \label{eq:me-factor}
  \lambda_d \equiv \frac{l_d}{w} \sum_{\nu' \ne 0}
  \frac{\braket{\nu' | z | 0}
        \braket{0 | \{ z, k_z^2 - \av{k_z^2} \} | \nu'}}
       {E_0^{(0)} - E_{\nu'}^{(0)}} \; ,
\end{equation}
where $l_d\equiv 2 m_0 d/\hbar^2$ is the length scale associated
with Dresselhaus spin splitting~\cite{misc:length}, and
\begin{equation}
  \xi (\Zc) \equiv \left\{ \begin{array}{cs{1em}l@{}}
      \Ds \frac{m}{\pi \hbar^2}\,\frac{\Zc}{\Ns} , &
      \Zc < E_\mathrm{F}^0 \\[2.0ex]
      1, &
      \Zc \ge E_\mathrm{F}^0
    \end{array}\right.
\end{equation}
with $E_\mathrm{F}^0 = (\pi \hbar^2 / m) \, \Ns$
distinguishes between a partially and a fully spin-polarized
(half-metallic) system.  For $\varphi_\Zc = n\pi/2$ ($n$ integer),
the $\Ec_z$-induced magnetization $\vekc{M}_\|^\mathrm{o}$ is
oriented perpendicular to the field $\vekc{Z}$.  More generally, a
clockwise rotation of $\vekc{Z}$ implies a counterclockwise rotation
of $\vekc{M}_\|^\mathrm{o}$.

\begin{figure}[b]
  \includegraphics[width=\linewidth]{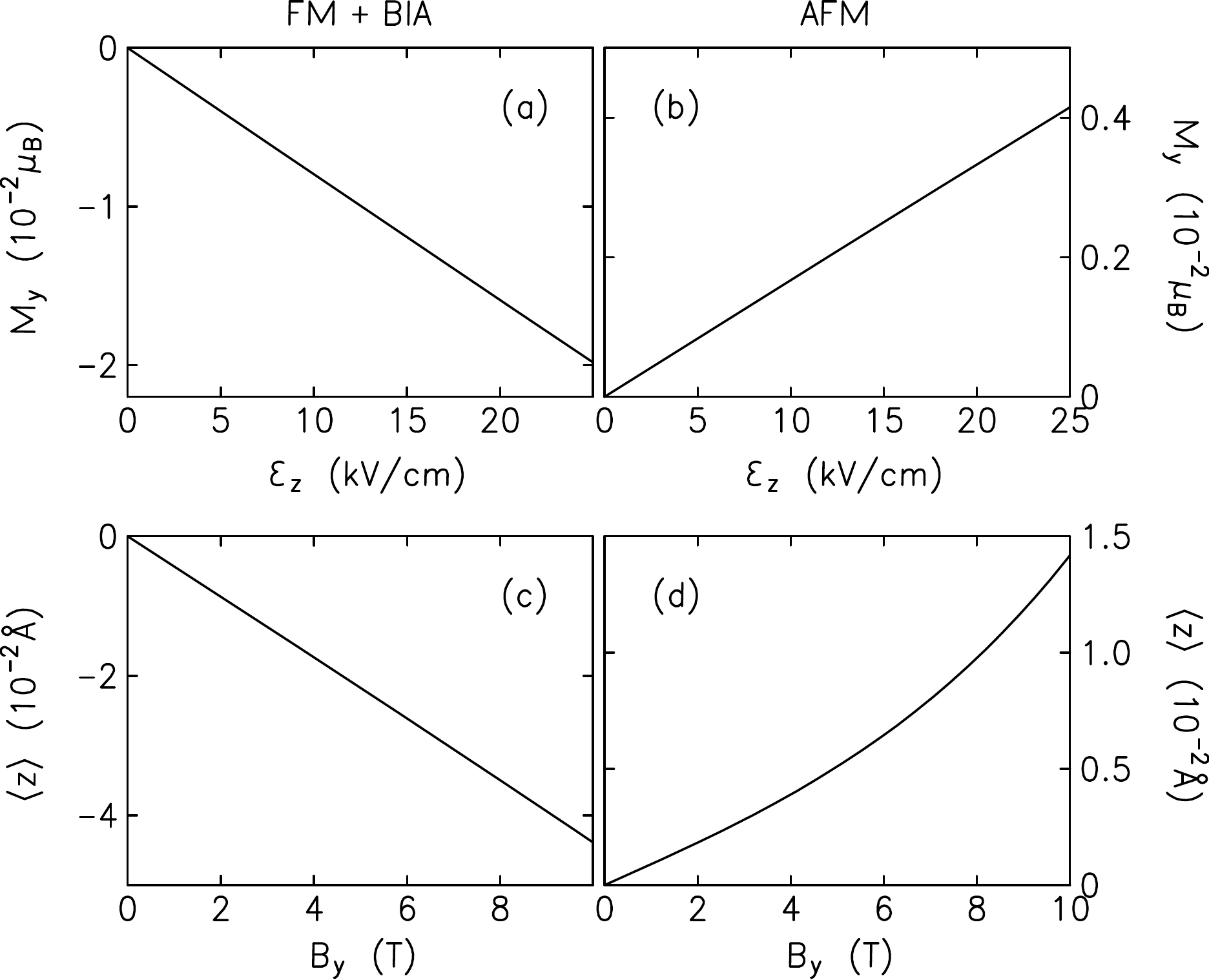}
  \caption{$\Ec_z$-induced orbital magnetic moment per particle
  $\bohrmag\, M_y^\mathrm{o}$ [(a),(b)] and $\Bc_y$-induced
  displacement $\av{z}$ representing the electrostatic polarization
  via Eq.\ (\ref{eq:polToAvZ}) [(c),(d)] in a quantum well with width
  $w = 150\,$\AA, barrier height $V_0 = 1.2\,$eV, and electron
  density $N_s = 1.0 \times 10^{11}\,$cm$^{-2}$.  Left column
  [panels (a), (c)]: Ferromagnetic InSb with $\Xc_x =
  8\,$meV and BIA.  Right column [panels (b), (d)]: Diamond
  antiferromagnet with InSb band-structure parameters (without BIA)
  and $\mathcal{\SX}_x = 50\,$meV.}
  \label{fig:e-mom-150}
\end{figure}

The value obtained for the sum in Eq.\ (\ref{eq:me-factor}) depends
on particularities of the quantum-well confinement. Peculiarly, the
sum vanishes for a parabolic (i.e., harmonic-oscillator)
potential. In contrast, assuming an infinitely deep square well of
width $w$, we get
\begin{equation}
  \label{eq:magic-sum}
  \lambda_d \equiv \frac{\pi^2 - 6}{6 \pi^2} \, \frac{m w \, l_d}{\hbar^2} \; .
\end{equation}
Figure~\ref{fig:e-mom-150}(a) illustrates the $\Ec_z$-induced
orbital magnetic moment per particle for a ferromagnetic InSb
quantum well with width $w = 150\,$\AA\ and electron density $N_s =
1.0 \times 10^{11}\,$cm$^{-2}$.  Results in Fig.~\ref{fig:e-mom-150}
are based on the more accurate multiband Hamiltonian $\HK$.

The magnetization (\ref{eq:EtoM:final}) complements the more trivial
magnetization $\vekc{M}_\Zc^\mathrm{tot} = \vekc{S}_\Zc +
\vekc{M}_\Zc$ that we get already in the absence of a field $\Ec_z$,
which is oriented (anti)parallel to $\vekc{Z}$.  The spin
magnetization $\vekc{S}_\Zc$ is due to an imbalance between spin
eigenstates induced by the Zeeman term (\ref{eq:ham-cb:0Z})
[see Eq.\ (\ref{eq:gen:free-en:pauli}) below].  The
orbital magnetization $\vekc{M}_\Zc$ is due to spin-orbit
coupling.  Just like $\vekc{S}_\Zc$, the orbital contribution
is already present in inversion-symmetric diamond structures, i.e.,
it is a manifestation of spin-orbit coupling beyond the Dresselhaus
term (\ref{eq:ham-cb:1:D}) and beyond the simple $2 \times 2$ model
studied in this section.  Therefore, $\vekc{M}_\Zc$ is always
present in the numerical calculations based on $\HK$.  An analytical
model for $\vekc{M}_\Zc$ based on $\HK$ is discussed in
Appendix~\ref{app:orb-magnet}.

The numerical calculations presented in Fig.~\ref{fig:e-mom-150}
also include higher-order contributions to the $\Ec_z$-induced
magnetization beyond the mechanism underlying the perturbative
calculation yielding Eq.\ (\ref{eq:EtoM:final}).  Such contributions
arise, e.g., from the interplay of Rashba spin-orbit coupling with
the in-plane Zeeman field \cite{misc:rashba}.  The pattern of
the numerically calculated $\Ec_z$-induced
magnetization including, e.g., the dependence on the orientation of the
Zeeman field $\vekc{Z}$, is dictated by symmetry so that the
more complete numerical calculations are in line with the qualitative
predictions of the analytical calculations.

It is illuminating to relate the magnetization (\ref{eq:EtoM:final})
to the equilibrium current distribution (\ref{eq:2D-current:def}).
Using $\phi_\nu (z) \equiv \braket{z|\nu}$, the perturbed wave
functions read
\begin{equation}
  \Phi_{\nu\sigma} (z) \equiv \braket{z | \nu\sigma^{(1)}}
   = \Bigl[ \phi_\nu (z) + e\Ec_z \sum_{\nu' \ne \nu} c_{\nu'\nu} \,
   \phi_{\nu'} (z) \Bigr] \ket{\sigma} \; .
\end{equation}
In the following, we suppress the argument $z$ of $\phi_\nu$ for the
sake of brevity .  Using the velocity operator (\ref{eq:vel:k}), we get in
first order of $\Ec_z$ and $d$
\begin{subequations}
  \begin{align} \hspace*{-0.6em}
  \vek{j}_\| (z, \nu \sigma \kk_\|)
  & = \Phi_{\nu\sigma}^\ast (z) \, \vek{v}_\| (\kk_\|) \, \Phi_{\nu\sigma} (z) \,\, , \\
  & = \braket{\nu\sigma | \vek{v}_\| (\kk_\|) | \nu\sigma} \, |\phi_\nu|^2
  + \sigma \sum_{\nu' \ne \nu} \vek{\kappa}_{\nu'\nu} \, \phi_\nu^\ast \phi_{\nu'}
  \nonumber \\*[-1ex] & \hspace{1em} {}
  + e\Ec_z \sigma \!\!\sum_{\nu',\nu''} \!
  \bigl[ \left( c_{\nu''\nu'} \, \vek{\kappa}_{\nu'\nu}
    + \vek{\kappa}_{\nu''\nu'} c_{\nu'\nu} \right) \phi_\nu^\ast \phi_{\nu''}
  \nonumber \\*[-1ex] & \hspace{7.0em} {}
  + c_{\nu''\nu}^\ast \vek{\kappa}_{\nu'\nu} \, \phi_{\nu''}^\ast \phi_{\nu'} \bigr]
  \,\, ,   \label{eq:EtoM:curr}
\end{align}
\end{subequations}
with matrix elements (spin $\sigma=+$)
\begin{equation}
  \label{eq:EtoM:curr:mel}
  \vek{\kappa}_{\nu'\nu}
  \equiv \frac{\hbar}{m} \, \braket{\nu' {+} | \kk_0 | \nu {+}}
  = \frac{d}{\hbar}
      \begin{pmatrix}
        \cos \varphi_\Zc \\ - \sin \varphi_\Zc
      \end{pmatrix}
       \, \braket{\nu' | k_z^2 | \nu} \; .
\end{equation}
In thermal equilibrium, the first term in Eq.\ (\ref{eq:EtoM:curr})
averages to zero in Eq.\ (\ref{eq:2D-current:def:av}).  The
remaining terms are independent of $\kk_\|$ so that, for $\Zc \ne
0$, they do not average to zero in Eq.\
(\ref{eq:2D-current:def:av}).

\begin{figure}
  \includegraphics[width=\linewidth]{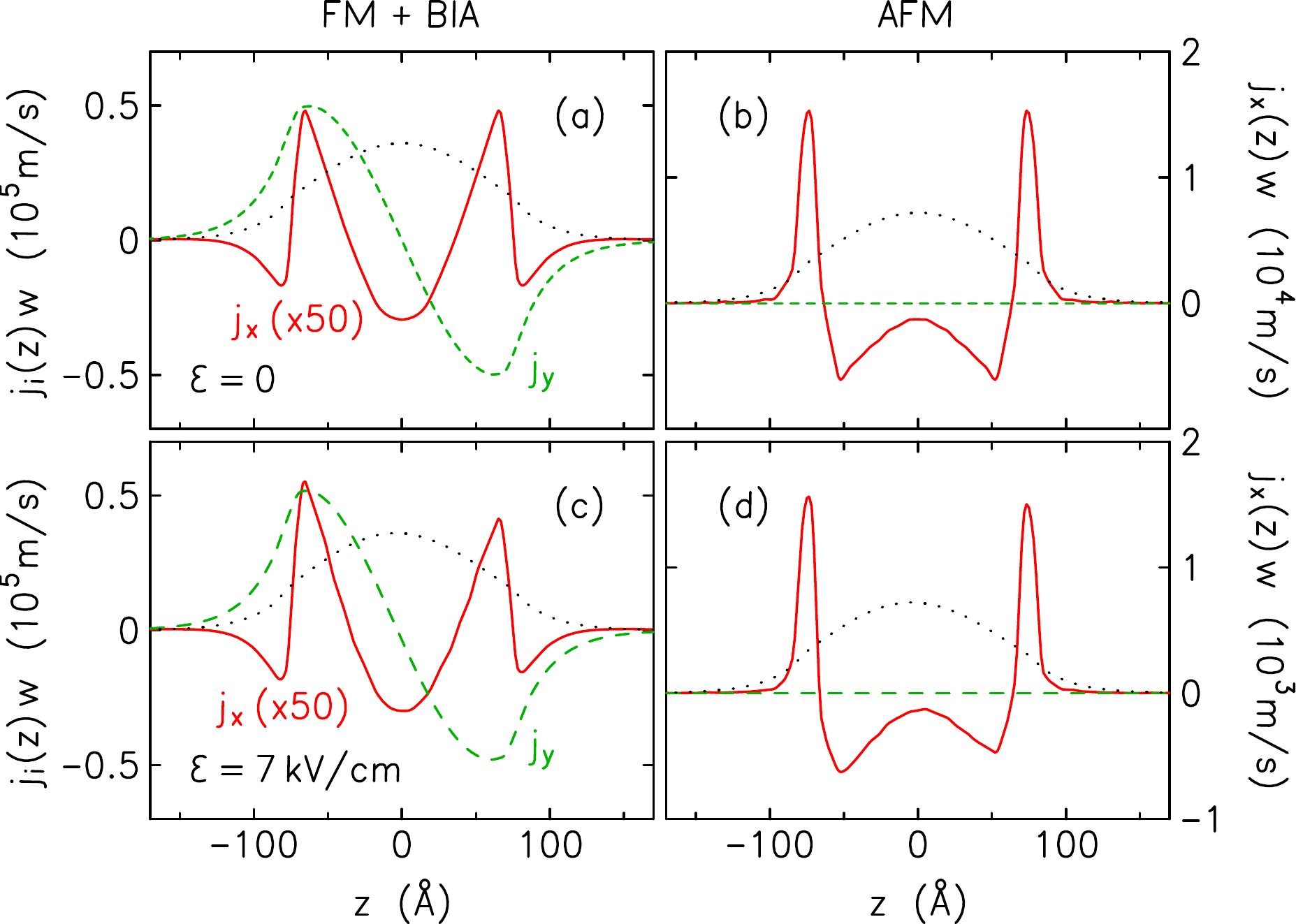}
  \caption{Equilibrium current distribution $\vek{j}_\| (z)$ in a
  quantum well with width $w = 150\,$\AA, barrier height $V_0 =
  0.12\,$eV, and electron density $N_s = 1.0 \times
  10^{11}\,$cm$^{-2}$.  Left column [panels (a), (c)]: Ferromagnetic
  InSb with $\Xc_x = 8\,$meV and BIA.  Right column [panels
  (b), (d)]: Diamond antiferromagnet with InSb band-structure
  parameters (without BIA) and $\mathcal{\SX}_x = 50\,$meV.  Upper
  row [panels (a), (b)]: symmetric quantum well ($\Ec_z = 0$).
  Lower row [panels (c), (d)]: tilted quantum well ($\Ec_z =
  7\,$kV/cm).  In each panel, the dotted line shows for comparison
  the charge distribution $\rho(z)$ (arbitrary units).  In both
  configurations (FM+BIA and AFM), the applied electric field
  distorts the, at zero field purely quadrupolar,
  equilibrium-current-density component $j_x$, thus inducing a
  finite magnetization in $y$ direction
  [Figs.~\ref{fig:e-mom-150}(a) and~\ref{fig:e-mom-150}(b)].}
  \label{fig:e-vel}
\end{figure}

The matrix elements contributing to the second term in Eq.\
(\ref{eq:EtoM:curr}) are nonzero independently of an electric field
$\Ec_z$ (provided the product $\nu' \nu$ is also even).  For
$\nu'=2$, we get equilibrium currents proportional to $\phi_0 (z)
\phi_2 (z)$ that give rise to a magnetic quadrupole $\vekc{Q}$ [Eq.\
(\ref{eq:quadmom:def}) below].  The quadrupolar currents are
illustrated in numerical calculations for a quantum well with finite
barriers and using the more complete multiband Hamiltonian $\HK$,
see Figs.~\ref{fig:e-vel}(a) and~\ref{fig:e-vel}(c).  The
quadrupolar currents and the magnetic quadrupole $\vekc{Q}$ are
indicative of orbital antiferromagnetic order that is induced
parallel to the Zeeman field $\vekc{Z}$ by the interplay of
$\vekc{Z}$, the Dresselhaus term (\ref{eq:ham-cb:1:D}), and
confinement [the potential $V(z)$]. The quadrupolar currents are odd
both under spatial inversion and time inversion, consistent with the
general discussion of antiferromagnetic order in
Sec.~\ref{sec:neel-op} below.  The orbital antiferromagnetic order
can be quantified using the N\'eel operator $\vek{\tau}$ defined
below [Eq.\ (\ref{eq:AFM:op})].  The Hamiltonian
(\ref{eq:ham-cb:EtoM:tot}) (with $\Ec_z = 0$) yields a nonzero
expectation value
\begin{equation}
  \label{eq:EtoM:curr:afm}
  \av{\vek{\tau}} = 2 \pi \, q_\tau \, d  \, \Ns \, \xi (\Zc)
  \begin{pmatrix} \cos\varphi_\Zc \\ \sin \varphi_\Zc \end{pmatrix}
  \sum_{\nu' \ne 0} \frac{|\braket{\nu' | k_z^2 | 0}|^2}
  {E_0^{(0)} - E_{\nu'}^{(0)}} \; ,
\end{equation}
where we assumed, as before, that only the lowest subband $\nu=0$ is
occupied.  [Here $q_\tau$ is a band-structure parameter whose properties
are discussed in greater detail below Eq.~(\ref{eq:AFM:op}).] As we have
$\av{\vek{\tau}} \parallel \vekc{Z}$ we can interpret such a scenario as
ferrimagnetic order.  This classification proposed here applies, in particular,
to Mn-doped semiconductors such as GaMnAs and InMnSb
\footnote{Drawing conclusions about hole-doped systems such as
GaMnAs and InMnSb from our analytical considerations of quasi-2D
electron systems is possible because itinerant quasi-2D holes are
shown in Sec.~\ref{sec:fm-holes} to exhibit qualitatively similar features
as quasi-2D electrons.}.
It is a peculiarity of an infinitely
deep square well that $\vek{\kappa}_{\nu'\nu} \propto
\delta_{\nu'\nu}$ so that within this model we do not obtain
quadrupolar equilibrium currents and orbital antiferromagnetic order.

The last term in Eq.\ (\ref{eq:EtoM:curr}) (with $\nu''=1$)
describes $\Ec_z$-induced dipolar currents that contribute to the
magnetization [Fig.~\ref{fig:e-vel}(c)].  For $\varphi_\Zc = n\pi/2$
($n$ integer), the quadrupolar and dipolar currents flow
(anti)parallel to the field $\vekc{Z}$, consistent with Eq.\
(\ref{eq:EtoM:final}).  As to be expected, the total current
$\vekc{J}_\| = \int dz \, \vekc{j}_\| (z)$ always vanishes.  The
fact that the coupling of the currents $\vekc{j}_\| (z)$ to a
perpendicular electric field $\Ec_z$ is dissipationless resembles
the Lorentz force.  However, it needs to be emphasized that the
equilibrium currents $\vekc{j}_\|(z)$ and their manipulation via
electric fields are pure quantum effects with no classical analogue.

The numerical calculations for a ferromagnetic quantum well based on
the multiband Hamiltonian $\HK$ and presented in
Figs.~\ref{fig:e-vel}(a) and~\ref{fig:e-vel}(c) assume that the
exchange field $\vekc{X}_\|$ is oriented in $x$ direction.  In this
case, the equilibrium currents $\vekc{j}_\| (z)$ represented by Eq.\
(\ref{eq:EtoM:curr}) are oriented likewise in $x$ direction.  These
currents are complemented by equilibrium currents $\mathcal{j}_y$
representing the orbital magnetization $\vekc{M}_\Zc$
induced by the exchange field $\Xc_z$ and discussed in more
detail in Appendix~\ref{app:orb-magnet}\@.

\subsection{$\Bc$-induced electric polarization}
\label{sec:BtoP}

To calculate the equilibrium electric polarization induced by a
magnetic field $\vekc{B}_\|$, we start again from the Hamiltonian
(\ref{eq:ham-cb:rot}).  Specializing to $\Ec_z = 0$ yields
\begin{equation}
  \label{eq:ham-cb:BtoP}
  \Hcb_\BtoP = \Hcb^{(0)} + \Hcb_\Zc^{(0)}
   + \Hcb_\mathrm{D}^{(1)} + \Hcb_\Bc^{(1)} \,\, ,
\end{equation}
with $\Hcb^{(0)}$, $\Hcb_\Zc^{(0)}$, and
$\Hcb_\mathrm{D}^{(1)}$ given by Eqs.\ (\ref{eq:ham-cb:0}),
(\ref{eq:ham-cb:0Z}), and (\ref{eq:ham-cb:1:D}), respectively, and
[ignoring terms $\mathcal{O} (\Bc_\|^2)$]
\begin{widetext}
\begin{subequations}
  \label{eq:ham-cb:B1}
  \begin{align}
    \Hcb_\Bc^{(1)}
    & = \frac{e\hbar}{2m}
    \left[ \left(\kk_\| - \kk_0 \right) \cdot \vekc{A}
      + \vekc{A} \cdot \left(\kk_\| - \kk_0 \right) \right] \,\, , \\
    & = \frac{e\hbar}{2m} \left[ 2\left(\kk_\| - \Av{\kk_0} \right) \cdot\vekc{A}
  - \left( \kk_0 - \Av{\kk_0} \right) \cdot \vekc{A}
  - \vekc{A} \cdot \left( \kk_0 - \Av{\kk_0} \right) \right] \,\, , \\
    & = \frac{e\hbar}{m} \left(\kk_\| - \Av{\kk_0} \right) \cdot \vekc{A}
    - \frac{e}{\hbar}  \, d \, \Bc_\|
     \left[ \sin ( \varphi_\Zc + \varphi_\Bc ) \, \sigma_z
        + \cos ( \varphi_\Zc + \varphi_\Bc ) \, \sigma_x \right]
      \{ z, k_z^2 - \av{k_z^2} \} \,\, ,
    \label{eq:ham-cb:B1:final}
  \end{align}
\end{subequations}
\end{widetext}
where $\varphi_\Bc$ is the angle between the direction of the applied
magnetic field $\vekc{B}_\|$ and the $[100]$ crystallographic
direction.  The perturbation $\Hcb_\Bc^{(1)}$ yields the
perturbed states
\begin{equation}
  \ket{\nu\sigma^{(1)}} = \ket{\nu\sigma}
  + \sum_{\sigma', \nu'\ne \nu}
  \frac{\Braket{\nu' \sigma' | \Hcb_\Bc^{(1)} | \nu\sigma}}
  {E_\nu^{(0)} - E_{\nu'}^{(0)}} \, \ket{\nu'\sigma'} \,\, .
\end{equation}
We get
\begin{subequations}
  \label{eq:z:av}
  \begin{align}
    \Av{z} & = \av{z}_{\nu\sigma} + 2 \sum_{\nu' \ne \nu} c_{\nu'\nu}
    \Braket{\nu \sigma | \Hcb_\Bc^{(1)} | \nu' \sigma} \,\, , \\
    & = \av{z}_{\nu\sigma} + 2 \sum_{\nu' \ne \nu} c_{\nu'\nu} \biggl[
      \frac{e\hbar}{m} (\kk_\| - \Av{\kk_0}) \cdot \braket{\nu | \vekc{A} | \nu'}
      \nonumber \\* & \hspace{1em} {}
      - \sigma \, \sin (\varphi_\Zc + \varphi_\Bc) \,
      \frac{e}{\hbar}  \mathcal{d} \, \Bc_\|
      \, \braket{\nu | \{ z, k_z^2 - \av{k_z^2} \} | \nu'} \biggr] \,\, .
  \end{align}
\end{subequations}
Here, the first term $\av{z}_{\nu\sigma}$ vanishes for a symmetric
potential $V(z)$.  The first term in the square brackets describes a
$\kk_\|$-dependent shift \cite{smr95a, lay97, kam12a, mue15} that
yields a vanishing contribution to $\mathcal{P}_z^\mathrm{e}$ when
summed over the equilibrium Fermi sea.  Therefore, a nonzero
polarization is due to the second term in the square brackets, which
yields a contribution independent of the wave vector $\kk_\|$.
Summing over the Fermi sea, we obtain \cite{misc:gfak}
\begin{equation}
  \label{eq:BtoP:final}
  \mathcal{P}_z^\mathrm{e} = \mathcal{P}_0 \,\, \bohrmag \Bc_\| \,\, \lambda_d
  \,\, \xi (\Zc)\,\, \sin (\varphi_\Zc + \varphi_\Bc)   \,\, ,
\end{equation}
where $\lambda_d$ is given by Eq.\ (\ref{eq:me-factor}).  We see
that the induced magnetoelectric effects are most pronounced when
$\sin (\varphi_\Zc + \varphi_\Bc) = \pm 1 \equiv \zeta$.  This
situation is realized in ferromagnetic systems when
$\vekc{X}_\| \parallel \vek{\hat{x}} \equiv [100]$ and
$\vekc{B}_\| \parallel \vek{\hat{y}} \equiv [010]$.  Here the
magnetization scales linearly with $\vekc{B}_\|$ [for $(g/2)
\bohrmag \Bc_\| \ll \Xc$].  In paramagnetic systems with
$\Xc=0$ and $\vekc{Z} = (g/2) \bohrmag \vekc{B}_\|$, we have
$\zeta = \pm 1$ when $\vekc{B}_\| \parallel [110]$.  In this case
the polarization $\mathcal{P}_z^\mathrm{e}$ depends quadratically on
$\vekc{B}_\|$, consistent with Eq.\ (\ref{eq:moments:gen-def}). Thus
the system exhibits a higher-order magnetoelectric
effect \cite{asc68, gri94} that is the nondissipative counterpart of
the previously discussed magnetically induced electric polarization
in a multi-quantum-well system \cite{gor93, gor94, gor09}.
Figure~\ref{fig:e-mom-150}(c) illustrates the polarization
(\ref{eq:BtoP:final}) for a ferromagnetic InSb quantum well.

The mechanism for the $\vekc{B}$-induced polarization can be
understood as follows: the vector potential $\vekc{A}$ of a magnetic
field $\vekc{B}_\|$ has previously been used as a tool to manipulate
the charge density $\rho(z)$ in quasi-2D systems such as
semiconductor quantum wells. Ordinarily, a field $\vekc{B}_\|$
makes the charge distribution $\rho(z)$ bilayer-like by pushing
$\rho(z)$ towards the barriers, but $\rho(z)$ still preserves
the mirror symmetry of a symmetric quantum well \cite{smr95a, lay97,
kam12a, mue15}.  This effect stems from terms quadratic in
$\vekc{A}$ that we have ignored in the above analytical model.  In a
low-symmetry configuration [indicated here by the presence of the
Dresselhaus term $\Hcb_\mathrm{D}$ given in
Eq.\ (\ref{eq:ham-cb:dressel})], odd powers of the vector potential
$\vekc{A}$ can change $\rho(z)$ in a way that no longer preserves the
mirror symmetry of the confining potential $V(z)$. This effect resembles
the Coulomb force, where the scalar potential is replaced by the vector
potential $\vekc{A}$.  However, it needs to be emphasized that, similar
to Landau diamagnetism, we have here a pure quantum effect; it has no
classical analogue.  This effect is orbital in nature; it does not require a
spin degree of freedom. For example, it exists also in spinless 2D hole
systems that have a purely orbital Dresselhaus term.

\subsection{Magnetoelectric contribution to the free energy}
\label{sec:free-en}

We evaluate the change $\delta F$ in the free-energy density due to the
presence of both $\Hcb_\Ec^{(1)}$ [Eq.\ (\ref{eq:ham-cb:E})] and
$\Hcb_\Bc^{(1)}$ [Eq.\ (\ref{eq:ham-cb:B1:final})] as
\begin{equation}
  \label{eq:ME:free-en:def}
  \delta F = \frac{1}{w} \sum_{\nu,\sigma} \int \frac{d^2 k_\|}{(2\pi)^2}
  \; f (E_{\nu\sigma \kk_\|})
  \; \Av{\Hcb_\Ec^{(1)} + \Hcb_\Bc^{(1)}}_{\nu\sigma\kk_\|}
  \, ,
\end{equation}
using second-order perturbation theory
\begin{widetext}
\begin{subequations}
  \label{eq:ME:free-en:matrix-elmnt}
  \begin{align}
    \Av{\Hcb_\Ec^{(1)} + \Hcb_\Bc^{(1)}}_{\nu\sigma\kk_\|}
    & = 2 \Re \sum_{\nu'\ne\nu} \frac{\braket{\nu| \Hcb_\Ec^{(1)} | \nu'}
    \braket{\nu'| \Hcb_\Bc^{(1)} | \nu}}{E_\nu^{(0)} - E_{\nu'}^{(0)}} \\
    & = 2 \Re \sum_{\nu'\ne\nu} \frac{
    \braket{\nu| e \Ec_z \, z | \nu'}
    \Braket{\nu'| \frac{e\hbar}{m} \left(\kk_\parallel - \Av{\vekc{k}_0} \right)
    \cdot \vekc{A}
    - \frac{e}{\hbar} \, \sigma \mathcal{d} \, \Bc_\parallel \,
     \sin ( \varphi_\Zc + \varphi_\Bc ) \, \{ z, k_z^2 - \av{k_z^2} \} | \nu}}
     {E_\nu^{(0)} - E_{\nu'}^{(0)}} \; ,
  \end{align}
\end{subequations}
\end{widetext}
where we ignored terms $\mathcal{O}(\Ec_z^2)$ and
$\mathcal{O}(\Bc_\|^2)$.  When averaging over all occupied states,
the terms $\propto \vekc{A}$ drop out.  Using Eq.\
(\ref{eq:me-factor}), we get \cite{misc:gfak}
\begin{equation}
  \label{eq:free-en:final}
  \delta F = \Ns \,\, e \Ec_z \,\, \bohrmag \Bc_\| \,\, \lambda_d \,\,
  \xi (\Zc) \, \sin ( \varphi_\Zc + \varphi_\Bc) \; ,
\end{equation}
consistent with Eqs.\ (\ref{eq:EtoM:final}) and
(\ref{eq:BtoP:final}).  Hence, within the present model, we have
$\mathcal{P}_z = \mathcal{P}_z^\mathrm{e}$ and $\vekc{M}_\| =
\vekc{M}_\|^\mathrm{e}$.

The expression (\ref{eq:free-en:final}) can be written as a sum of
terms of the type appearing in the third line of the general
expansion (\ref{eq:Fdens}).  More specifically, we find
\begin{subequations}\label{eq:expandFmag}
\begin{equation}
\delta F = -\alpha_{z x}\,\Ec_z \Bc_x -\alpha_{z y}\,\Ec_z \Bc_y
- \beta_{z x y}\, \Ec_z \Bc_x \Bc_y \quad ,
\end{equation}
with
\begin{eqnarray}
  \vek{\alpha}_{z \|} & \equiv
  & \begin{pmatrix} \alpha_{zx} \\ \alpha_{zy} \end{pmatrix}
  = -e \bohrmag\, \lambda_d \times \left\{ \! \begin{array}{@{}>{\Ds}cl@{}}
      \frac{m}{\pi\hbar^2} \, \Xc
      \begin{pmatrix} \sin \varphi_\Xc \\
        \cos \varphi_\Xc \end{pmatrix} ,
      & \mathcal Z < E_\mathrm{F}^0 \\[2ex]
      \Ns \begin{pmatrix} \sin \varphi_\mathcal{Z} \\
        \cos \varphi_\mathcal{Z} \end{pmatrix} ,
      & \mathcal Z \ge E_\mathrm{F}^0
    \end{array} \right. \nonumber \\ && \label{eq:alphaVec} \\
  \label{eq:beta} \hspace{-1ex}
  \beta_{z x y} & = & \left\{ \begin{array}{>{\Ds}cl}
      -e \bohrmag^2\, \lambda_d\, \frac{g m}{\pi\hbar^2}\, ,
      & \mathcal Z < E_\mathrm{F}^0 \\[1ex]
      0 \, , & \mathcal Z \ge E_\mathrm{F}^0 \; .
    \end{array} \right.
\end{eqnarray}
\end{subequations}
Clearly, $\alpha_{i j} \ne 0$ requires spontaneous ferromagnetic
order due to a finite exchange field $\Xc_\|$ or full spin
polarization (i.e., half-metallicity), and the
particular form of the tensor $\alpha_{ij}$ with two nonzero entries
$\alpha_{zx}$ and $\alpha_{zy}$ is consistent with the magnetic
point group symmetry $22'2'$ of a ferromagnetic
symmetric quantum well on a zincblende $(001)$ surface.  A tensor
$\alpha_{i j} \ne 0$ will generally also facilitate higher-order
terms of the type $\propto \gamma_{ijk}$ in Eq.\
(\ref{eq:Fdens}). In contrast, $\beta_{i j k} \ne 0$ occurs even in
paramagnets, which is consistent with basic symmetry considerations
\cite{sch73, asc68, gri94, riv94, wat18a} as zincblende structures
allow for piezoelectricity.

\begin{figure}
  \includegraphics[width=0.8\linewidth]{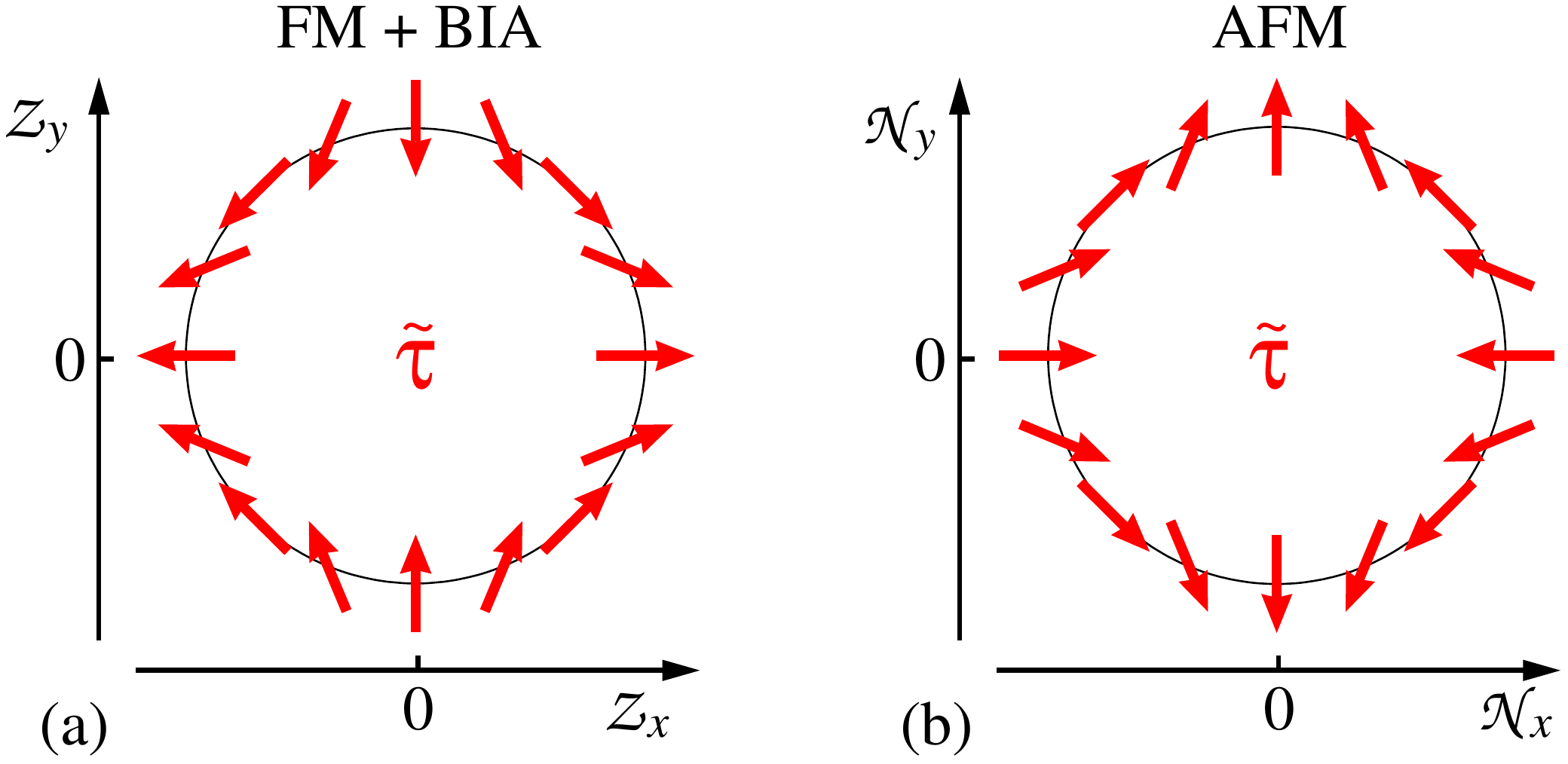}
  \caption{(a) Angular dependence of the orientation of the
  magnetoelectric vector $\tilde{\vek{\tau}}$ [Eq.\
  (\ref{eq:free-en:ME:toroid:all})] on the orientation of the Zeeman
  field $\vekc{Z}$ in ferromagnets with BIA.  (b) Angular dependence
  of the orientation of the magnetoelectric vector
  $\tilde{\vek{\tau}}$ [Eq.\ (\ref{eq:AFM:free-en:toroid})], which is
  parallel to the vector $-\vekc{k}_0$ [Eq.\ (\ref{eq:AFM:k0})], on the
  orientation of the N\'eel vector $\vekc{N}$.  The orientation of
  $\tilde{\vek{\tau}}$ in (b) is antiparallel to the orientation of
  $\tilde{\vek{\tau}}$ in (a) \cite{misc:minus}.}
  \label{fig:orient}
\end{figure}

The magnetoelectric contribution (\ref{eq:free-en:final}) to the
free energy can also be expressed as \cite{misc:toroid}
\begin{subequations}
\label{eq:free-en:ME:toroid:all}
\begin{equation}
\label{eq:free-en:ME:toroid}
\delta F = - \tilde{\vek{\tau}} \cdot (\Ec_z \hat{\vek{z}} \times \vekc{B}_\|)
\end{equation}
in terms of a magnetoelectric vector
\begin{equation}\label{eq:free-en:ME:toroid:vec}
\tilde{\vek{\tau}} = \Ns \, e \bohrmag\, \lambda_d\,
\xi(\Zc) \begin{pmatrix} \cos \varphi_\Zc \\ -\sin \varphi_\Zc
\end{pmatrix} \quad .
\end{equation}
\end{subequations}
The angular dependence of the magnetoelectric effect is governed by
the orientation of the vector $\tilde{\vek{\tau}}$, which in turn is
determined by the orientation of the Zeeman field $\vekc{Z}$.
In particular, there is a one-to-one correspondence between the
orientation of the vector $\vekc{Z}$ in position space and the vector
$\kk_0$ in reciprocal space; specifically the part $\kk_0^{(z)}$ of
$\kk_0$ proportional to $\sigma_z$ that turned out to be relevant for
the magnetoelectric effect in the above analysis.  This vector
$\kk_0^{(z)}$ is collinear with the vector $\tilde{\vek{\tau}}$.
Figure~\ref{fig:orient}(a) shows the relation between the
orientation of $\vekc{Z}$ and the orientation of $\tilde{\vek{\tau}}$.
A similar pattern exists for the current-induced spin magnetization (cf.\ Appendix~\ref{app:current-induced-magnetization})
in systems with Dresselhaus spin-orbit coupling
(\ref{eq:ham-cb:dressel}) for the orientation of the induced spin
polarization as a function of the orientation of an in-plane
electric field \cite{win03}.

The vector $\Ec_z \hat{\vek{z}} \times \vekc{B}_\|$ in Eq.\
(\ref{eq:free-en:ME:toroid}) is a toroidal vector, i.e., it is odd
under both space inversion and time reversal \cite{dub90}.  On the
other hand, Eq.\ (\ref{eq:free-en:ME:toroid:vec}) shows that the vector
$\tilde{\vek{\tau}}$ transforms like a magnetic field, i.e., it is
even under space inversion and odd under time reversal.  The
different transformational properties of the vectors $\Ec_z
\hat{\vek{z}} \times \vekc{B}_\|$ and $\tilde{\vek{\tau}}$ in Eq.\
(\ref{eq:free-en:ME:toroid}) reflect the broken space-inversion
symmetry in a zincblende structure.

The term $\delta F \propto \Ec_z\, \Bc_\|$ in Eq.\
(\ref{eq:free-en:final}) is generally complemented by a second
magnetoelectric term $\propto \Ec_z\, \Bc_\|$.  This is because the
Hamiltonian $\Hcb$ also includes a term
\begin{equation}
  \label{eq:ME:bulk}
  \Hcb_\mathcal{EB} = b \, e \Ec_z \, \bohrmag
  (\Bc_x \sigma_y + \Bc_y \sigma_x) \; ,
\end{equation}
characterizing the bulk zincblende structure that underlies the
quasi-2D systems studied here.  The prefactor $b$ is given in Eq.\
(7.5) of Ref.\ \cite{win03} in terms of momentum matrix elements and
energy gaps appearing in the larger Hamiltonian $\HK$, yielding $b =
-221\,$\AA/eV for InSb and $-1.36\,$\AA/eV for GaAs.  The term
(\ref{eq:ME:bulk}) produces a second magnetoelectric term in
the free energy,
\begin{equation}
  \delta F_\mathrm{bulk} = \Ns\,\, e\Ec_z \,\, \bohrmag \Bc_\|\,
  \frac{b}{w}\,\,  \xi (\Zc) \, \sin ( \varphi_\Zc + \varphi_\Bc) \; ,
\end{equation}
that complements $\delta F$ in Eq.\ (\ref{eq:free-en:final}).  Their
ratio is given by
\begin{equation}\label{eq:bulk2Dratio}
  \left| \frac{\delta F_\mathrm{bulk}}{\delta F}\right| \equiv
  \left| \frac{b}{w \lambda_d}\right| = \frac{6\pi^2}{\pi^2-6}\,
  \frac{\hbar^2 b}{m w^2 l_d} \; ,
\end{equation}
where the expression on the far r.h.s.\ of
Eq.\ (\ref{eq:bulk2Dratio}) is obtained using
Eq.\ (\ref{eq:me-factor}) for a hard-wall confinement $V(z)$. This
ratio evaluates to $9300 / (w[\mathrm{\AA}])^2$ in InSb and $330 /
(w[\mathrm{\AA}])^2$ in GaAs, and it is consequently much smaller
than $1$ for typical quantum-well widths $w \gtrsim
150\,$\AA. Experimental signatures of an $\Ec_z\, \Bc_\|$-type
magnetoelectric coupling have recently been observed for charge
carriers in deformed donor bound states~\cite{web18}.

\subsection{Magnetic quadrupole moment}
\label{sec:fm:qmag}

The magnetoelectric tensor $\alpha_{ij}$ behaves under symmetry
transformations like a magnetic quadrupole moment \cite{spa08},
i.e., both of these second-rank material tensors require broken
space-inversion symmetry and broken time-reversal symmetry and these
tensors share the same pattern of nonzero components.  Similar to
the magnetization $\vekc{M} = - \partial F / \partial \vekc{B}$, the
components of the magnetic quadrupole moment can be obtained from
the free energy density $F$ via the relations~\cite{raa05}
\begin{equation}
  \label{eq:quadmom:def}
  \mathcal{Q}_{ij}
  = - 2 \frac{\partial F}{ \partial (\nabla_i \mathcal{B}_j)} \; ,
\end{equation}
where $\nabla_i \mathcal{B}_j$ denotes field gradients.  [Note that,
for the purpose of discussing the magnetic quadrupole moment,
$\vekc{B}_\|\equiv \vekc{B}_\|(\rr)$ necessarily denotes an
inhomogeneous magnetic field, in contrast to the other parts of this
Article where $\vekc{B}_\|$ is assumed to be homogeneous.]  For the
quasi-2D systems studied here, $\mathcal{Q}_{ij}$ is the sum of two
contributions
\begin{equation}
  \mathcal{Q}_{ij} = \mathcal{Q}_{ij}^\mathrm{e}
   + \mathcal{Q}_{ij}^\mathrm{q} \,\, ,
\end{equation}
with
\begin{subequations}
\begin{align}
  \mathcal{Q}_{ij}^\mathrm{e} & = -\frac{2}{w} \sum_{n} \int
  \frac{d^2 k_\|}{(2\pi)^2} \; f (E_{n \kk_\|}) \, \frac{\partial
  E_{n\kk_\|}}{\partial (\nabla_i \mathcal{B}_j)} \; , \\
  \mathcal{Q}_{ij}^\mathrm{q} & = -\frac{2}{w} \sum_{n} \int
  \frac{d^2 k_\|}{(2\pi)^2} \; E_{n \kk_\|} \, \frac{\partial
  f(E_{n\kk_\|})}{\partial (\nabla_i \mathcal{B}_j)} \; ,
\end{align}
\end{subequations}
that represent the electromagnetic and the quantum-kinetic effects of the
field gradients $\nabla_i \mathcal{B}_j$, respectively. Given that
$\vekc{B}_\|$ generally enters the Hamiltonian $H$ via both the
vector potential $\vekc{A}$ and also via the Zeeman term, the
contribution $\mathcal{Q}_{ij}^\mathrm{e}$ can be split further into
orbital and spin contributions,
\begin{equation}
  \mathcal{Q}_{ij}^\mathrm{e} = \mathcal{Q}_{ij}^\mathrm{o}
  + \mathcal{Q}_{ij}^\mathrm{s} \; ,
\end{equation}
where the spin quadrupole moment
\begin{equation}
  \mathcal{Q}_{ij}^\mathrm{s}
  = - \frac{2}{w} \sum_{n} \int \frac{d^2 k_\|}{(2\pi)^2}
  \; f (E_{n \kk_\|}) \,
  \Big\langle \frac{\partial H}{\partial (\nabla_i \mathcal{B}_j)}
  \Big\rangle_{n \kk_\|}
\end{equation}
vanishes for the Hamiltonian (\ref{eq:ham-cb}) studied here.
For the orbital part
\begin{equation}
  \mathcal{Q}_{ij}^\mathrm{o}
  = - \frac{2}{w} \sum_{n}\! \int\! \frac{d^2 k_\|}{(2\pi)^2}
  \, f (E_{n \kk_\|})\,
  \Big\langle \sum_{l=x,y} \frac{\partial H}{\partial \mathcal{A}_l} \frac{\partial\mathcal{A}_l}{\partial(\nabla_i \mathcal{B}_j)}
  \Big\rangle_{n \kk_\|} ,
\end{equation}
we consider the inhomogenous magnetic field to have the particular form
\begin{equation}
  \vekc{B}_\|(z) = \vekc{B}_\|^0 + z\, \vekc{b}_\|
\end{equation}
with constant vectors $\vekc{B}_\|^0$ and $\vekc{b}_\|$, so that
$\nabla_z \vekc{B}_\| = \vekc{b}_\|$, and we choose the vector
potential
\begin{equation}
  \vekc{A} = \bigl( z\, \vekc{B}_\|^0 + \frack{1}{2}\, z^2\, \vekc{b}_\| \bigr)
   \times \hat{\vek{z}} \; .
\end{equation}
Similar to the discussion in Appendix~\ref{app:mag-mom-op}, this
gauge yields for the components $(\mathcal{Q}_{zx}^\mathrm{o},
\mathcal{Q}_{zy}^\mathrm{o}) \equiv \vekc{Q}_{z \|}^\mathrm{o}$ of
the orbital quadrupole moment
\begin{equation}
  \label{eq:quadmom:def:2d}
  \vekc{Q}_{z \|}^\mathrm{o} = -\frac{e}{w} \sum_n \int
  \frac{d^2 k_\|}{(2\pi)^2} \,\, f (E_{n \kk_\|}) \,
  \hat{\vek{z}} \times \av{\{ z^2\, , \vek{v}_\| (\kk_\|) \}}_{n \kk_\|} \, .
\end{equation}
Similar to the polarization $\mathcal{P}_z$ and magnetization
$\vekc{M}_\|$ discussed in Sec.~\ref{sec:responses}, the orbital
quadrupole moment (\ref{eq:quadmom:def:2d}) avoids the
technical problems arising for these quantities in bulk (3D)
systems.  The orbital quadrupole moment in 3D systems has recently
been discussed in Ref.~\cite{shi18}.

We evaluate the matrix elements $\av{\{ z^2\, , \vek{v}_\| (\kk_\|)
\}}$ similar to Eq.\ (\ref{eq:z-v:av}).  We get in first order
perturbation theory
\begin{widetext}
  \begin{subequations}
    \label{eq:z2-v:av}
    \begin{align}
      \av{\{ z^2\, , \vek{v}_\| (\kk_\|) \}}
      & = \av{z^2} \av{\vek{v}_\| (\kk_\|)}
          + \left( \av{\{ z^2, \vek{v}_\| (\kk_\|) \}}
                   - \av{z^2} \av{\vek{v}_\|(\kk_\|)} \right) \\
      & = \av{z^2} \av{\vek{v}_\| (\kk_\|)}
      - \frac{\hbar}{m} \bigl( \av{\{ z^2, \kk_0 \}}
                   - \av{z^2} \av{\kk_0} \bigr) \\
      & = \av{z^2} \av{\vek{v}_\| (\kk_\|)}
      - \frac{d}{\hbar} \left[
      \begin{pmatrix}
        \cos \varphi_\Zc \\ - \sin \varphi_\Zc
      \end{pmatrix}
      \av{\{ z^2, k_z^2 - \av{k_z^2} \} \, \sigma_z}
      - \begin{pmatrix}
        \sin \varphi_\Zc \\ \cos \varphi_\Zc
      \end{pmatrix}
      \av{\{ z^2, k_z^2 - \av{k_z^2} \} \, \sigma_x}
      \right] \\
      & = \av{z^2} \av{\vek{v}_\| (\kk_\|)}
      - \sigma \frac{d}{\hbar}
      \begin{pmatrix}
        \cos \varphi_\Zc \\ - \sin \varphi_\Zc
      \end{pmatrix}
      \av{\{ z^2, k_z^2 - \av{k_z^2} \}}
      \; .  \label{eq:z2-v:av:final}
    \end{align}
  \end{subequations}
\end{widetext}
In the last step, we kept only terms linear in the Dresselhaus
coefficient $d$.  The first term in Eq.\ (\ref{eq:z2-v:av:final})
yields a vanishing contribution when summed over the equilibrium
Fermi sea, as it is proportional to the system's total equilibrium
current.  Therefore, a nonzero quadrupole moment is due to the
second term in Eq.\ (\ref{eq:z2-v:av:final}), which yields a
contribution independent of the wave vector $\kk_\|$.  Consistent
with the above discussion of equilibrium currents, we have $\av{\{
z^2, k_z^2 - \av{k_z^2} \}} = 0$ for an infinitely deep square
well and $\av{\{ z^2, k_z^2 - \av{k_z^2} \}} =
-\nu-\frac{1}{2}$ for the $\nu$th subband in a harmonic-oscillator
potential.
Summing over the Fermi sea and assuming a small density
$\Ns$ such that only the lowest subband $\nu = 0$ is
occupied, we obtain
\begin{equation}
  \vekc{Q}_{z \|}^\mathrm{o} = \mathcal{Q}_0 \,\, \frac{l_d}{w}
  \, \av{\{ z^2, k_z^2 - \av{k_z^2} \}} \,\, \xi (\Zc)
  \begin{pmatrix} \sin\varphi_\Zc \\ \cos \varphi_\Zc
  \end{pmatrix} \,\, ,
  \label{eq:qmag:final}
\end{equation}
with $\mathcal{Q}_0 = -\bohrmag \Ns$.  The quadrupole
moment $\vekc{Q}_{z \|}^\mathrm{o}$ shows the same dependence on the
orientation of the Zeeman field $\vekc{Z}$ as the magnetization
$\vekc{M}_\|^\mathrm{o}$ [Eq.\ (\ref{eq:EtoM:final})], where
$\vekc{Z}$ may be due to an exchange field $\vekc{X}_\|$ or due to
an external field $\vekc{B}_\|$.  On the other hand, the vector
$\vekc{Q}_{z \|}^\mathrm{o}$ shows a different dependence on the
orientation of $\vekc{Z}$ than is exhibited by the expectation value
(\ref{eq:EtoM:curr:afm}) of the N\'eel operator $\vek{\tau}$.  This
is similar to how a spin magnetization $\vekc{S}$ and the spin
polarization $\av{\vek{\sigma}}$ can show different dependences on
the orientation of an external magnetic field $\vekc{B}$ when the
Zeeman coupling in the field $\vekc{B}$ is characterized by a $g$
tensor. See Eq.\ (\ref{eq:spin-mag:def}) and Ref.~\cite{gra18}.

Alternatively, we can obtain the result (\ref{eq:qmag:final}) by evaluating
the free energy $F$ in the presence of the vector potential $\vekc{A}$
for the field gradient $\vekc{b}_\|$. To first order in $\vekc{b}_\|$
and the Dresselhaus coefficient $d$, we get an energy shift of the
occupied states given by the expectation value of
\begin{widetext}
\begin{equation}
  H_\mathcal{b}^{(1)}
  = \frac{e\hbar}{m} \left(\kk_\| - \av{\kk_0} \right) \cdot \vekc{A}
    - \frac{e}{\hbar}  \, d \, \frac{\mathcal{b}_\|}{2}
     \left[ \sin ( \varphi_\Zc + \varphi_\mathcal{b} ) \, \sigma_z
        + \cos ( \varphi_\Zc + \varphi_\mathcal{b} ) \, \sigma_x \right]
      \{ z^2, k_z^2 - \av{k_z^2} \} \,\, ,
\end{equation}
\end{widetext}
where $\varphi_\mathcal{b}$ is the angle between the direction of
the field gradient $\vekc{b}_\|$ and the $[100]$ crystallographic
direction; compare Eq.\ (\ref{eq:ham-cb:B1}).  When
averaging over all occupied states, the first term $\propto
\vekc{A}$ drops out.  We get
\begin{equation}
  \label{eq:free-en:qmag:final}
  \delta F = -\mathcal{Q}_0 \, \frac{\mathcal{b}_\|}{2} \,
  \frac{l_d}{w} \av{\{ z^2, k_z^2 - \av{k_z^2} \}} \, \xi (\Zc) \,
  \sin ( \varphi_\Zc + \varphi_\mathcal{b}) \, ,
\end{equation}
consistent with Eqs.\ (\ref{eq:quadmom:def}) and (\ref{eq:qmag:final}).

The contribution (\ref{eq:free-en:qmag:final}) to the free energy can
also be expressed as \cite{misc:toroid}
\begin{subequations}
\label{eq:free-en:qmag:toroid:all}
\begin{equation}
\label{eq:free-en:qmag:toroid}
\delta F = - \bar{\vek{\tau}} \cdot (\vek{\nabla}_{\!z} \times \vekc{B}_\|)
\end{equation}
in terms of a vector
\begin{equation}\label{eq:free-en:qmag:toroid:vec}
  \bar{\vek{\tau}} = \mathcal{Q}_0 \,
  \frac{l_d}{2w} \av{\{ z^2, k_z^2 - \av{k_z^2} \}} \, \xi (\Zc)
  \begin{pmatrix} -\cos \varphi_\Zc \\ \sin \varphi_\Zc
  \end{pmatrix}
\end{equation}
\end{subequations}
that characterizes the orientation of the magnetic quadrupole,
similar to how the angular dependence of the magnetoelectric effect
is governed by the orientation of the vector $\tilde{\vek{\tau}}$,
compare Eq.\ (\ref{eq:free-en:ME:toroid:all}).

It has recently been suggested \cite{gao18, shi18, gao18a} that the
components $\mathcal{Q}_{ij}$ of the magnetic quadrupole moment
are connected with the components $\alpha_{ij}$ of the magnetoelectric
tensor via the relation $e \, \partial \mathcal{Q}_{ij}/ \partial \mu =
- \alpha_{ij}$, where $\mu$ is the chemical potential.  For the metallic
quasi-2D systems studied here, this relation is fulfilled neither for an
infinitely deep square well, where $\mathcal{Q}_{ij} = 0$ but
$\alpha_{ij} \ne 0$, nor for a parabolic well, where
$\mathcal{Q}_{ij} \ne 0$ but $\alpha_{ij} = 0$.  The magnetic quadrupole
moment is found to arise solely from the energy change of the
confined-charge-carrier states due to the magnetic-field gradient
$\vekc{b}_\|$ obtained by first-order perturbation theory [Eq.\
(\ref{eq:free-en:qmag:final})]; it has none of the additional contributions
derived for bulk systems~\cite{gao18a}. On the other hand, the
magnetoelectric tensor $\alpha_{ij}$ requires second-order perturbation
theory for two perturbations $\vekc{E}$ and $\vekc{B}$, see, e.g., Eq.\
(\ref{eq:ME:free-en:def}) and Ref.~\cite{spa08}. These results suggest
that $\mathcal{Q}_{ij}$ and $\alpha_{ij}$ should generally be viewed as
independent coefficients in a Taylor expansion of the free energy $F$
as a function of the external fields $\vekc{E}$ and $\vekc{B}$.

\subsection{Magnetoelectricity in ferromagnetic hole systems}
\label{sec:fm-holes}

\begin{figure}
  \includegraphics[width=\linewidth]{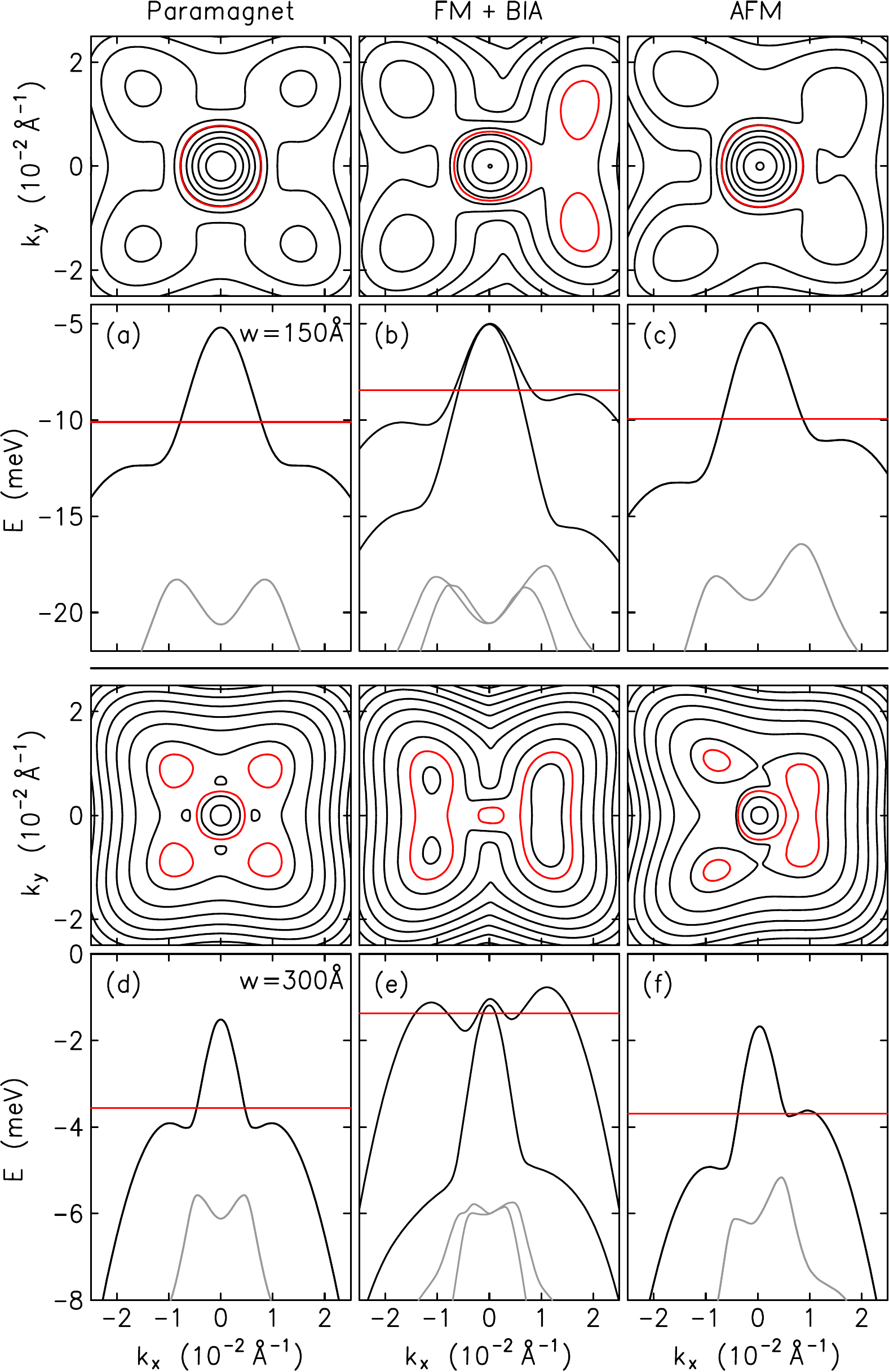}
  \caption{Subband dispersion (lower panels) of the HH subband
  (black) and LH subband (gray) of a quantum well with barrier
  height $V_0 = 0.12\,$eV, width $w = 150\,$\AA\ [upper row (a),
  (b), (c)] and $w = 300\,$\AA\ [lower row (d), (e), (f)] for $\Ec_z
  = \Bc_\| = 0$. The upper panels show contour plots of the
  same dispersion with line increments of 0.1~meV.  Red lines
  indicate the Fermi energy (lower panels) [Fermi contour (upper
  panels)] for a hole density $N_s = 1.0 \times
  10^{11}\,$cm$^{-2}$. Left column [panels (a), (d)]: Paramagnetic
  InSb ignoring BIA. Center column [panels (b), (e)]: Ferromagnetic
  InSb with $\Xc_x = 8\,$meV and BIA.  Right column [panels
  (c), (f)]: Diamond antiferromagnet with InSb band-structure
  parameters (without BIA) and $\mathcal{\SX}_x = 50\,$meV.}
  \label{fig:h-disp}
\end{figure}

The magnetoelectric response obtained in the realistic calculations
for electron systems in ferromagnetic InSb quantum wells is
small [Figs.~\ref{fig:e-mom-150}(a) and~\ref{fig:e-mom-150}(c)].
The response can be greatly enhanced by a suitable engineering of
the band structure $E_{n\kk_\|}$ of the quasi-2D systems.  Here
quasi-2D hole systems have long been known as a versatile playground
for bandstructure engineering, where the dispersion of the first
heavy-hole (HH) subband is strongly affected by the coupling to the
first light-hole (LH) subband \cite{eke84, bro85, win03}.
Figures~\ref{fig:h-disp}(a) and~\ref{fig:h-disp}(d) illustrate this
for quasi-2D hole systems in paramagnetic InSb quantum wells with
width $w = 150\,$\AA\ [Fig.~\ref{fig:h-disp}(a)] and $w = 300\,$\AA\
[Fig.~\ref{fig:h-disp}(d)], where HH-LH coupling results in a highly
nonparabolic dispersion $E_{0\kk_\|}$ of the doubly degenerate
ground HH subband.  Furthermore, the dispersion is also highly
anisotropic, which reflects the cubic symmetry of the underlying
crystal structure.

\begin{figure}
  \includegraphics[width=\linewidth]{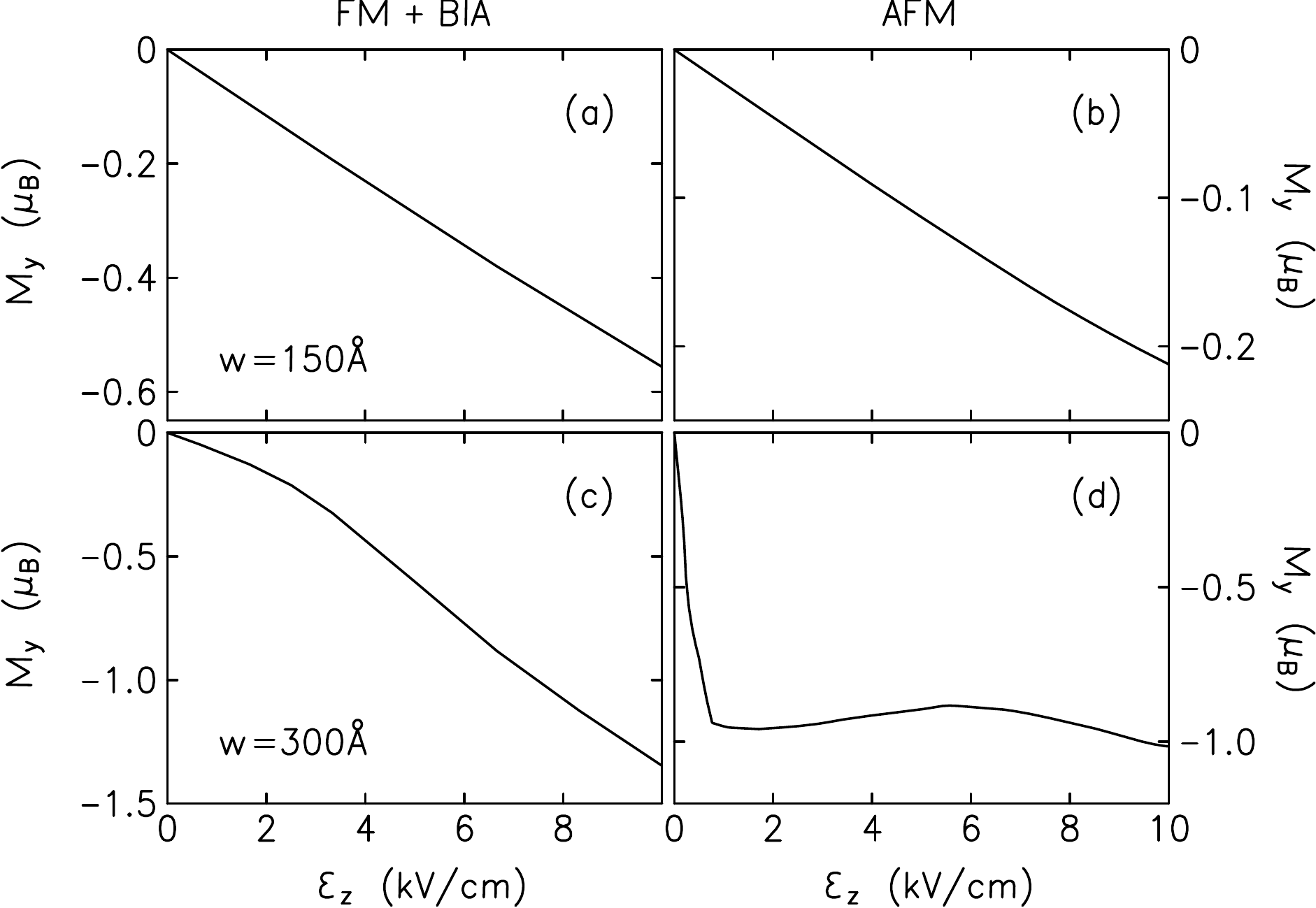}
  \caption{$\Ec_z$-induced orbital magnetic moment per particle
  $\bohrmag M_y^\mathrm{o}$ in a quantum well with barrier height
  $V_0 = 0.12\,$eV, width $w = 150\,$\AA\ [upper row (a), (b)] and
  $w = 300\,$\AA\ [lower row (c), (d)] and hole density $N_s = 1.0
  \times 10^{11}\,$cm$^{-2}$. Left column [panels (a), (c)]:
  Ferromagnetic InSb with $\Xc_x = 8\,$meV and BIA.  Right
  column [panels (b), (d)]: Diamond antiferromagnet with InSb
  band-structure parameters (without BIA) and $\mathcal{\SX}_x =
  50\,$meV.}
  \label{fig:h-mom}
\end{figure}

\begin{figure}
  \includegraphics[width=\linewidth]{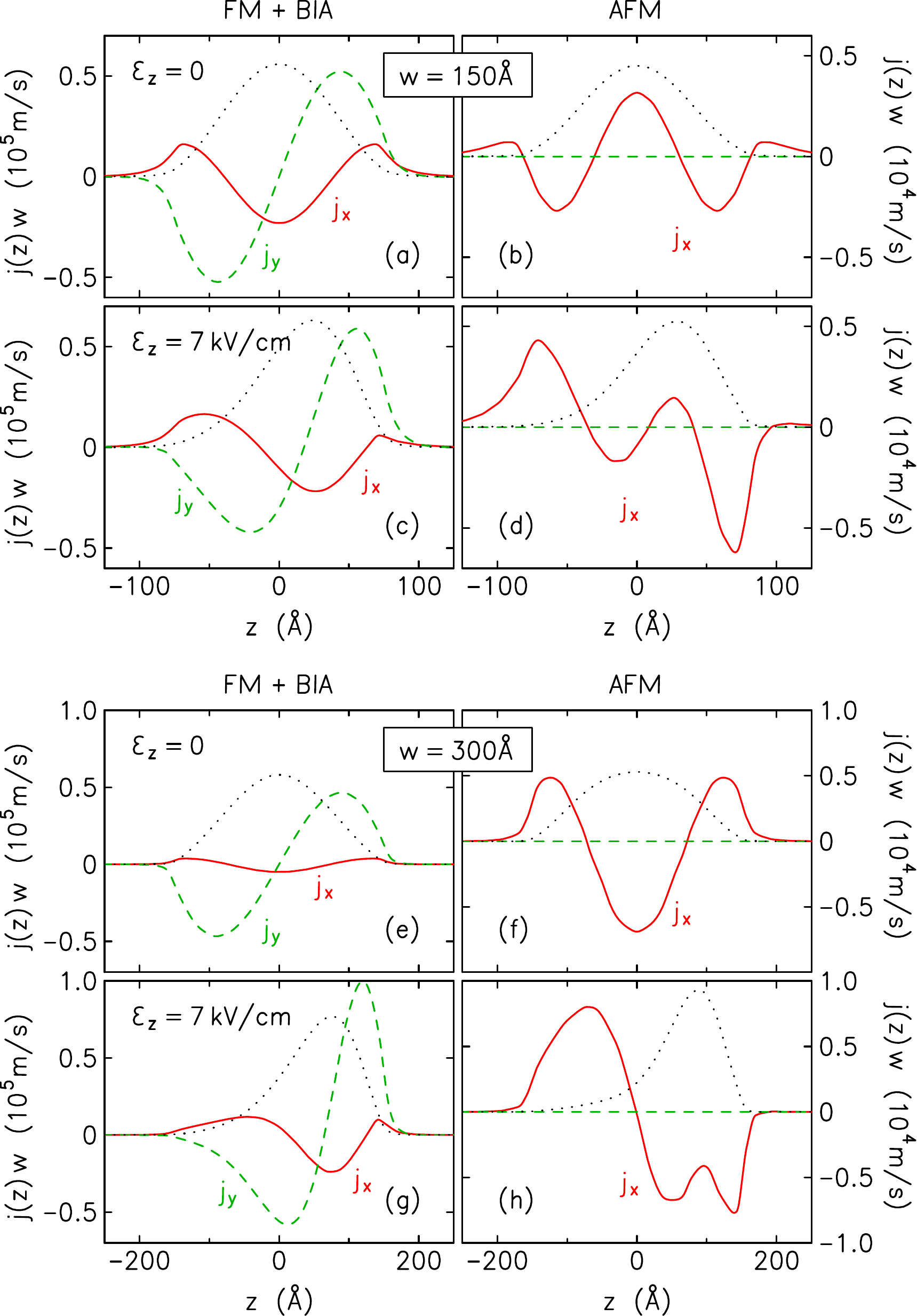}
  \caption{Equilibrium current distribution $\vek{j}_\| (z)$ in a
  quantum well with barrier height $V_0 = 0.12\,$eV, width $w =
  150\,$\AA\ [panels (a), (b), (c), (d)] and $w = 300\,$\AA\ [panels
  (e), (f), (g), (h)] and hole density $N_s = 1.0 \times
  10^{11}\,$cm$^{-2}$.  Left column [panels (a), (c), (e), (g)]:
  Ferromagnetic InSb with $\Xc_x = 8\,$meV and BIA.  Right
  column [panels (b), (d), (f), (h)]: Diamond antiferromagnet with
  InSb band-structure parameters (without BIA) and $\mathcal{\SX}_x
  = 50\,$meV.  Panels (a), (b), (e), (f): symmetric quantum well
  ($\Ec_z = 0$).  Panels (c), (d), (g), (h): tilted quantum well
  ($\Ec_z = 7\,$kV/cm).  In each panel, the dotted line shows for
  comparison the charge distribution $\rho(z)$ (arbitrary units).}
  \label{fig:h-vel}
\end{figure}

\begin{figure}
  \includegraphics[width=\linewidth]{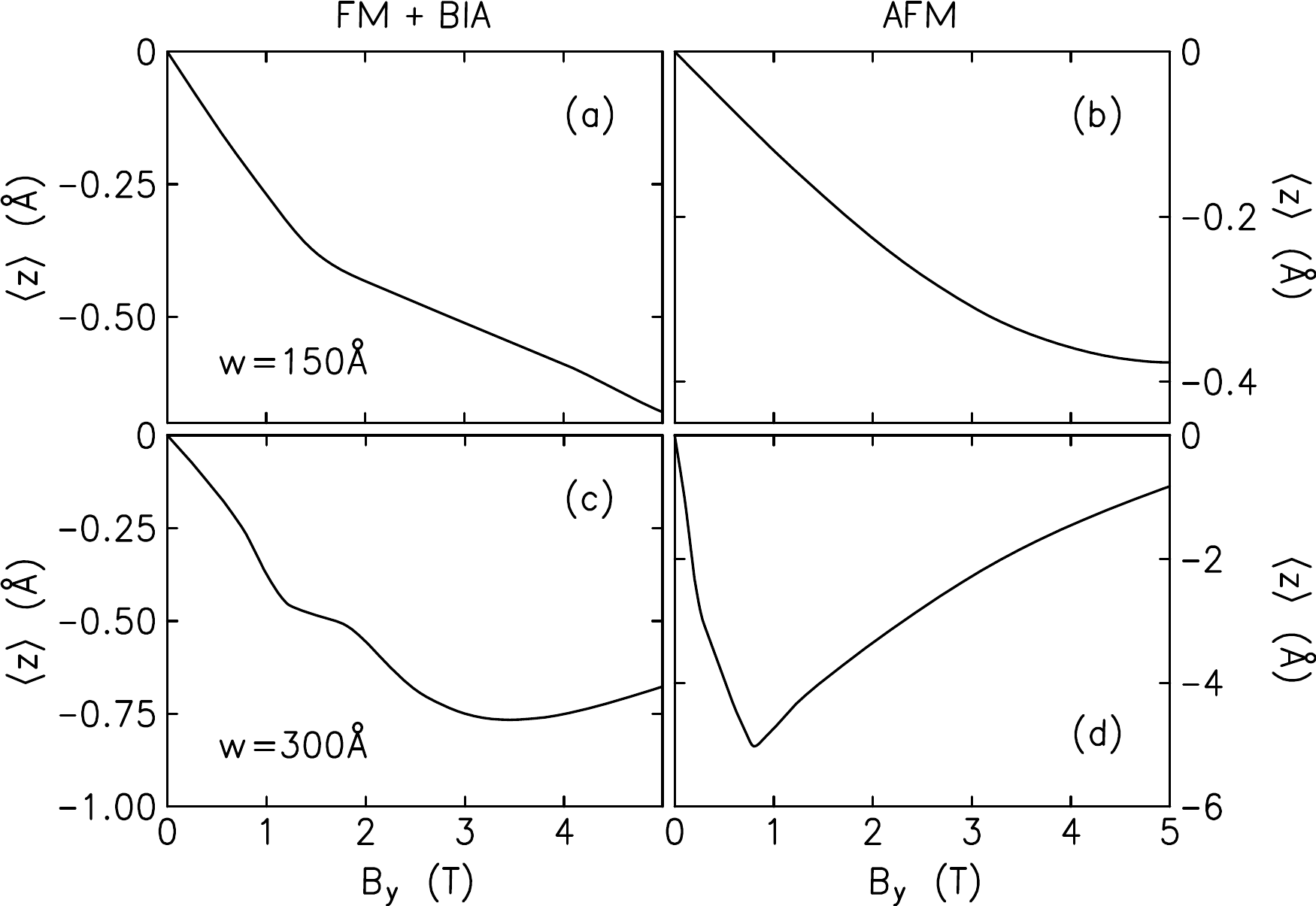}
  \caption{$\Bc_y$-induced displacement $\av{z}$ representing the
  electrostatic polarization via Eq.\ (\ref{eq:polToAvZ}) in a
  quantum well with barrier height $V_0 = 0.12\,$eV, width $w =
  150\,$\AA\ [upper row (a), (b)] and $w = 300\,$\AA\ [lower row
  (c), (d)] and hole density $N_s = 1.0 \times 10^{11}\,$cm$^{-2}$.
  Left column [panels (a), (c)]: Ferromagnetic InSb with
  $\Xc_x = 8\,$meV and BIA.  Right column [panels (b), (d)]:
  Diamond antiferromagnet with InSb band-structure parameters
  (without BIA) and $\mathcal{\SX}_x = 50\,$meV.}
  \label{fig:h-dip}
\end{figure}

An important aspect for the magnetoelectric response is the breaking
of time-reversal symmetry so that $E_{n, -\kk_\|} \ne E_{n\kk_\|}$.
The interplay between a ferromagnetic exchange field $\vekc{X}_\|$
and HH-LH coupling can result in a highly asymmetric band structure
of quasi-2D HH systems with multiple disconnected parts of the Fermi
surface, as illustrated in Figs.~\ref{fig:h-disp}(b) and
\ref{fig:h-disp}(e) for ferromagnetic InSb quantum wells
\cite{woj03}.  Figures~\ref{fig:h-mom}(a) and \ref{fig:h-mom}(c)
exemplify the $\Ec_z$-induced orbital magnetic moment per particle,
which can rise as high as $\sim 1\,\bohrmag$ for moderate electric
fields $\Ec_z$.  Figures~\ref{fig:h-vel}(a), \ref{fig:h-vel}(c),
\ref{fig:h-vel}(e), and \ref{fig:h-vel}(g) show the equilibrium
currents.  Finally, Figs.~\ref{fig:h-dip}(a) and \ref{fig:h-dip}(c)
show the $\vekc{B}_\|$-induced displacement $\av{z}$ that represents
the electrostatic polarization via Eq.\ (\ref{eq:polToAvZ}).

The large magnetoelectric response of quasi-2D hole systems
can be ascribed to the strong asymmetry $E_{n, -\kk_\|} \ne
E_{n\kk_\|}$ of the band structure.  With increasing fields, the
disconnected parts of the Fermi sea that are located away from
$\kk_\| = \vek{0}$ get depopulated and eventually disappear.  The
field-induced response drops again when finally only the central
part of the Fermi sea around $\kk_\| = \vek{0}$ accommodates all
charge carriers.  Thus, unlike the electron case discussed above,
the hole systems show a strongly nonlinear dependence of the
magnetoelectric response as a function of the applied fields.

\section{Magnetoelectricity in diamond antiferromagnets}
\label{sec:AFM}

Space-inversion symmetry of a diamond structure is broken in the
zincblende structure [Figs.~\ref{fig:diamond}(a)
and~\ref{fig:diamond}(b)].  Opposite magnetic moments placed
alternatingly on the atomic sites of a diamond structure result in
an antiferromagnetic structure [Fig.~\ref{fig:diamond}(c)].  Both
time reversal $\Theta$ and space inversion $I$ are broken symmetries
in such a diamond antiferromagnet.  The joint operation $\Theta I$,
however, remains a good symmetry so that, similar to paramagnetic
diamond, a two-fold spin degeneracy is preserved throughout the
Brillouin zone. Nonetheless, as these symmetries are broken
individually, invariants proportional to the N\'eel vector $\vekc{N}$
appear in the Kane Hamiltonian that are forbidden in paramagnetic
systems because of time-reversal symmetry.  These invariants are
derived in Sec.~\ref{sec:AFM:model}.

The diamond structure is realized by the A atoms of intermetallic
cubic (C-15) Laves phases AB$_2$, and it has been demonstrated that
NpCo$_2$ is an itinerant antiferromagnet, where the magnetic moments
on the Np atoms are ordered as shown in Fig.~\ref{fig:diamond}(c)
(magnetic space group $I4_1{}'/a'm'd$) \cite{ald75, san13a}.  The
diamond structure is also realized by the A atoms of spinels
AB$_2$X$_4$.  Frequently, spinels with magnetic A atoms give rise to
highly frustrated magnetic order \cite{tak11}.  Beyond that, a
recent study combining experiment and theory \cite{ge17} identified
CoRh$_2$O$_4$ as a canonical diamond-structure antiferromagnet,
where the magnetic moments on the Co atoms are ordered as shown in
Fig.~\ref{fig:diamond}(c).

\subsection{The model}
\label{sec:AFM:model}

Our goal is to incorporate the effect of antiferromagnetic order
into the $\kk \cdot \pp$ envelope-function theory \cite{bir74,
win03} underlying multiband Hamiltonians as in
Eq.\ (\ref{eq:Kane-ham}).  To this end, we start from the well-known
$sp^3$ tight-binding model for diamond and zincblende structures
with spin-orbit coupling included \cite{cha75b, cha77}.  This model
includes the $s$-bonding valence band $\Gamma_6^v$, the $p$-bonding
valence bands $\Gamma_8^v$ and $\Gamma_7^v$, the $s$-antibonding
conduction band $\Gamma_6^c$, and the $p$-antibonding conduction
bands $\Gamma_8^c$ and $\Gamma_7^c$.  Except for the low-lying
valence band $\Gamma_6^v$, these bands are also the basis states for
the $14 \times 14$ extended Kane model \cite{roe84, win03}.

We add a staggered exchange field $\vekc{\SX}$ on the two sublattices
of the diamond structure as depicted in Fig.~\ref{fig:diamond}(c).
Using the phase conventions for the basis functions of $\HK$ that
are described in detail in Appendix C of Ref.\ \cite{win03}, the
field $\vekc{\SX}$ yields terms in the off-diagonal blocks
$\mm[]{8c}{8v}$, $\mm[]{8c}{7v}$, and $\mm[]{7c}{7v}$ of $\HK$ that
are listed in Table~\ref{tab:kane-afm} using the notation of Table
C.5 in Ref.\ \cite{win03}.  The vector $\vekc{N}$ denotes the N\'eel unit
vector with components $\Nc_i = \mathcal{\SX}_i / \mathcal{\SX}$.
Within the $14 \times 14$ extended Kane model, the off-diagonal
invariants $\mm{8c}{8v}$, $\mm{8c}{7v}$, and $\mm{7c}{7v}$ provide a
complete account of the antiferromagnetic order shown in
Fig.~\ref{fig:diamond}(c).

The off-diagonal invariants $\mm{8c}{8v}$, $\mm{8c}{7v}$, and
$\mm{7c}{7v}$ appear already for $\kk = \vek{0}$.  In the diagonal
blocks $\mm[]{6c}{6c}$, $\mm[]{8v}{8v}$, and $\mm[]{7v}{7v}$, a
Taylor expansion of the tight-binding Hamiltonian about $\kk =
\vek{0}$ yields mixed terms proportional to powers of components of
$\vekc{\SX}$ and powers of components of $\kk$.  The lowest-order
invariants obtained in this way are also listed in
Table~\ref{tab:kane-afm}.  Alternatively, these terms can be derived
by means of quasi-degenerate perturbation theory \cite{win03}
applied to $\HK$ with $\mm{8c}{8v}$, $\mm{8c}{7v}$, and
$\mm{7c}{7v}$ included.  The latter approach yields explicit, albeit
lengthy, expressions for the prefactors $\mathcal{d}$ and
$\mathcal{D}^i_{jj}$ as a function of $\mathcal{\SX} = |\vekc{\SX}|$
that are omitted here.
As to be expected for antiferromagnetic diamond, the
$\vekc{\SX}$-dependent invariants in Table~\ref{tab:kane-afm} break
time-reversal symmetry, but they do not lift the spin degeneracy.
Using quasi-degenerate perturbation theory, we also obtain several
invariants in the valence band block $\mm{8v}{8v}$ that are
proportional to both $\vekc{\SX}$ and an external electric field
$\vekc{E}$.  These invariants are listed in Table~\ref{tab:kane-afm}
as well. They describe a spin splitting proportional to the field
$\vekc{E}$ (but independent of the wave vector $\kk$) that is
induced by the antiferromagnetic exchange coupling.  All invariants
listed in Table~\ref{tab:kane-afm} can also be derived by means of
the theory of invariants \cite{bir74} using the fact that the
staggered exchange field $\vekc{\SX}$ is a polar vector that is odd
under time reversal.

\begin{table}
  \caption{Lowest-order invariants describing antiferromagnetic
  order within the $14\times 14$ extended Kane model.  The notation
  follows Table C.5 of Ref.\ \cite{win03}, and $\Nc_i \equiv
  \mathcal{\SX}_i/\mathcal{\SX}$ denotes the Cartesian components of
  the unit vector parallel to the staggered exchange field
  $\vekc{\SX}$ on the sublattices of the diamond structure [see
  Fig.~\ref{fig:diamond}(c)].}
  \label{tab:kane-afm}
  \renewcommand{\arraystretch}{1.2}
  \begin{tabular}{CRL} \hline\hline
    \mm{8c}{8v} & = & (2i/3) \, \mathcal{\SX} (\Nc_x J_x + \cp) \\
    \mm{8c}{7v} & = & -2i \, \mathcal{\SX} (\Nc_x U_x + \cp) \\
    \mm{7c}{7v} & = & (-i/3) \, \mathcal{\SX} (\Nc_x \sigma_x + \cp) \\ \hline
    \mm{6c}{6c} & = &
    \mathcal{d} ( \{ \kc_x, \kc_y^2 - k_z^2 \} \Nc_x + \cp ) \\
    \mm{8v}{8v} & = &
    \mathcal{D}_{88}^1 ( \{ \kc_x, \kc_y^2 - k_z^2 \} \Nc_x + \cp ) \\ && {}
    + \mathcal{D}_{88}^2 [(\Nc_y \kc_y - \Nc_z \kc_z) J_x^2 + \cp] \\ && {}
    + \mathcal{D}_{88}^3 [(\Nc_x \kc_y - \Nc_y \kc_x) \{ J_x, J_y \} + \cp] \\ && {}
    + \mathcal{D}_{88}^4 [ (\Nc_y \Ec_z - \Nc_z \Ec_y) \{ J_x, J_y^2 - J_z^2 \} + \cp] \\ && {}
    + \mathcal{D}_{88}^5 [ (\Nc_y \Ec_z + \Nc_z \Ec_y) J_x + \cp] \\ && {}
    + \mathcal{D}_{88}^6 [ (\Nc_y \Ec_z + \Nc_z \Ec_y) J_x^3 + \cp] \\ && {}
    + \mathcal{D}_{88}^7 (\Nc_x \Ec_x + \cp) (J_x J_y J_z + J_z J_y J_x) \\
    \mm{7v}{7v} & = &
    \mathcal{D}_{77}^1 ( \{ \kc_x, \kc_y^2 - k_z^2 \} \Nc_x + \cp )
    \\ \hline\hline
  \end{tabular}
\end{table}

According to Table~\ref{tab:kane-afm}, in lowest order the
$\Gamma_6$ conduction band in a diamond antiferromagnet is described
by the Hamiltonian
\begin{subequations}
  \begin{equation}
    \label{eq:ham-cb:AFM:tot}
    \Hcb = \Hcb_k + V(z) + \Hcb_\Nc + e\Ec_z z \,\, ,
  \end{equation}
  with $\Hcb_k$ given in Eq.\ (\ref{eq:ham-cb:kin}), and
  \begin{equation}
    \label{eq:ham-cb:AFM}
    \Hcb_\Nc = \mathcal{d} \left( \left\{ \kc_x, \kc_y^2
        - k_z^2 \right\} \Nc_x + \cp \right) \,\, ,
  \end{equation}
\end{subequations}
where $\mathcal{d}$ is a prefactor proportional to $\mathcal{\SX}$.
Formally, $\Hcb_\Nc$ has the same structure as the Dresselhaus term
(\ref{eq:ham-cb:dressel}), with the spin operators $\sigma_i$
replaced by the numbers $\Nc_i$ and $d$ replaced by $\mathcal{d}$.
Therefore, the following study of magnetoelectric coupling in
antiferromagnetic diamond proceeds in remarkable analogy to the
study of magnetoelectric coupling in a paramagnetic or ferromagnetic
zincblende structure presented in Sec.~\ref{sec:para-ferro}
\cite{misc:minus}.  As $\Hcb_\Nc$ and, in fact, the entire
Hamiltonian (\ref{eq:ham-cb:AFM:tot}), do not depend on the
charge carriers' spin, the latter will be a silent degree of freedom
in the following considerations.

For the analytical model studied below, it is easy to see that a
purely in-plane N\'eel unit vector $\vekc{N}$ yields the largest
magnetoelectric coupling.  Assuming therefore that $\vekc{N}$ is
oriented in-plane, the full Hamiltonian becomes [including terms up
to second order in $\vekc{k}_\|$, compare Eq.~(\ref{eq:ham-cb:rot})]
\begin{subequations}
  \label{eq:AFM:ham-cb}
  \begin{align}
  \Hcb &= \frac{\hbar^2  \kc^2}{2 m} + V(z)
  - \mathcal{d} \, k_z^2 \left( \kc_x \cos \varphi_\Nc
      - \kc_y \sin \varphi_\Nc \right) + e\Ec_z z , \nonumber \\
      \\
          &= \frac{\hbar^2  k_z^2}{2 m} + V(z)
          + \frac{\hbar^2}{2 m} \left( \vekc{k}_\| - \vekc{k}_0 \right)^2 -
          \frac{\hbar^2 \kc_0^2}{2 m} + e\Ec_z z \,\, . \nonumber \\
\end{align}
\end{subequations}
Here $\varphi_\Nc$ denotes the angle that $\vekc{N}$ makes
with the $x$ axis, and we introduced the operator
\begin{equation}
  \label{eq:AFM:k0}
  \vekc{k}_0 = \frac{m}{\hbar^2} \: \mathcal{d} k_z^2
  \begin{pmatrix}
    \cos \varphi_\Nc \\ - \sin \varphi_\Nc
  \end{pmatrix} \; .
\end{equation}
For $\Ec_z = \Bc_\| = 0$ and treating $\Hcb_\Nc$ in first
order, the subband dispersions become
\begin{equation}
  \label{eq:AFM:disp-cb}
  E_{\nu\sigma, \kk_\|} \equiv E_{\nu \kk_\|}
  = E_\nu + \frac{\hbar^2}{2m} \left[ \left( \kk_\| - \av{\vekc{k}_0}_\nu
    \right)^2 - \av{\vekc{k}_0}_\nu^2 \right] \, ,
\end{equation}
which are spin-degenerate parabolae that are shifted in the $(k_x,
k_y)$ plane by $\av{\vekc{k}_0}_\nu$.  The shift
$\av{\vekc{k}_0}_\nu$ is a fingerprint for the broken time-reversal
symmetry in the antiferromagnet.

Figures~\ref{fig:e-disp}(b) and \ref{fig:e-disp}(d) illustrate the
lowest-subband dispersion $E_{0 \kk_\|}$ for a quasi-2D electron
system in an antiferromagnetic InSb quantum well with
$\mathcal{\SX}_x = 50\,$meV, width $w = 150\,$\AA\ and with an
electron density $N_s = 1.0 \times 10^{11}\,$cm$^{-2}$.  The numerical
calculations are based on the $14 \times 14$ extended Kane model
including the terms $\mm{8c}{8v}$, $\mm{8c}{7v}$, and $\mm{7c}{7v}$
from Table~\ref{tab:kane-afm}.  One can approximate these results with
the smaller Hamiltonian (\ref{eq:AFM:ham-cb}) using $\mathcal{d}
\approx 80\,$eV\AA$^3$.  In the Hamiltonian for the extended Kane
model, we preclude the Dresselhaus terms $\HK_\mathrm{D}$ due to BIA
by setting to zero the band parameters $P'$ and $C_k$ defined in
Table C.5 of Ref.~\cite{win03}.

\subsection{The N\'eel operator}
\label{sec:neel-op}

We now digress to discuss a few general properties of the model for
antiferromagnetic order proposed here. It is well-known that the
Zeeman term (\ref{eq:ham-cb:zeeman}) with an exchange field
$\vekc{X}$ provides a simple mean-field model for itinerant-electron
ferromagnetism. Similarly, $\Hcb_\Nc$ is a phenomenological model
for collinear (two-sublattice) itinerant-electron
antiferromagnetism.

The operator conjugate to the ferromagnetic exchange field
$\vekc{X}$ is the (dimensionless) spin-polarization operator
$\vek{\sigma} = \partial \Hcb / \partial \vekc{X}$. In the
mean-field theory underlying the present work, a nonzero expectation
value $\av{\vek{\sigma}}$ indicates ferromagnetic order of spins.
Similarly, the operator conjugate to the staggered exchange field
$\vekc{\SX}$ is the (again dimensionless) N\'eel operator for the
staggered magnetization,
\begin{equation}
  \label{eq:AFM:op}
  \vek{\tau} = \frac{\partial \Hcb}{\partial \vekc{\SX}}
  = \frac{\mathcal{d}}{\mathcal{\SX}} \,\, k_z^2
  \begin{pmatrix}
    k_x \\ - k_y
  \end{pmatrix}
  = q_\tau \, k_z^2
  \begin{pmatrix}
    k_x \\ - k_y
  \end{pmatrix} \; ,
\end{equation}
where the prefactor $q_\tau \equiv \mathcal{d} / \mathcal{Y}$
depends on the momentum matrix elements and energy gaps
characterizing the Hamiltonian $\HK$, but it is independent of the
exchange field $\mathcal{Y}$.
A nonzero expectation value $\av{\vek{\tau}}$ indicates collinear
orbital (itinerant-electron) antiferromagnetic order.  Like the
staggered exchange field $\vekc{\SX}$, the N\'eel operator
$\vek{\tau}$ is a polar vector that is odd under time reversal. Thus
$\av{\vek{\tau}} \ne \vek{0}$ represents a (polar) toroidal moment
\cite{art85, dub90, spa08}.  On the other hand, $\vekc{X}$ and
$\av{\vek{\sigma}}$ are axial vectors that are odd under time
reversal.  In that sense, $\av{\vek{\sigma}}$ and $\av{\vek{\tau}}$
quantify complementary aspects of itinerant-electron collinear
magnetic order in solids \cite{misc:currents}.

In systems with spin-orbit coupling such as the ones studied here,
the spin magnetization $\av{\vekc{S}}$ associated with the
expectation value $\av{\vek{\sigma}}$ is augmented by an
orbital-magnetization contribution, yielding the total magnetization
$\vekc{M}$.  A magnetization $\vekc{M}$ arises due to the presence
of an exchange field $\vekc{X}$ or an external magnetic field
$\vekc{B}$, but it may also arise due to, e.g., an electric field
$\vekc{E}$ (the magnetoelectric effect studied here) or a strain
field (piezomagnetism \cite{tav56, dzi58, lan84}). Similarly, a
nonzero expectation value $\av{\vek{\tau}}$ can be due to a
staggered exchange field $\vekc{\SX}$.  But it may also arise due
to, e.g., the interplay of an exchange field $\vekc{X}$, spin-orbit
coupling, and confinement [Eq.\ (\ref{eq:EtoM:curr})].

\subsection{$\Ec$-induced magnetization}
\label{sec:AFM:EtoM}

To calculate the equilibrium magnetization, we start from the
Hamiltonian (\ref{eq:AFM:ham-cb}).  Treating the electric field
$\Ec_z$ in first-order perturbation theory, the eigenstates become
\begin{equation}
  \ket{\nu^{(1)}} = \ket{\nu} + e\Ec_z \, \sum_{\nu' \ne \nu} c_{\nu'\nu} \ket{\nu'} \; ,
\end{equation}
compare Eq.\ (\ref{eq:eigen:z}).  For the equilibrium magnetization
(\ref{eq:orb-mag:def}), we need to evaluate expectation values
$\Av{\{ z\, , \vek{v}_\| (\kk_\|) \}}$ using the velocity operator
associated with the Hamiltonian (\ref{eq:AFM:ham-cb})
($\vekc{B}_\|=0$)
\begin{equation}
  \label{eq:AFM-vel:k}
  \vek{v}_\| (\kk_\|) = \frac{\hbar}{m} \left( \kk_\| - \vekc{k}_0 \right)
  \,\, .
\end{equation}
We get
\begin{widetext}
  \begin{subequations}
    \label{eq:AFM:z-v:av}
    \begin{align}
      \Av{\{ z\, , \vek{v}_\| (\kk_\|) \}}
      & = \Av{z} \Av{\vek{v}_\| (\kk_\|)}
          + \left( \Av{\{ z, \vek{v}_\| (\kk_\|) \}}
                   - \Av{z} \Av{\vek{v}_\|} \right) \,\, , \\
      & = \Av{z} \Av{\vek{v}_\| (\kk_\|)}
      - \frac{\hbar}{m} \bigl( \Av{\{ z, \vekc{k}_0 \}}
                   - \Av{z} \Av{\vekc{k}_0} \bigr) \,\, , \\
      & = \Av{z} \Av{\vek{v}_\| (\kk_\|)}
      - \frac{\mathcal{d}}{\hbar}
      \begin{pmatrix}
        \cos \varphi_\Nc \\ - \sin \varphi_\Nc
      \end{pmatrix}
      \Av{\{ z, k_z^2 - \Av{k_z^2} \}} \,\, , \\
      & = \Av{z} \Av{\vek{v}_\| (\kk_\|)}
      - \frac{\mathcal{d}}{\hbar}
      \begin{pmatrix}
        \cos \varphi_\Nc \\ - \sin \varphi_\Nc
      \end{pmatrix}
      \biggl[ \av{ \{ z, k_z^2 - \av{k_z^2} \}}
        + 2 e\Ec_z \sum_{\nu' \ne \nu} c_{\nu'\nu} \,
      \braket{\nu | \{ z, k_z^2 - \av{k_z^2} \} - \av{z} k_z^2 | \nu'}
      \biggr] \,\, , \\[-1ex]
      & = \Av{z} \Av{\vek{v}_\| (\kk_\|)}
      - \frac{2 \mathcal{d} \, e\Ec_z}{\hbar}
      \begin{pmatrix}
        \cos \varphi_\Nc \\ - \sin \varphi_\Nc
      \end{pmatrix}
        \sum_{\nu' \ne \nu} c_{\nu'\nu} \, \braket{\nu | \{ z, k_z^2 - \av{k_z^2} \} | \nu'}
      \; , \label{eq:AFM:z-v:av:final}
    \end{align}
  \end{subequations}
\end{widetext}
compare Eq.\ (\ref{eq:z-v:av}).  Again, we ignored any $\kk_\|$ or
$\vekc{k}_0$ dependence of the perturbed states $\ket{\nu^{(1)}}$,
which is a higher-order effect.  The first term in Eq.\
(\ref{eq:AFM:z-v:av:final}) yields a vanishing contribution when
summed over the equilibrium Fermi sea, as it is proportional to the
system's total equilibrium current.  Therefore, a nonzero
magnetization is due to the second term in Eq.\
(\ref{eq:AFM:z-v:av:final}), which is independent of the wave vector
$\kk_\|$.  We can obtain Eq.\ (\ref{eq:AFM:z-v:av:final}) from Eq.\
(\ref{eq:z-v:av:final}) by replacing $\varphi_\Zc$ with
$\varphi_\Nc$ and putting $\sigma=1$ for all states.  The
latter implies that the effect described by Eq.\
(\ref{eq:AFM:z-v:av:final}) is maximized compared with Eq.\
(\ref{eq:z-v:av:final}) because both spin orientations in the
antiferromagnet contribute constructively.

Summing over the Fermi sea, we obtain for the
magnetization~(\ref{eq:orb-mag:def})
\begin{equation}
  \vekc{M}_\|^\mathrm{o} = -\mathcal{M}_0 \,\, e\Ec_z w\,\, \lambda_\mathcal{d}
  \begin{pmatrix} \sin\varphi_\Nc \\ \cos \varphi_\Nc
  \end{pmatrix} \,\, ,
  \label{eq:AFM:EtoM:final}
\end{equation}
with
\begin{equation}
  \label{eq:AFM:me-factor}
  \lambda_\mathcal{d} \equiv \frac{l_\mathcal{d}}{w} \sum_{\nu' \ne 0}
  \frac{\braket{\nu' | z | 0}
        \braket{0 | \{ z, k_z^2 - \av{k_z^2} \} | \nu'}}
       {E_0^{(0)} - E_{\nu'}^{(0)}}
\end{equation}
and $l_\mathcal{d} \equiv 2 m_0 \mathcal{d} / \hbar^2$, in complete
analogy with Eqs.\ (\ref{eq:EtoM:final}) and (\ref{eq:me-factor})
\cite{misc:minus}.  For $\varphi_\Nc = n\pi/2$ ($n$ integer) the
induced magnetization is oriented perpendicular to the N\'eel vector
$\vekc{N}$.  More generally, a clockwise rotation of $\vekc{N}$
implies a counterclockwise rotation of $\vekc{M}_\|^\mathrm{o}$.
Figure~\ref{fig:e-mom-150}(b) illustrates the $\Ec_z$-induced
magnetization for an antiferromagnetic InSb quantum well with
width $w = 150\,$\AA\ and electron density $N_s = 1.0 \times
10^{11}\,$cm$^{-2}$.

Again, it is illuminating to compare Eq.\ (\ref{eq:AFM:EtoM:final}) with
the equilibrium current distribution (\ref{eq:2D-current:def}).
Using $\phi_\nu (z) \equiv \braket{z|\nu}$, the perturbed wave functions
read
\begin{equation}
  \Phi_\nu (z) \equiv \braket{z | \nu^{(1)}}
   = \phi_\nu (z) + e\Ec_z \sum_{\nu' \ne \nu} c_{\nu'\nu} \, \phi_{\nu'} (z) \; .
\end{equation}
Using the velocity operator (\ref{eq:AFM-vel:k}), we get in first
order of $\Ec_z$ and $\mathcal{d}$
\begin{subequations}
  \label{eq:AFM-EtoM:curr}
  \begin{align}
  \vek{j}_\| (z, \nu \kk_\|)
  & = \Phi_\nu^\ast (z) \, \vek{v}_\| (\kk_\|) \, \Phi_\nu (z) \,\, , \\
  & = \av{ \vek{v}_\| (\kk_\|)} \, |\phi_\nu|^2
  + \sum_{\nu' \ne \nu} \vek{\kappa}_{\nu'\nu} \, \phi_\nu^\ast \phi_{\nu'}
  \nonumber \\*[-1ex] & \hspace{0.5em} {}
  + e\Ec_z \!\!\sum_{\nu',\nu'' \ne \nu} \!
  \bigl[ \left( c_{\nu''\nu'} \, \vek{\kappa}_{\nu'\nu}
    + \vek{\kappa}_{\nu''\nu'} c_{\nu'\nu} \right) \phi_\nu^\ast \phi_{\nu''}
  \nonumber \\*[-1ex] & \hspace{7.0em} {}
  + c_{\nu''\nu}^\ast \vek{\kappa}_{\nu'\nu} \, \phi_{\nu''}^\ast \phi_{\nu'} \bigr] \,\, ,
\end{align}
\end{subequations}
where the matrix elements $\vek{\kappa}_{\nu'\nu}$ are given by Eq.\
(\ref{eq:EtoM:curr:mel}) with $\varphi_\Zc$ replaced by
$\varphi_\Nc$.  Equation (\ref{eq:AFM-EtoM:curr}) is obtained from
Eq.\ (\ref{eq:EtoM:curr}) by putting $\sigma = +1$ so that the
interpretation of Eq.\ (\ref{eq:AFM-EtoM:curr}) proceeds similarly.
In thermal equilibrium, the first term in Eq.\
(\ref{eq:AFM-EtoM:curr}) averages to zero in Eq.\
(\ref{eq:2D-current:def:av}).  The remaining terms are independent
of $\kk_\|$ so that they do not average to zero in Eq.\
(\ref{eq:2D-current:def:av}).  The second term ($\nu'=2$) describes
a quadrupolar equilibrium current proportional to $\phi_0 (z) \phi_2
(z)$ independent of the electric field $\Ec_z$.  Such quadrupolar
orbital currents are a generic feature of antiferromagnets; they
are the counterpart of dipolar orbital currents representing the
orbital magnetization in ferromagnets (see
Appendix~\ref{app:orb-magnet})
\footnote{Ferromagnetic order also gives rise to higher multipoles
($l>1$) in the current distribution beyond dipolar currents ($l=1$).
Similarly, Eq.\ (\ref{eq:AFM-EtoM:curr}) also yields multipoles $l >
2$.  However, such higher multipoles generally depend on the origin
of the coordinate system~\cite{jac99}.}.
Similar to Eq.\ (\ref{eq:EtoM:curr:afm}), the orbital
antiferromagnetic order can be quantified using the N\'eel operator
$\vek{\tau}$.  The Hamiltonian (\ref{eq:AFM:ham-cb}) (with $\Ec_z =
0$) yields
\begin{equation}
  \label{eq:AFM:EtoM:curr:afm}
  \av{\vek{\tau}} = - 2 \pi \, q_\tau \, d \, \Ns
  \begin{pmatrix} \cos\varphi_\Nc \\ \sin \varphi_\Nc \end{pmatrix}
  \sum_{\nu' \ne 0} \frac{|\braket{\nu' | k_z^2 | 0}|^2}
  {E_0^{(0)} - E_{\nu'}^{(0)}} \; .
\end{equation}
As to be expected, we have $\av{\vek{\tau}} \parallel \vekc{N}$.

The last term in Eq.\ (\ref{eq:AFM-EtoM:curr}) ($\nu''=1$) describes
$\Ec_z$-induced dipolar currents, i.e., a magnetization.
In a quantum well of width $w$, the equilibrium currents $\vek{j}_\|
(z, \nu \kk_\|)$ occur on a length scale of order $w$, which is typically
much larger than the lattice constant of the underlying crystal
structure.  The magnetic multipoles associated with the current
distribution may thus be accessible experimentally.  They may even
open up new avenues to manipulate the magnetic order in
antiferromagnets.
Figures~\ref{fig:e-vel}(b) and~\ref{fig:e-vel}(d) illustrate the
equilibrium currents for antiferromagnetic InSb quantum wells.


It is illuminating to study a second mechanism for an $\Ec$-induced
magnetization based on the antiferromagnetic exchange term
(\ref{eq:ham-cb:AFM}) that manifests itself as a spin magnetization
(\ref{eq:spin-mag:def}).
Generally, an electric field $\Ec_z$ applied to a quantum well gives
rise to a Rashba term \cite{byc84a, win03}
\begin{equation}
  \label{eq:AFM-E+Rashba}
  H_\mathrm{R} = a_\mathrm{R} \, \Ec_z
  \left(k_y \sigma_x - k_x \sigma_y\right)
\end{equation}
with Rashba coefficient $a_\mathrm{R} \, \Ec_z$, resulting in
spin-split eigenstates
\begin{equation}
  \ket{\nu\kk_\|\pm} = \ket{\nu} \otimes \frac{1}{\sqrt{2}}
  \Ket{\begin{array}{c} 1 \\ \mp i e^{i \varphi_{\kk_\|}}
  \end{array}} \; ,
\end{equation}
where $\varphi_{\kk_\|}$ is the angle between $\kk_\|$ and the $x$
axis, and we assumed as before that the orbital part $\ket{\nu}$ of
the eigenstates is independent of $\kk_\|$.  Thus we have
\begin{equation}
  \label{eq:Rashba:spin-orient}
 \av{\vek{\sigma}}_{\nu\kk_\| \pm} = \pm
 \begin{pmatrix}
   \cos (\varphi_{\kk_\|} - \pi/2) \\ \sin (\varphi_{\kk_\|} - \pi/2)
 \end{pmatrix} \; .
\end{equation}
Also, Rashba spin-orbit coupling gives rise to an imbalance between the two
spin subbands $\pm$, which can be characterized by Fermi wave
vectors $k_{\mathrm{F} \pm} \approx \sqrt{2\pi \Ns} \mp
a_\mathrm{R} \Ec_z m/ \hbar^2$.
Performing the average (\ref{eq:spin-mag:def}) over all occupied
states in these spin subbands [assuming a dispersion
(\ref{eq:AFM:disp-cb}) with small $\av{\vekc{k}_0} \ne \vek{0}$ and
slightly different Fermi wave vectors $k_{\mathrm{F} \pm}$], we
obtain a nonzero equilibrium spin polarization
\begin{subequations}
  \label{eq:AFM-edelstein}
  \begin{align}
  \vek{S}_\| & = \frac{m}{\hbar^2} \,
  \frac{a_\mathrm{R} \Ec_z}{\pi \Ns}
  \, \av{\vekc{k}_0} \times \hat{\vek{z}} \,\, , \\
  & = - \left(\frac{m}{\hbar^2}\right)^2
  \frac{a_\mathrm{R} \Ec_z \, \mathcal{d} \, \av{k_z^2}_\nu}
  {\pi \Ns}
   \begin{pmatrix}
   \sin \varphi_\Nc \\ \cos \varphi_\Nc
   \end{pmatrix} \; .
\end{align}
\end{subequations}
Inserting this result into (\ref{eq:total-mag:def}) yields a spin
magnetization that complements the orbital magnetization
(\ref{eq:AFM:EtoM:final}).  As to be expected, both terms have the
same dependence on the direction of the vector $\vekc{N}$.  The
mechanism described by Eq.\ (\ref{eq:AFM-edelstein}) contributes
to the numerically calculated magnetization presented in
Fig.~\ref{fig:e-mom-150}(b) \cite{misc:rashba}.  

We can interpret the spin polarization (\ref{eq:AFM-edelstein}) as
follows. The Rashba term (\ref{eq:AFM-E+Rashba}) yields a spin
orientation (\ref{eq:Rashba:spin-orient}) of individual states
$\ket{\nu\kk_\|\pm}$.  Nonetheless, for nonmagnetic systems in
thermal equilibrium, the net spin polarization is zero because
time-reversal symmetry implies that we have equal probabilities for
the occupation of time-reversed states $\ket{\nu\kk_\|\pm}$ and
$\ket{\nu,-\kk_\|\pm}$ with opposite spin orientations.  This
argument for nonmagnetic systems is closely related to the fact that
thermal equilibrium in a time-reversal-symmetric system requires
that the Fermi sea is centered symmetrically about $\bar{\kk} =
\vek{0}$.  A nonzero shift $\bar{\kk}$ of the Fermi sea, and thus a
nonzero average spin polarization, are permitted in nonmagnetic
systems as a quasistationary nonequilibrium configuration in the
presence of a driving electric field $\vekc{E}_\|$, which is an
important mechanism for the current-induced magnetization reviewed
in Appendix~\ref{app:current-induced-magnetization}.  The spin
polarization (\ref{eq:AFM-edelstein}), on the other hand, is
entirely an equilibrium effect. It can occur in antiferromagnetic
systems, where time-reversal symmetry is already broken in thermal
equilibrium as expressed by the shift $\bar{\kk} = \av{\vekc{k}_0}$.

It follows from Table~\ref{tab:kane-afm} that we generally get a
spin splitting proportional to $\Ec_z$ even at \mbox{$\vek{k}_\| =
\vek{0}$}, which yields a third, Zeeman-like contribution to the
total magnetization (\ref{eq:total-mag:def}).  For quasi-2D hole
systems, this effect can be substantial.  For quasi-2D electron
systems, this effect is of second order in the staggered exchange
field $\vekc{\SX}$.

\subsection{$\Bc$-induced electric polarization}
\label{sec:AFM:BtoP}

Our goal is to evaluate the polarization (\ref{eq:pol:def}) in the
presence of an in-plane magnetic field $\vekc{B}_\|$.  The starting
point is again the Hamiltonian (\ref{eq:AFM:ham-cb}).  An in-plane
magnetic field $\vekc{B}_\|$ represented via the vector potential
$\vekc{A}$ gives rise to the perturbation [ignoring terms
$\mathcal{O} (\Bc_\|^2)$]
\begin{widetext}
\begin{subequations}
  \label{eq:AFM:ham-cb:BtoP:perturb}
  \begin{align}
    \Hcb_\Bc^{(1)}
    & = \frac{e\hbar}{2m}
    \left[ \left(\kk_\| - \vekc{k}_0 \right) \cdot \vekc{A}
      + \vekc{A} \cdot \left(\kk_\| - \vekc{k}_0 \right) \right] \,\, , \\
    & = \frac{e\hbar}{2m} \left[ 2\left(\kk_\| - \Av{\vekc{k}_0} \right) \cdot\vekc{A}
  - \left( \vekc{k}_0 - \Av{\vekc{k}_0} \right) \cdot \vekc{A}
  - \vekc{A} \cdot \left( \vekc{k}_0 - \Av{\vekc{k}_0} \right) \right] \,\, , \\
    & = \frac{e\hbar}{m} \left(\kk_\| - \Av{\vekc{k}_0} \right) \cdot \vekc{A}
    - \frac{e}{\hbar}  \mathcal{d} \, \Bc_\| \,
     \sin ( \varphi_\Nc + \varphi_\Bc) \, \{ z, k_z^2 - \av{k_z^2} \} \,\, .
    \label{eq:AFM:BtoP:ham:final}
  \end{align}
\end{subequations}
\end{widetext}
The perturbation $\Hcb_\Bc^{(1)}$ yields perturbed states
$\ket{\nu^{(1)}}$.  We get
\begin{subequations}
  \begin{align}
    \Av{z} & = \av{z}_\nu + 2 \sum_{\nu' \ne \nu} c_{\nu'\nu}
    \Braket{\nu | \Hcb_\Bc^{(1)} | \nu'} \,\, , \\
    & = \av{z}_\nu + 2 \sum_{\nu' \ne \nu} c_{\nu'\nu} \biggl[
      \frac{e\hbar}{m} (\kk_\| - \av{\vekc{k}_0})
      \cdot \braket{\nu | \vekc{A} | \nu'}
      \nonumber \\* & \hspace{2em} {}
      - \frac{e}{\hbar}  \mathcal{d} \, \Bc_\| \,
      \sin (\varphi_\Nc + \varphi_\Bc)
      \, \braket{\nu | \{ z, k_z^2 - \av{k_z^2} \} | \nu'} \biggr] \,\, .
  \end{align}
\end{subequations}
As before [Eq.\ (\ref{eq:z:av})], the first term $\av{z}_\nu$
vanishes for a symmetric potential $V(z)$.  The first term in the
square brackets describes a $\kk_\|$-dependent shift \cite{smr95a,
lay97, kam12a, mue15} that yields a vanishing contribution to
$\mathcal{P}_z^\mathrm{e}$ when summed over the equilibrium Fermi sea.
Therefore, a nonzero polarization is due to the second term in the
square brackets, which is independent of the wave vector $\kk_\|$.
Summing over the Fermi sea, we obtain
\begin{equation}
  \label{eq:AFM:BtoP:final}
  \mathcal{P}_z^\mathrm{e} = -\mathcal{P}_0\,\, \bohrmag \Bc_\|\,\, \lambda_\mathcal{d} \,
  \sin (\varphi_\Nc + \varphi_\Bc) \; ,
\end{equation}
compare Eq.\ (\ref{eq:BtoP:final}) \cite{misc:minus}.
Figure~\ref{fig:e-mom-150}(b) illustrates the $\vekc{B}_\|$-induced
polarization for an antiferromagnetic InSb quantum well.

\subsection{Magnetoelectric contribution to the free energy}

As before, we evaluate the change $\delta F$ in the free-energy density
in the presence of both $\Hcb_\Ec^{(1)}$ [Eq.\ (\ref{eq:ham-cb:E})]
and $\Hcb_\Bc^{(1)}$ [Eq.\ (\ref{eq:AFM:BtoP:ham:final})] using Eq.\
(\ref{eq:ME:free-en:def}) and second-order perturbation theory;
\begin{widetext}
\begin{subequations}
  \label{eq:AFM:free-en:matrix-elmnt}
  \begin{align}
    \Av{\Hcb_\Ec^{(1)} + \Hcb_\Bc^{(1)}}_{\nu\sigma\kk_\|}
    & = 2 \Re \sum_{\nu'\ne\nu} \frac{\braket{\nu| \Hcb_\Ec^{(1)} | \nu'}
    \braket{\nu'| \Hcb_\Bc^{(1)} | \nu}}{E_\nu^{(0)} - E_{\nu'}^{(0)}} \\
    & = 2 \Re \sum_{\nu'\ne\nu} \frac{
    \braket{\nu| e \Ec_z \, z | \nu'}
    \Braket{\nu'| \frac{e\hbar}{m} \left(\kk_\parallel - \Av{\vekc{k}_0} \right)
    \cdot \vekc{A}
    - \frac{e}{\hbar}  \mathcal{d} \, \Bc_\parallel \,
     \sin ( \varphi_\Nc + \varphi_\Bc) \, \{ z, k_z^2 - \av{k_z^2} \} | \nu}}
     {E_\nu^{(0)} - E_{\nu'}^{(0)}} \; ,
  \end{align}
\end{subequations}
\end{widetext}
where we ignored terms $\mathcal{O}(\Ec_z^2)$ and
$\mathcal{O}(\Bc_\|^2)$.  When averaging over all occupied
states, the terms $\propto \vekc{A}$ drop out.  Using Eq.\
(\ref{eq:AFM:me-factor}), we get
\begin{equation}
  \label{eq:AFM:free-en:final}
  \delta F = - \Ns\, e \Ec_z \,\, \bohrmag \Bc_\| \,\, \lambda_\mathcal{d} \,
   \sin ( \varphi_\Nc + \varphi_\Bc) \; ,
\end{equation}
consistent with Eqs.\ (\ref{eq:AFM:EtoM:final}) and
(\ref{eq:AFM:BtoP:final}).  Decomposing $\delta F$ into
terms present in the third line of Eq.\ (\ref{eq:Fdens}) yields
\begin{subequations}
\label{eq:AFM:expandFmag}
\begin{equation}
  \label{eq:AFM:free-en:expand}
  \delta F = -\alpha_{zx}\, \Ec_z \Bc_x -\alpha_{zy}\, \Ec_z \Bc_y \quad ,
\end{equation}
with
\begin{equation}\label{eq:AFMalphaVec}
  \vek{\alpha}_{z \|}
  \equiv \begin{pmatrix} \alpha_{zx} \\ \alpha_{zy} \end{pmatrix}
  = \Ns\, e\bohrmag\, \lambda_\mathcal{d}
  \begin{pmatrix} \sin \varphi_\mathcal{N} \\ \cos \varphi_\mathcal{N} \end{pmatrix} \quad .
\end{equation}
\end{subequations}
Thus similar to the ferromagnetic case [Eqs.\
(\ref{eq:expandFmag})], antiferromagnetic order gives rise to
$\alpha_{ij}\ne 0$, and two nonzero entries $\alpha_{zx}$ and
$\alpha_{zy}$ are consistent with the magnetic point group symmetry
$m'mm$ of an antiferromagnetic symmetric quantum well on
a diamond $(001)$ surface.  The antiferromagnetic order could also
generate higher-order magnetoelectric contributions of the type
$\propto \beta_{ijk}$ and $\propto \gamma_{ijk}$ in Eq.\
(\ref{eq:Fdens}). However, unlike the paramagnetic zincblende
structure where $\beta_{ijk}\ne 0$, the high symmetry of a
paramagnetic diamond structure precludes the existence of any
magnetoelectric effects.

Equation~(\ref{eq:AFM:free-en:expand}) can also be expressed as
$\delta F = - \tilde{\vek{\tau}}\cdot (\Ec_z
\hat{\vek{z}}\times\vekc{B}_\|)$ in terms of the magnetoelectric
vector
\begin{equation}
  \label{eq:AFM:free-en:toroid}
\tilde{\vek{\tau}} = \Ns\, e\bohrmag\, \lambda_\mathcal{d}
\begin{pmatrix} -\cos \varphi_\Nc \\ \sin \varphi_\Nc
\end{pmatrix} \quad ,
\end{equation}
which is analogous to the magnetoelectric vector
(\ref{eq:free-en:ME:toroid:vec}) found for the ferromagnetic case
\cite{misc:toroid}.  We have $\tilde{\vek{\tau}} \parallel
-\vek{k}_0$, and, like $\vekc{N}$, the vector $\tilde{\vek{\tau}}$
is a toroidal vector.  Figure~\ref{fig:orient}(b) shows the angular
dependence of the orientation of the vector $\tilde{\vek{\tau}}$ on
the orientation of the vector~$\vekc{N}$.

\subsection{Magnetic quadrupole moment}
\label{sec:afm:qmag}

Similar to Sec.~\ref{sec:fm:qmag}, we can evaluate the magnetic
quadrupole moment in antiferromagnetic systems.  We evaluate the
matrix elements $\av{\{ z^2\, , \vek{v}_\| (\kk_\|) \}}$ similar to
Eq.\ (\ref{eq:z2-v:av}).  We get in first order perturbation theory
\begin{equation}
  \begin{array}[b]{>{\Ds}r>{\Ds}l}
    \av{\{ z^2\, , \vek{v}_\| (\kk_\|) \}}
     = & \av{z^2} \av{\vek{v}_\| (\kk_\|)} \\[1.5ex]
     & {} - \frac{\mathcal{d}}{\hbar}
      \begin{pmatrix}
        \cos \varphi_\Nc \\ - \sin \varphi_\Nc
      \end{pmatrix}
      \av{\{ z^2, k_z^2 - \av{k_z^2} \}}
      \; .
    \end{array}
    \label{eq:afm:z2-v:av}
    \end{equation}
Summing over the Fermi sea, we obtain
\begin{equation}
  \vekc{Q}_{z \|}^\mathrm{o} = - \mathcal{Q}_0 \,\, \frac{l_\mathcal{d}}{w}
  \, \av{\{ z^2, k_z^2 - \av{k_z^2} \}}
  \begin{pmatrix} \sin\varphi_\Nc \\ \cos \varphi_\Nc
  \end{pmatrix} \,\, .
  \label{eq:afm:qmag:final}
\end{equation}

Alternatively, we can obtain the result (\ref{eq:afm:qmag:final}) by
evaluating the free energy $F$ in the presence of the vector
potential $\vekc{A}$ for the field gradient $\vekc{b}_\|$. To first
order in $\vekc{b}_\|$ and the coefficient $\mathcal{d}$, we get an
energy shift of the occupied states given by the expectation value
of
\begin{equation}
  \begin{array}[b]{>{\Ds}r>{\Ds}l}
    H_\mathcal{b}^{(1)}
    = & \frac{e\hbar}{m} \left(\kk_\| - \av{\vekc{k}_0} \right) \cdot \vekc{A}
    \\[1.5ex] & {}
    - \frac{e}{\hbar}  \, \mathcal{d} \, \frac{\mathcal{b}_\|}{2}
     \sin ( \varphi_\Nc + \varphi_\mathcal{b} )
      \{ z^2, k_z^2 - \av{k_z^2} \} \,\, .
    \end{array}
  \end{equation}
When averaging over all occupied states, the first term $\propto
\vekc{A}$ drops out.  We get
\begin{equation}
  \label{eq:free-en:afm:qmag}
  \delta F = \mathcal{Q}_0 \, \frac{\mathcal{b}_\|}{2} \,
  \frac{l_\mathcal{d}}{w} \av{\{ z^2, k_z^2 - \av{k_z^2} \}} \,
  \sin ( \varphi_\Nc + \varphi_\mathcal{b}) \, ,
\end{equation}
consistent with Eqs.\ (\ref{eq:quadmom:def}) and (\ref{eq:afm:qmag:final}).

The contribution (\ref{eq:free-en:afm:qmag}) to the free energy can also be expressed as \cite{misc:toroid}
\begin{subequations}
\label{eq:free-en:afm:qmag:toroid:all}
\begin{equation}
\label{eq:free-en:afm:qmag:toroid}
\delta F = - \bar{\vek{\tau}} \cdot (\vek{\nabla}_{\!z} \times \vekc{B}_\|)
\end{equation}
in terms of the vector
\begin{equation}\label{eq:free-en:afm:qmag:toroid:vec}
  \bar{\vek{\tau}} = \mathcal{Q}_0 \,
  \frac{l_\mathcal{d}}{2w} \av{\{ z^2, k_z^2 - \av{k_z^2} \}}
  \begin{pmatrix} \cos \varphi_\Nc \\ -\sin \varphi_\Nc
  \end{pmatrix} \; ,
\end{equation}
\end{subequations}
compare Eq.\ (\ref{eq:free-en:qmag:toroid:all}).

\subsection{Magnetoelectricity in antiferromagnetic hole systems}
\label{sec:afm-holes}

As was the case in the ferromagnetic configuration, the
magnetoelectric response obtained in the realistic calculations for
electron systems in antiferromagnetic InSb quantum wells is small
[Figs.~\ref{fig:e-mom-150}(b) and~\ref{fig:e-mom-150}(d)].  However,
as before, antiferromagnetic hole systems show much larger
magnetoelectric effects.  The physical origin of this enhancement can
again be traced to the more pronounced asymmetry $E_{n, -\kk_\|} \ne
E_{n\kk_\|}$ and nonparabolicity of quasi-2D hole subbands.
Figures~\ref{fig:h-disp}(c) and
\ref{fig:h-disp}(f) show the energy dispersion and energy contours
for quasi-2D hole systems in antiferromagnetic InSb quantum wells
with width $w = 150\,$\AA\ [Fig.~\ref{fig:h-disp}(c)] and $w =
300\,$\AA\ [Fig.~\ref{fig:h-disp}(f)].  The $\Ec_z$-induced orbital
magnetic moment per particle is plotted in Figs.~\ref{fig:h-mom}(b)
and \ref{fig:h-mom}(d).  Figures~\ref{fig:h-vel}(b),
\ref{fig:h-vel}(d), \ref{fig:h-vel}(f), and \ref{fig:h-vel}(h) show
the equilibrium currents. Finally, Figs.~\ref{fig:h-dip}(b) and
\ref{fig:h-dip}(d) illustrate the $\vekc{B}_\|$-induced
polarization.  Once again, the nonlinear dependence of the
magnetoelectric response on the applied fields is due to the
depopulation of the disconnected parts of the Fermi sea that are
located away from $\kk_\| = \vek{0}$.

\section{Upper bound on magnetoelectric couplings in quasi-2D systems}
\label{sec:bound}

In this section, we derive an upper bound on the magnitude
of the magnetoelectric couplings in 2D quantum-well systems based on
the change $\delta F$ in the free-energy density due to the electric
field $\Ec_z$ and the magnetic field $\vekc{B}_\|$ \cite{bro68a}.
This will illustrate the versatility of the system studied here.  In
generalization of Eq.\ (\ref{eq:ME:free-en:def}) we consider
\begin{equation}
  \label{eq:gen:free-en:def}
  \delta F = \frac{1}{w} \sum_{\nu,\sigma}
  \int \! \frac{d^2 k_\|}{(2\pi)^2}
  \, f (E_{\nu\sigma \kk_\|})
  \, \Av{\Hcb_\Ec^{(1)} + \Hcb_\Bc^{(1)}
  + \Hcb_\Bc^{(2)}}_{\nu\sigma\kk_\|}
  \, ,
\end{equation}
where $\Hcb_\Ec^{(1)}$ [Eq.\ (\ref{eq:ham-cb:E})] and
$\Hcb_\Bc^{(1)}$ [Eq.\ (\ref{eq:ham-cb:B1})] represent the
perturbations linear in the fields $\Ec_z$ and $\vekc{B}_\|$, and
\begin{equation}
  \Hcb_\Bc^{(2)} = \frac{e^2 z^2 \Bc_\|^2}{2m}
  = \frac{m_0^2}{m^2} \, \bohrmag^2 \Bc_\|^2
  \, \frac{2m}{\hbar^2} \, z^2
\end{equation}
is the perturbation quadratic in $\vekc{B}_\|$ appearing in the
Hamiltonian (\ref{eq:ham-cb}).  In generalization of Eq.\
(\ref{eq:ME:free-en:matrix-elmnt}), we obtain up to second order in
the fields $\Ec_z$ and $\vekc{B}_\|$
\begin{widetext}
\begin{subequations}
  \label{eq:free-en:matrix-elmnt}
  \begin{align}
    \Av{\Hcb_\Ec^{(1)} + \Hcb_\Bc^{(1)}
    + \Hcb_\Bc^{(2)}}_{\nu\sigma\kk_\|}
    & = \sum_{\nu'\ne\nu}
    \frac{\bigl| \braket{\nu| \Hcb_\Ec^{(1)}
    + \Hcb_\Bc^{(1)} | \nu'} \bigr|^2}
    {E_\nu^{(0)} - E_{\nu'}^{(0)}}
    + \braket{ \nu| \Hcb_\Bc^{(2)} | \nu}
    \label{eq:free-en:matrix-elmnt:gen} \\
    & = \sum_{\nu'\ne\nu} \frac{
    \bigl| \braket{\nu| \Hcb_\Ec^{(1)} | \nu'} \bigr|^2
    + \bigl| \braket{\nu| \Hcb_\Bc^{(1)} | \nu'} \bigr|^2
    + 2 \Re \braket{\nu| \Hcb_\Ec^{(1)} | \nu'}
    \braket{\nu' | \Hcb_\Bc^{(1)} | \nu} }
    {E_\nu^{(0)} - E_{\nu'}^{(0)}}
    + \braket{ \nu| \Hcb_\Bc^{(2)} | \nu}
    \; . \label{eq:free-en:matrix-elmnt:terms}
  \end{align}
\end{subequations}
\end{widetext}
The second term in Eq.\ (\ref{eq:free-en:matrix-elmnt:gen}) is
always positive, i.e., it describes a diamagnetic energy shift
proportional to $\Bc_\|^2$.  On the other hand, for the lowest
subband $\nu = 0$ the first term in Eq.\
(\ref{eq:free-en:matrix-elmnt:gen}) is always negative
\cite{bro68a}, i.e., it represents a negative definite quadratic
form in the fields $\Ec_z$ and $\vekc{B}_\|$.

We evaluate the different terms in Eq.\
(\ref{eq:free-en:matrix-elmnt}) assuming, as before, that only the
lowest subband $\nu = 0$ is occupied.  The dielectric contribution
to the free energy (\ref{eq:gen:free-en:def}) is
\begin{subequations}
\begin{align}
  \hspace{-1em}
  \delta F^{(2)}_\Ec & = \frac{1}{w} \sum_\sigma
  \int \! \frac{d^2 k_\|}{(2\pi)^2}
  \, f (E_{0 \sigma \kk_\|})
  \sum_{\nu' \ne 0}
  \frac{\bigl| \braket{0 | \Hcb_\Ec^{(1)} | \nu'} \bigr|^2}
    {E_0^{(0)} - E_{\nu'}^{(0)}} \\
  & = \frac{\Ns}{w} \, e^2 \Ec_z^2 \, \lambda_z
\end{align}
\end{subequations}
with
\begin{equation}
\lambda_z = \sum_{\nu'\ne 0} \frac{\left|
\braket{0 | z | \nu'}\right|^2}{E_0^{(0)} - E_{\nu'}^{(0)}}
\; .
\end{equation}
The paramagnetic contribution is
\begin{subequations}
\begin{align}
  \hspace{-0.5em}
  \delta F^{(2,\mathrm{p})}_\Bc & = \frac{1}{w} \sum_\sigma
  \int \! \frac{d^2 k_\|}{(2\pi)^2}
  \, f (E_{0 \sigma \kk_\|})
  \sum_{\nu' \ne 0}
  \frac{\bigl| \braket{0 | \Hcb_\Bc^{(1)} | \nu'} \bigr|^2}
    {E_0^{(0)} - E_{\nu'}^{(0)}} \\
  & = \frac{4\bohrmag^2}{w} \frac{m_0^2}{m^2}
  \sum_{\sigma} \int \! \frac{d^2 k_\|}{(2\pi)^2}
  \, f (E_{0 \sigma \kk_\|})\!
  \left[ \vekc{B}_\| \cdot (\hat{\vek{z}}
  \times \kk_\|)\right]^2 \! \lambda_z \\
  & = \frac{\Ns^2}{w} \, \frac{m_0^2}{m^2}
  \, \bohrmag^2 \Bc_\|^2\, 2\pi \lambda_z
  \times \left\{ \begin{array}{ll} 1 + \xi(\mathcal{Z})^2 & \mbox{FM}
  \\[0.5ex] 1 & \mbox{AFM} \, , \end{array}\right.
\end{align}
\end{subequations}
where we ignored higher-order corrections due to the Dresselhaus
term (\ref{eq:ham-cb:dressel}) (in ferromagnets) or the N\'eel term
(\ref{eq:ham-cb:AFM}) (in antiferromagnets).  The diamagnetic
contribution is \cite{ste68}
\begin{subequations}
\label{eq:free-en:diamagnetic}
\begin{align}
  \delta F^{(2,\mathrm{d})}_\Bc & = \frac{1}{w} \sum_{\sigma}
  \int \! \frac{d^2 k_\|}{(2\pi)^2}
  \, f (E_{0 \sigma \kk_\|})
  \, \braket{ 0 | \Hcb_\Bc^{(2)} | 0} \\
  & = \frac{\Ns}{w} \, \frac{m_0^2}{m^2}
      \, \bohrmag^2 \, \Bc_\|^2 \, \frac{2m}{\hbar^2}
      \, \braket{ 0| z^2 |0} \; .
  \end{align}
\end{subequations}
The magnetoelectric contribution
\begin{align}
  \delta F^{(2)}_{\Ec \Bc} & = \frac{2}{w} \sum_{\sigma}
  \int \! \frac{d^2 k_\|}{(2\pi)^2}
  \, f (E_{0 \sigma \kk_\|})
  \nonumber \\* & \hspace{2em} {} \times
    \Re \sum_{\nu' \ne 0} \frac{
    \braket{0 | \Hcb_\Ec^{(1)} | \nu'}
    \braket{\nu' | \Hcb_\Bc^{(1)} | 0} }
    {E_0^{(0)} - E_{\nu'}^{(0)}}
\end{align}
was evaluated in Eqs.\ (\ref{eq:ME:free-en:matrix-elmnt}) (for
ferromagnets) and (\ref{eq:AFM:free-en:matrix-elmnt}) (for
antiferromagnets).

Explicit evaluation of relevant matrix elements for an infinitely
deep square well and a parabolic (harmonic-oscillator) potential
yields \cite{end89}
\begin{subequations}
\begin{align}
  \label{eq:coeff:z2:explicit}
  \lambda_{z} & = - \frac{m w^4}{\hbar^2} \times \left\{ \begin{array}{>{\Ds}ls{1em}l}
   \frac{15-\pi^2}{24 \pi^4} & \mbox{square} \\[2ex]
   1/2 & \mbox{parabolic} \end{array} \right. \\[2ex]
  \braket{ 0| z^2 |0} & = w^2 \times \left\{ \begin{array}{>{\Ds}ls{1em}l}
    \frac{\pi^2-6}{12\pi^2} & \mbox{square} \\[2ex]
    1/2 & \mbox{parabolic} \end{array} \right.
\end{align}
\end{subequations}
using the relation $w = \sqrt{\hbar/(m\omega)}$ between well
width $w$ and harmonic-oscillator frequency $\omega$.

We write $\delta F$ in Eq.\ (\ref{eq:gen:free-en:def}) in the form
of Eq.\ (\ref{eq:Fdens}), restricting ourselves to terms quadratic in $\Ec_z$ and $\vekc{B}_\|$,
\begin{align}
  \delta F & = - \frack{1}{2} \, \chi^\Ec_{zz}\, \Ec_z^2
  - \frack{1}{2} \, (\chi^{\Bc, \mathrm{p}}_{i i}
  + \chi^{\Bc, \mathrm{d}}_{i i}) \, \Bc_i^2
  \nonumber \\*[1ex] & \hspace{1.0em} {}
  -\alpha_{z x}\,\Ec_z \Bc_x -\alpha_{z y}\,\Ec_z \Bc_y \; ,
  \label{eq:gen:free-en:EB}
\end{align}
where ($i=x,y$)
\begin{subequations}
  \begin{align}
    \chi^\Ec_{zz}
    & = - \frac{\partial^2 \bigl(\delta F^{(2)}_\Ec \bigr)}
    {\partial \Ec_z^2} \; , \\
    \chi^{\Bc,k}_{ii} & = - \frac{\partial^2 \bigl(\delta F^{(2,k)}_\Bc \bigr)}
    {\partial \Bc_i^2} \; , \\
    \alpha_{zi} & = - \frac{\partial^2 \bigl(\delta F^{(2,k)}_\Bc \bigr)}
    {\partial \Ec_z \, \partial \Bc_i} \; .
  \end{align}
\end{subequations}
We obtain the following explicit expressions for these
susceptibilities
\begin{subequations}
  \begin{align} \label{eq:chi:E:final}
  \chi^\Ec_{zz}
    & = - \frac{\Ns}{w} \, e^2 \, 2 \lambda_z
    \; , \\
    \label{eq:chi:para-B:final}
    \chi^{\Bc, \mathrm{p}}_{ii}
  & = - \frac{\Ns^2}{w} \, \frac{m_0^2}{m^2}
  \, \bohrmag^2 \, 4\pi \lambda_z
  \times \left\{ \begin{array}{ll} 1 + \xi(\mathcal{Z})^2 & \mbox{FM}
  \\[0.5ex] 1 & \mbox{AFM} \, , \end{array}\right. \\
    \chi^{\Bc, \mathrm{d}}_{ii}
  & = - \frac{\Ns}{w} \, \frac{m_0^2}{m^2}
      \, \bohrmag^2 \, \, \frac{4 m}{\hbar^2}
      \, \braket{ 0| z^2 |0} \; , \label{eq:chi:dia-B:final}
  \end{align}
\end{subequations}
and $\vek{\alpha}_{z \|}$ was given in Eq.\ (\ref{eq:alphaVec})
[Eq.\ (\ref{eq:AFMalphaVec})] for the ferromagnetic
(antiferromagnetic) case.

\begin{table*}[t]
  \caption{\label{tab:susceptibilities} Parametric dependences of
  the electromagnetic susceptibilities on well width $w$ and density
  $\Ns$ for a quasi-2D electron system in ferromagnetic and
  antiferromagnetic quantum wells.  In ferromagnetic systems, entries
  in the first row apply to a partially spin-polarized system, whereas
  entries in the second row (if present) apply to a fully spin-polarized
  (half-metallic) system.  All quantities are defined per volume.
  To estimate the relative importance of these terms, the last row
  gives numerical values of the susceptibilities for a 2D electron
  system in a 150-{\AA}-wide square quantum well with density $\Ns =
  1.0 \times 10^{11}$~cm$^{-2}$ and parameters $m = 0.0139 \, m_0$,
  $g = 51.5$, $d = 760$~eV\AA$^3$, $\mathcal{X}_\| = 8\,$meV (for
  $\alpha_{zi}^\mathrm{FM}$), and $\mathcal{d} = 80$~eV\AA$^3$ (for
  $\alpha_{zi}^\mathrm{AFM}$).}
  \renewcommand{\arraystretch}{1.1}
\begin{tabular}{@{}C*{7}{s{0.80em}C}@{}}
\hline\hline \rule{0pt}{3.5ex}
\chi^\Ec_{zz} & \chi^{\Bc, \mathrm{p}}_{ii} & \chi^{\Bc, \mathrm{d}}_{ii}
& \sqrt{\chi^\Ec_{zz} \, \chi^{\Bc, \mathrm{p}}_{ii}}
& \alpha_{zi}^\mathrm{FM} & \alpha_{zi}^\mathrm{AFM}
& \chi^\mathcal{Z}_{ii} & \mathcal{M}^\mathrm{s}_\| \\ \hline
\rule{0pt}{2.8ex}
\propto w^3 \Ns & \propto w^3 \Ns^2 & \propto w \Ns & \propto w^3 \Ns^{3/2} &
\propto w & \propto w \Ns & \propto w^{-1} & \propto w^{-1} \\
& & & &  \propto w \Ns & & & \propto w^{-1} \Ns \\
\rule{0pt}{2.8ex}
4.9 {\times} 10^{-2} \, \epsilon_0 &
7.2 {\times} 10^{-8} / \mu_0&
- 1.2 {\times} 10^{-6} / \mu_0 &
5.9 {\times} 10^{-5} \sqrt{\epsilon_0 / \mu_0} &
5.8 {\times} 10^{-6} \sqrt{\epsilon_0 / \mu_0} &
1.3 {\times} 10^{-6} \sqrt{\epsilon_0 / \mu_0} &
1.7 {\times} 10^{-6} / \mu_0 &
12 \, \mathcal{M}_0
\\ \hline \hline
\end{tabular}
\end{table*}

The first term in Eq.\ (\ref{eq:free-en:matrix-elmnt:gen})
yields a contribution to the free energy that can be written as
\begin{equation}
  \delta F^{(2)}_\Ec + \delta F^{(2,\mathrm{p})}_\Bc
  + \delta F^{(2)}_{\Ec \Bc}
  = - \frack{1}{2} \, \vekc{F}^\dagger \cdot \vek{\chi} \cdot \vekc{F}
\end{equation}
where $\vekc{F}^\dagger \equiv (\Ec_z, \Bc_x, \Bc_y)$, and
\begin{equation}
  \vek{\chi} =
  \begin{pmatrix}
    \chi^\Ec_{zz} & \alpha_{zx} & \alpha_{zy} \\
    \alpha_{zx} & \chi^{\Bc, \mathrm{p}}_{xx} & 0 \\
    \alpha_{zy} & 0 & \chi^{\Bc, \mathrm{p}}_{yy}
  \end{pmatrix}
\end{equation}
is a positive definite symmetric matrix.  It follows from
Sylvester's criterion for positive-definiteness of symmetric
matrices that we obtain upper bounds for the magnitude of the
components of the magnetoelectric tensor \cite{bro68a, misc:odell}
\begin{equation}
  \label{eq:alpha:qm:bound}
  |\alpha_{zi}| \le \sqrt{\chi^\Ec_{zz} \,
  \chi^{\Bc, \mathrm{p}}_{ii}} \; .
\end{equation}
In bulk materials, the electric and paramagnetic susceptibilities
$\chi^\Ec_{ij}$ and $\chi^{\Bc, \mathrm{p}}_{ij}$ represent
generally fixed properties of the underlying material, and Eq.\
(\ref{eq:alpha:qm:bound}) has previously been invoked in order to
explain why frequently the magnetoelectric coefficients
$\alpha_{ij}$ are small in magnitude \cite{bro68a}.  It is a unique
feature of the quasi-2D systems studied here that the properties
represented by the elements of the tensor $\vek{\chi}$ can be
engineered \cite{esa90, hei10}.  This is illustrated by the pronounced
dependence of the coefficient $\lambda_z$ on well width $w$ found in
Eq.\ (\ref{eq:coeff:z2:explicit}), which is matched by Eq.\
(\ref{eq:magic-sum}) for the coefficient $\lambda_d$ showing that
the magnetoelectric coefficients $\alpha_{zi}$ likewise increase
with increasing width of the quantum well.  (The tensor $\vek{\chi}$
gives the susceptibilities per volume.)  Hence the magnetoelectric
response can be maximized in a superlattice consisting of
wide quantum wells.  The susceptibilities $\chi^\Ec_{zz}$
and $\chi^{\Bc, \mathrm{p}}_{ii}$ scale also with the 2D density
$\Ns$ in the quantum well that can easily be tuned
experimentally over a wide range via doping and electric gates
\cite{hei10}.  Again, this is matched by the density dependence of
$\alpha_{zi}$ in the antiferromagnetic case [Eq.\
(\ref{eq:AFMalphaVec})] and in the half-metallic regime $\mathcal{Z}
\ge E_\mathrm{F}^0$ of the ferromagnetic case [Eq.\
(\ref{eq:alphaVec})].
Explicitly, we have [ignoring $|\xi(\mathcal{Z})| \le 1$ in Eq.\
(\ref{eq:chi:para-B:final})] \cite{misc:brown:dresselhaus}
\begin{equation}
  \sqrt{\chi^\Ec_{zz} \, \chi^{\Bc, \mathrm{p}}_{ii}}
  = \frac{\Ns^{3/2}}{w} \, e\, \bohrmag \, \frac{m_0}{m}
     \, \sqrt{8\pi} \, \left|\lambda_z \right| \; .
\end{equation}
To illustrate the tunability of Eq.\ (\ref{eq:alpha:qm:bound}), we
summarize in Table~\ref{tab:susceptibilities} the parametric
dependences of the susceptibilities on well width $w$ and density
$\Ns$ for a quasi-2D electron system in ferromagnetic and
antiferromagnetic quantum wells.  Furthermore, to estimate the
relative importance of these terms, Table~\ref{tab:susceptibilities}
also gives numerical values of the susceptibilities using the
analytical results derived above and considering a 2D electron
system in a 150-{\AA}-wide square quantum well with density $\Ns =
1.0 \times 10^{11}$~cm$^{-2}$

For completeness, we remark that the free energy also contains a
term representing the spin magnetization (\ref{eq:spin-mag:def}) due
to the Zeeman field $\vekc{Z} = (g/2) \, \bohrmag \, \vekc{B}_\| +
\vekc{X}_\|$,
\begin{equation}
  \label{eq:gen:free-en:pauli}
  \delta F_\mathcal{Z}
  = - \frac{\Ns}{w} \, \frac{g}{2} \, \bohrmag \Bc_\| \, \xi(\Zc)
  = \mathcal{M}_0 \, \frac{g}{2} \, \Bc_\| \, \xi(\Zc) \; .
\end{equation}
This term includes a contribution quadratic in the external field
$\vekc{B}_\|$ that corresponds to the paramagnetic Pauli spin
susceptibility
\begin{equation}
  \chi^\mathcal{Z}_{ii}
  = - \frac{\partial^2 \bigl(\delta F_\mathcal{Z} \bigr)}{\partial \Bc_i^2}
  = \frac{1}{w} \left( \frac{g}{2} \, \bohrmag \right)^2
   \frac{m}{\pi \hbar^2} \; .
\end{equation}
Our discussion of AFM diamond in Sec.~\ref{sec:AFM} ignored the
effect of $\chi^\mathcal{Z}_{ii}$.  In ferromagnets, the exchange
field $\vekc{X}_\|$ yields a contribution to $\delta F_\mathcal{Z}$
linear in the external field $\vekc{B}_\|$ that represents the
spontaneous magnetization
\begin{equation}
  \vekc{M}^\mathrm{s}_\|
  = - \mathcal{M}_0 \, \frac{g}{2} \times
  \left\{ \begin{array}{cs{1em}l@{}}
      \Ds \frac{m}{\pi \hbar^2} \, \frac{\vekc{X}_\|}{\Ns} , &
      \Xc_\| < E_\mathrm{F}^0 \\[2.0ex]
      \hat{\vekc{X}}_\| , & \Xc_\| \ge E_\mathrm{F}^0  \; ,
    \end{array}\right.
\end{equation}
compare Eq.\ (\ref{eq:Fdens}).

\section{Conclusions and outlook}
\label{sec:concl}

We present a detailed theoretical study of how magnetoelectricity
arises in magnetically ordered quantum wells with broken
time-reversal symmetry and broken space-inversion symmetry.
Quasi-2D systems based on zincblende ferromagnets
[Fig.~\ref{fig:diamond}(b)] and diamond-structure antiferromagnets
[Fig.~\ref{fig:diamond}(c)] exhibit an analogous linear
magnetoelectric response, i.e., an in-plane magnetization induced by
a perpendicular electric field [Eqs.\ (\ref{eq:EtoM:final}) and
(\ref{eq:AFM:EtoM:final})], as well as a perpendicular electric
polarization arising from an in-plane magnetic field [Eqs.\
(\ref{eq:BtoP:final}) and (\ref{eq:AFM:BtoP:final})].  In
realistic calculations, the magnitude of the magnetoelectric
response is small in quasi-2D electron system
(Fig.~\ref{fig:e-mom-150}), but it is sizable for quasi-2D hole
systems (Figs.~\ref{fig:h-mom} and \ref{fig:h-dip}).  See
Table~\ref{tab:compareME} for a comparison of benchmark values for
our systems of interest with other known magnetoelectric materials.
While typical magnitudes of the magnetoelectric-tensor components
are comparable to the those of Cr$_2$O$_3$, the maximum
electric-field-induced magnetization per particle reaches the same
large order of magnitude ($\sim 1\bohrmag$) as demonstrated for the
giant magnetoelectric effect in FeRh/BTO.
Our findings suggest that bandstructure engineering and
nanostructuring are fruitful avenues for generating and tailoring
magnetoelectricity in a host of materials.

Our study yields a new unified picture of magnetic order.
Ferromagnetic order is characterized by a magnetic-moment density
$\vekc{M}$ (a magnetization).  In itinerant-electron systems,
orbital ferromagnetic order is associated with dipolar equilibrium
currents.  On the other hand, collinear orbital antiferromagnetic
order is characterized by a toroidal-moment density
$\av{\vek{\tau}}$ for the N\'eel operator $\vek{\tau}$ that is
associated with quadrupolar equilibrium currents.  For the
itinerant-electron systems studied in the present work, the
equilibrium current distributions are slowly varying on the length
scale of the lattice constant (Figs.~\ref{fig:e-vel}
and~\ref{fig:h-vel}).  The magnetization $\vekc{M}$ and the
toroidal-moment density $\av{\vek{\tau}}$ quantify complementary
aspects of itinerant-electron collinear magnetic order in solids.
Ferrimagnetic systems are characterized by both expectation values
$\vekc{M}$ and $\av{\vek{\tau}}$ being finite simultaneously.
Generally, the manipulation of itinerant-electron ferromagnetic or
antiferromagnetic order via external perturbations can be viewed as
manipulating the underlying equilibrium current distribution
(Figs.~\ref{fig:e-vel} and~\ref{fig:h-vel}).

Ferromagnetic order $\vekc{M}$ arises due to the presence of an
exchange field or an external magnetic field, but it may also arise
due to, e.g., an electric field (the magnetoelectric effect studied
here) or a strain field (piezomagnetism \cite{tav56, dzi58, lan84}).
Similarly, antiferromagnetic order $\av{\vek{\tau}}$ can be due to a
staggered exchange field.  But it may also arise due to, e.g., the
interplay of ferromagnetic order, spin-orbit coupling, and
confinement [Eq.\ (\ref{eq:EtoM:curr})].  The explicit form of the
N\'eel operator $\vek{\tau}$ and how it can be manipulated depends
on the symmetry of the system under investigation.  In the present
work, we used the envelope-function theory to derive explicit
expressions for $\vek{\tau}$ in antiferromagnetic diamond
structures.  The theory for how $\vekc{M}$ and $\av{\vek{\tau}}$ are
induced by external perturbations can be phrased very generally
using the theory of material tensors taking advantage of crystal
symmetry \cite{nye57, bir64, bir74, new05}.  In the magnetoelectric
effect an $I$-odd $\Theta$-even electric field $\vekc{E}$ induces an
$I$-even $\Theta$-odd magnetization $\vekc{M}$, which is permitted
in thermal equilibrium for magnetic media breaking both space
inversion symmetry $I$ and time-reversal symmetry $\Theta$
(Appendix~\ref{app:current-induced-magnetization}).  Similarly, an
electric field $\vekc{E}$ can induce antiferromagnetic order
represented via the $I$-odd $\Theta$-odd toroidal moment
$\av{\vek{\tau}}$ if the medium breaks time-reversal symmetry
$\Theta$, while the medium may preserve space-inversion symmetry
$I$.  Such a nondissipative \emph{antiferromagnetoelectric effect}
$\av{\tau_i} = \zeta_{ij} \Ec_j$ is characterized via an $I$-even
$\Theta$-odd second-rank tensor $\zeta_{ij}$.  The constraints due
to crystal symmetry for a nonzero tensor $\zeta_{ij}$ are fulfilled,
e.g., by antiferromagnetic MnF$_2$ (magnetic point group $4'/mm'm$);
this effect will be discussed in more detail in a future
publication.  It expands recent efforts geared towards an electric
manipulation of antiferromagnetic order \cite{zel14, wad16, zel17,
wat18b, man19}.

Beyond that, the theoretical formalism and fundamental understanding
of antiferromagnetic order presented in this work can be applied to
undertake more comprehensive studies of itinerant-electron
antiferromagnets.  Reliable modeling of
antiferromagnetic-spintronics devices \cite{jun16, bal18} requires
the level of detail and realism provided by our envelope-function
theory.  Basic questions concerning magnetization dynamics in
metallic antiferromagnets that are attracting current interest
\cite{sim20} can also be addressed.

\begin{acknowledgments}
  RW and UZ acknowledge stimulating discussions with A.~Hoffmann and
  H.~Saglam.  In addition, they thank A.~Hoffmann for support.  RW
  also benefitted from discussions with D.~Cahill, D.~M.\ Ceperley,
  M.~Gilbert, T.~Hughes, K.~Kang, E.~I.\ Rashba, A.~Schleife,
  M.~Shayegan, D.~Shoemaker, J.~Sipe, and G.~Vignale.  UZ's interest
  in the magnetoelectricity of quantum wells was initiated by
  interesting conversations with B.~Weber, and he also thanks J.~B.\
  Curtis, R.~A.\ Duine, J.~C.\ Egues, I.~Garate, L.~I.\ Glazman,
  A.~Kamra, and B.~I.\ Shklovskii for useful discussions.  This work
  was supported by the NSF under Grant No.\ DMR-1310199 and by the
  Marsden Fund Council (Contract No.\ VUW1713) from New Zealand
  government funding managed by the Royal Society Te Ap\=arangi.
  Work at Argonne was supported by DOE BES under Contract No.\
  DE-AC02-06CH11357.  Research at UIUC was supported by the Illinois
  Materials Research Science and Engineering Center, supported by
  the NSF MRSEC program under Grant No.\ DMR-1720633.  RW
  acknowledges the kind hospitality of the collaborative Research
  Center CRC 1277 during a 5-months stay at the Physics Department
  in Regensburg.  This stay and parts of the work were funded by the
  Deutsche Forschungsgemeinschaft (DFG, German Research Foundation)
  Project-ID 314695032 through CRC 1277.  Work at the Kavli
  Institute for Theoretical Physics was supported by the NSF under
  Grant No.\ PHY-1748958.
\end{acknowledgments}

\appendix

\section{Comparison of magnetoelectricity with current-induced
magnetization}
\label{app:current-induced-magnetization}

It is the hallmark of the magnetoelectric effect that an
electric-field-induced magnetization and a magnetic-field-induced
polarization arise in thermal equilibrium, and that these responses
are duals of each other in that they are characterized by the same
magnetoelectric tensor $\alpha_{ij}$, see Eqs.\ (\ref{eq:Fdens}) and
(\ref{eq:moments:gen-def}) \cite{lan84}.  The central requirement
for the occurrence of magnetoelectricity is that space-inversion
symmetry $I$ and time-reversal symmetry $\Theta$ are both broken;
hence, magnetoelectricity is only possible in certain magnetic
systems.  More precisely, using the theory of material tensors
\cite{nye57, bir64, bir74, new05} magnetoelectricity is permitted
for those magnetic crystal classes characterized by a magnetic point
group $\mathcal{G}$, where the polar ($I$-odd), $\Theta$-even
vectors $\vekc{E}$ and $\vekc{P}$ and the axial ($I$-even),
$\Theta$-odd vectors $\vekc{B}$ and $\vekc{M}$ transform according
to the same representation of $\mathcal{G}$, i.e., $\alpha_{ij}$
must be an axial, $\Theta$-odd second-rank tensor. For each of the
58 groups $\mathcal{G}$ permitting an axial $\Theta$-odd second-rank
tensor \cite{ode70}, the patterns of nonzero components
$\alpha_{ij}$ allowed by symmetry have been tabulated, e.g., in
Ref.~\cite{new05}.

An at first glance closely related effect is a magnetization
$\mathcal{M}_i$ induced by a spin-unpolarized electric-charge
current $\mathcal{J}_j$, characterized by the relation \cite{ivc78,
bel78, aro91, ede90}
\begin{equation}
  \label{eq:current-induced-magnetization}
  \mathcal{M}_i = \eta_{ij} \, \mathcal{J}_j \; ,
\end{equation}
where $\eta_{ij}$ is a second-rank tensor.  The current $\vekc{J}$
is a polar vector, whereas the magnetization $\vekc{M}$ is an axial
vector (and both quantities are $\Theta$-odd).  Accordingly, a
\emph{current-induced magnetization}
(\ref{eq:current-induced-magnetization}) is permitted for those
nonmagnetic crystal classes characterized by a nonmagnetic point
group $G$, where the polar vector $\vekc{J}$ and the axial vector
$\vekc{M}$ transform according to the same representation of $G$,
i.e., $\eta_{ij}$ must be an axial (and $\Theta$-even) second-rank
tensor.  The 18 groups $G$ that permit a nonzero axial tensor
$\eta_{ij}$ are known as gyrotropic point groups
\cite{misc:gyro}.  Current-induced magnetization
(\ref{eq:current-induced-magnetization}) is forbidden for the
nonmagnetic bulk zincblende structure [point group $G = T_d =
\bar{4}3m$, Fig.~\ref{fig:diamond}(b)], despite the fact that
inversion symmetry is broken in the zincblende structure
\cite{misc:nongyrotropic}.  Current-induced magnetization in
nonmagnetic media has been reviewed, e.g., in Refs.~\cite{ivc08,
gan08, gan12}.

The symmetry requirements permitting a current-induced
magnetization are fundamentally distinct from those permitting
magnetoelectricity.  While magnetoelectricity is forbidden for
nonmagnetic media, current-induced magnetization is already
allowed in nonmagnetic media.  Extending the discussion to magnetic
media, magnetoelectricity is allowed, e.g., for systems with the bulk
antiferromagnetic diamond structure [magnetic point group
$\mathcal{G} = 4'/m'm'm$, Fig.~\ref{fig:diamond}(c)] \cite{new05}.
At the same time, the axial $\Theta$-even second-rank tensor $\eta_{ij}$
describing current-induced magnetization must vanish for systems
with $\mathcal{G} = 4'/m'm'm$.  Contrasting that, a ferromagnetic bulk
zincblende structure magnetized in $z$ direction (point group
$\mathcal{G} = \bar{4}m'2'$) permits both magnetoelectricity and
current-induced magnetization.

If the current $\mathcal{J}_j$ is induced by an external electric
field $\Ec_k$ via Ohm's law $\mathcal{J}_j = \sigma_{jk} \Ec_k$ (in
studies of current-induced magnetization, the conductivity tensor
$\sigma_{jk}$ is often treated within a simple Drude model
\cite{gan08, gan12}), such a dissipative current breaks
time-re\-ver\-sal symmetry even in nonmagnetic media, and we can
rewrite Eq.\ (\ref{eq:current-induced-magnetization}) as
\begin{equation}
  \label{eq:kinetic-ME}
  \mathcal{M}_i = \eta_{ij} \, \sigma_{jk} \, \Ec_k
  = \eta_{ik}' \, \Ec_k \; ,
\end{equation}
with $\eta_{ik}' \equiv \eta_{ij} \, \sigma_{jk}$, compare Eq.\
(\ref{eq:moments:gen-def:mag}).  Accordingly, the current-induced
magnetization has sometimes been called the kinetic magnetoelectric
effect \cite{lev85, ede90}, though it is clear from the above
discussion that the physics expressed by Eq.\ (\ref{eq:kinetic-ME})
is fundamentally distinct from magnetoelectricity.  In particular,
Refs.~\cite{lev85, wat18a} stressed the dissipative character of
Eq.\ (\ref{eq:kinetic-ME}), whereas the actual magnetoelectric
effect constitutes an equilibrium phenomenon.  In contrast to the
magnetoelectric effect, Eq.\ (\ref{eq:kinetic-ME}) has no dual
whereby a magnetic field $\vekc{B}$ could induce a polarization
$\vekc{P}$ in a nonmagnetic medium.

\section{Gauge dependence of the magnetic-moment operator for
itinerant electrons}
\label{app:mag-mom-op}

The magnetic-moment operator for a system with Hamiltonian $H$ is
generally defined as~\cite{whi07}
\begin{equation}
  \label{eq:mag-mom-op:def}
  \vekc{m} = - \frac{\partial H}{\partial \vekc{B}} \; .
\end{equation}
In a single-particle picture for itinerant electrons with kinetic
momentum $\hbar\vekc{k} = \hbar\kk + e \vekc{A}$, where $\vekc{A}$
is the vector potential for the magnetic field $\vekc{B} =
\vek{\nabla} \times \vekc{A}$, we get
\begin{equation}
  \mathcal{m}_i = - \left\{ \frac{\partial H}{\partial \mathcal{k}_j}
  , \frac{\partial \mathcal{k}_j}{\partial \Bc_i} \right\}
  = - e \left\{ v_j ,
    \frac{\partial \mathcal{A}_j}{\partial \Bc_i} \right\} \; ,
\end{equation}
where $\vek{v} = \partial H / (\partial \hbar \vekc{k}) = \partial H
/ (\partial \hbar \kk)$ is the velocity operator and we took the
symmetrized product of noncommuting operators.  Repeated indices are
summed over.  For the symmetric gauge $\vekc{A}^\mathrm{sym} =
\frac{1}{2} \, \vekc{B} \times \vek{r}$, we have
\begin{equation}
  \frac{\partial \mathcal{A}_j^\mathrm{sym}}{\partial \Bc_i}
  = - \frac{1}{2} \, \epsilon_{ijk} \, r_k \; ,
\end{equation}
where $\epsilon_{ijk}$ denotes the totally antisymmetric tensor.
Thus
\begin{equation}
   \label{eq:mag-mom-op:sym}
  \mathcal{m}^\mathrm{sym}_i
  = - \frac{e}{2} \, \epsilon_{ijk} \, \{ r_j , v_k \} \; ,
\end{equation}
which is the conventional formula for the magnetization \cite{res10,
whi07} consistent with classical electromagnetism \cite{jac99}.  On
the other hand, we get for the asymmetric gauge $\vekc{A} = z \,
\vekc{B}_\| \times \hat{\vek{z}}$ employed in the present work
\begin{equation}
  \vekc{m} = - e \, \hat{\vek{z}} \times \{z, \vek{v}_\| \} \; ,
\end{equation}
whose components differ by a factor of $2$ from corresponding terms
with $r_j = z$ in Eq.\ (\ref{eq:mag-mom-op:sym}).  Both expressions
for $\vekc{m}$ are consistent with \cite{jac99}
\begin{equation}
  \vekc{j} = - e \, \vek{v} = - \vek{\nabla} \times \vekc{m} \; .
\end{equation}

Similar to the definition (\ref{eq:mag-mom-op:def}) of the
magnetic-moment operator, the operator of the electric dipole moment
can be defined as $\vekc{p} = -\partial H / \partial \vekc{E}$.  The
electric field $\vekc{E}$ can be introduced into $H$ via a scalar
potential as in Eq.\ (\ref{eq:ham-cb:E}), or via a time-dependent
vector potential.  Therefore, the explicit form of the
electric-dipole moment operator $\vekc{p}$ is also gauge-dependent.

\section{Orbital magnetization induced by Zeeman coupling}
\label{app:orb-magnet}

Very generally, even in the absence of an electric field $\Ec_z$,
the Zeeman term induces a spin magnetization $\vekc{S}_\Zc$
(anti)parallel to the Zeeman field $\vekc{Z}$ and proportional to
the $g$ factor, see Eq.\ (\ref{eq:spin-mag:def}).  However, in a
more complete multiband description, the $g$ factor for an explicit
Zeeman term may be greatly reduced or completely absent
\cite{bow59}.  In such an approach, we obtain instead an orbital
magnetization $\vekc{M}_\Zc$ due to equilibrium spin-polarized
currents, for which spin-orbit coupling plays an essential role.  We
demonstrate in the following that the spin magnetization
$\vekc{S}_\Zc$ in a single-band model with $g$-factor $g$ is equal
to the orbital magnetization $\vekc{M}_\Zc$ in the corresponding
multiband model.  While we focus for conceptual clarity on the
simpler case of a magnetization due to Zeeman coupling to a magnetic
field, the arguments apply also to a magnetization induced, e.g., by
an electric field in the magnetoelectric effect.

It is well-known \cite{blo62} that the multiband description of
Bloch electrons is analogous to a fully relativistic description of
electrons based on the Dirac equation.  Accordingly, the observable
predictions of multiband theories embrace those of single-band
theories similar to how the observable predictions of fully
relativistic theories based on the Dirac equation embrace those of
weakly relativistic theories based on the Pauli equation.  The Pauli
equation includes a Zeeman term with $g$-factor $g$ that appears as
a prefactor for the spin magnetization (\ref{eq:spin-mag:def}).  On
the other hand, the Dirac equation does not contain a Zeeman term;
but the interaction of the electrons with a magnetic field is
entirely accounted for via the minimal coupling to the vector
potential for the magnetic field (i.e., we have $g=0$ in the Dirac
equation).  Accordingly, $\vekc{S}_\Zc$ must vanish in a fully
relativistic theory; and the observable magnetization is entirely
orbital even for strongly localized magnetic moments on the atoms
that are commonly modeled as spin magnetic moments.  This is a
direct consequence of the Dirac theory.

Working in a multiband theory, we demonstrate the equivalence of
$\vekc{S}_\Zc$ and $\vekc{M}_\Zc$ for the $8 \times 8$ Kane model
\cite{kan56, kan57, win03}, where the orbital magnetization
$\vekc{M}_\Zc$ is due to the off-diagonal coupling between the
conduction and valence bands linear in $\kk$.  The physics that is
essential for $\vekc{M}_\Zc$ is thus contained in the simplified
Kane Hamiltonian \cite{bow59, win03}
\begin{widetext}
\begin{equation}
 \label{eq:kane:simple}
  \tilde{\HK} =
  \begin{pmatrix}
    E_c + h_c
&   0
&   - \frac{1}{\sqrt{2}} P k_+
&   \sqrt{\frac{2}{3}} P k_z
&   \frac{1}{\sqrt{6}} P k_-
&   0
&   - \frac{1}{\sqrt{3}} P k_z
&   - \frac{1}{\sqrt{3}} P k_- \\[1ex]
    0
&   E_c + h_c
&   0
&   - \frac{1}{\sqrt{6}} P k_+
&   \sqrt{\frac{2}{3}} P k_z
&   \frac{1}{\sqrt{2}} P k_-
&   - \frac{1}{\sqrt{3}} P k_+
&   \frac{1}{\sqrt{3}} P k_z \\
  - \frac{1}{\sqrt{2}} P k_-
& 0
& E_v + h_v & 0 & 0 & 0 & 0 & 0 \\[1ex]
    \sqrt{\frac{2}{3}} P k_z
&   - \frac{1}{\sqrt{6}} P k_-
& 0 & E_v + h_v & 0 & 0 & 0 & 0 \\[1ex]
    \frac{1}{\sqrt{6}} P k_+
&   \sqrt{\frac{2}{3}} P k_z
& 0 & 0 & E_v + h_v & 0 & 0 & 0 \\[1ex]
    0
&   \frac{1}{\sqrt{2}} P k_+
& 0 & 0 & 0 & E_v + h_v & 0 & 0 \\[1ex]
    - \frac{1}{\sqrt{3}} P k_z
&   - \frac{1}{\sqrt{3}} P k_-
& 0 & 0 & 0 & 0 & E_v - \Delta_0 + h_v & 0 \\[1ex]
    - \frac{1}{\sqrt{3}} P k_+
&   \frac{1}{\sqrt{3}} P k_z
& 0 & 0 & 0 & 0 & 0 & E_v - \Delta_0 + h_v
\end{pmatrix} \,\, .
\end{equation}
\end{widetext}
Here $E_c$ denotes the conduction band edge ($\Gamma_6^c$), $E_v
\equiv E_c - E_0$ is the valence band edge ($\Gamma_8^v$) with
fundamental gap $E_0$, $\Delta_0$ is the spin-orbit gap between the
topmost valence band $\Gamma_8^v$ and the spin split-off valence
band $\Gamma_7^v$, and $P$ denotes Kane's momentum matrix element.
The terms $h_c = \mu_c k_z^2 + V_c (z)$ and $h_v = - \mu_v k_z^2 -
V_v (z)$ embody remote-band contributions quadratic in $k_z$ with
\mbox{$\mu_c, \mu_v > 0$} and confining potentials $V_c (z), V_v(z)
\ge 0$.

While $g=0$ for the Hamiltonian $\tilde{\HK}$, a spin magnetization
$\vekc{S}_\Zc$ is obtained when $\tilde{\HK}$ is projected on the
$\Gamma_6^c$ conduction band, yielding a $2 \times 2$ Hamiltonian as
in Eq.\ (\ref{eq:ham-cb}) including a Zeeman term $\Hcb_\Zc$ with
$g$-factor $g$.  To express $g$ in terms of the parameters of
$\tilde{\HK}$, we decompose $\tilde{\HK} = \tilde{\HK} {}^{(0)} +
\tilde{\HK} {}^{(1)}$, where $\tilde{\HK} {}^{(0)}$ contains the
diagonal elements of $\tilde{\HK}$, while $\tilde{\HK} {}^{(1)}$
contains the off-diagonal terms linear in $\kk$.  The eigenstates of
$\tilde{\HK} {}^{(0)}$ are bound states $\ket{\beta,\nu\sigma}
\equiv \ket{\beta,\nu} \otimes \ket{\sigma}$ in the conduction band
$\Gamma_6^c$ ($\beta = c$), in the light-hole valence band
$\Gamma_8^v$ ($\beta = l$) and in the spin split-off valence band
$\Gamma_7^v$ ($\beta = s$) with eigenenergies $E_{\nu\sigma}^\beta
\equiv E_\nu^\beta + \sigma \Zc$.  As before, we introduce an
in-plane magnetic field $\vekc{B}_\|$ via the vector potential
$\vekc{A} = z \, \vekc{B}_\| \times \hat{\vek{z}}$.  Second-order
quasi-degenerate perturbation theory for $\vekc{B}_\|$ then yields
Roth's formula~\cite{rot59, win03},
\begin{align}
  \frac{g}{2} & =
  \frac{2i}{3} \frac{2 m_0}{\hbar^2} P^2 \sum_{\nu'} \biggl[
  \frac{\braket{c,\nu |z| l,\nu'} \braket{l,\nu' |k_z| c,\nu}}
       {E_\nu^c - E_{\nu'}^l}
  \nonumber \\* & \hspace{7em} {}
  - \frac{\braket{c,\nu |z| s,\nu'} \braket{s,\nu' |k_z| c,\nu}}
         {E_\nu^c - E_{\nu'}^s}
  \biggr] \,\, .
  \label{eq:roth}
\end{align}
This calculation is similar to how the $g$-factor in the Zeeman term
of the Pauli equation is derived from the Dirac equation.  An
imbalance between spin-up and spin-down states (due to an exchange
field $\vekc{X}_\|$ or due to an external field $\vekc{B}_\|$) thus
implies a spin magnetization (\ref{eq:spin-mag:def}) proportional
to~$g$.

For comparison, we now evaluate the orbital magnetization
(\ref{eq:orb-mag:def}) from $\tilde{\HK}$ without projecting on the
subspace $\Gamma_6^c$.  In the following discussion,
$\tilde{\vekc{Z}}$ stands for an exchange field $\vekc{X}$ or a
magnetic field $\vekc{B}$ that enters $\tilde{\HK}$ via the vector
potential $\vekc{A}$.  Focusing on the states in the conduction band
and treating $\tilde{\HK} {}^{(1)}$ in first order perturbation
theory, the perturbed eigenstates read
\begin{align}
  \ket{c,\nu\sigma^{(1)}} & = \ket{c,\nu\sigma}
  + P \sum_{\nu' \ne \nu} \biggl( \sqrt{\frac{2}{3}}
  \frac{\braket{l,\nu' |k_z| c,\nu}}{E_\nu^c - E_{\nu'}^l}
  \ket{l, \nu'\sigma} \nonumber \\*
  & \hspace{3em} {} - \frac{\sigma}{\sqrt{3}}
  \frac{\braket{s,\nu' |k_z| c,\nu}}{E_\nu^c - E_{\nu'}^s}
  \ket{s, \nu'\sigma} \biggr) \,\, ,
  \label{eq:o-m:state-z}
\end{align}
where we neglected contributions linear in $\kk_\|$ as these lead to
higher-order corrections in Eq.\ (\ref{eq:o-m:me}) below.  In the
absence of a field $\tilde{\vekc{Z}}$, the eigenstates
(\ref{eq:o-m:state-z}) are twofold degenerate ($\sigma = \pm$).  The
states (\ref{eq:o-m:state-z}) are also the appropriate unperturbed
states for first-order degenerate perturbation theory for a field
$\tilde{\vekc{Z}}$ oriented in $z$ direction.  If instead, we consider
a field $\tilde{\vekc{Z}}$ oriented in-plane, the appropriate
unperturbed states become
\begin{equation}
  \ket{c,\nu\sigma, \varphi_\Zc {}^{(1)}} = \frac{1}{\sqrt{2}} \left[
    \ket{c,\nu{+}^{(1)}} + \sigma \, \exp(i \varphi_\Zc) \,
    \ket{c,\nu{-}^{(1)}} \right] \,\, ,
  \label{eq:o-m:state-inplane}
\end{equation}
where $\varphi_\Zc$ is defined, as before, as the angle between
$\tilde{\vekc{Z}}$ and the crystallographic direction $[100]$.

The velocity operator
\begin{equation}
  \label{eq:o-m:velo}
  \tilde{\vek{v}}_\| = \frac{\partial \tilde{\HK}}{\partial \, \hbar \kk_\|}
\end{equation}
is independent of $\kk$ and independent of $\tilde{\vekc{Z}}$.
Using the states (\ref{eq:o-m:state-inplane}), the matrix elements
(\ref{eq:z-v:av:def}) of the orbital magnetization can be expressed
in the form
\begin{equation}
  \label{eq:o-m:me}
  \frac{2m_0}{\hbar} \hat{\vek{z}} \times
  \av{\{ z\, , \tilde{\vek{v}}_\| \}}_{\nu\sigma} = \sigma
  \begin{pmatrix}
    \cos \varphi_\Zc \\ \sin \varphi_\Zc
  \end{pmatrix}
  \frac{g}{2}
\end{equation}
with $g$ given in Eq.\ (\ref{eq:roth}).
The matrix elements of the orbital magnetization within the
multiband Hamiltonian $\tilde{\HK}$ are thus equal to the matrix
elements of the spin magnetization in the two-band Hamiltonian
$\Hcb$.  In lowest order of $\tilde{\vekc{Z}}$, these
Hamiltonians yield the same imbalance between the occupation numbers
for the respective spin states $\sigma = \pm$.  Thus it follows from
Eq.\ (\ref{eq:total-mag:def:split}) that, averaged over all occupied
states, the orbital magnetization within $\tilde{\HK}$ equals the
spin magnetization within $\Hcb$.  In both approaches, the
magnetization vanishes in the limit $\Delta_0 \rightarrow 0$.

Again, it is illuminating to compare the orbital magnetization
$\vekc{M}_\Zc$ with the equilibrium current distribution
(\ref{eq:2D-current:def}).  Using $\phi_\nu^\beta (z) \equiv
\braket{z|\beta, \nu}$ and $\Phi_{\nu\sigma}^c (z) \equiv
\braket{z | c, \nu\sigma^{(1)}}$, we get
\begin{widetext}
  \begin{subequations}
    \label{eq:o-m:cur}
    \begin{align}
  \vek{j}_\| (z, \nu\sigma)
  & = \Re \left[ \Phi_{\nu\sigma}^\ast (z) \, \tilde{\vek{v}}_\|
    \, \Phi_{\nu\sigma} (z) \right] \,\, , \\
  & = \sigma
  \begin{pmatrix}
    \cos (\varphi_\Zc - \pi/2) \\ \sin (\varphi_\Zc - \pi/2)
  \end{pmatrix}
  \frac{2}{3} \frac{P^2}{\hbar} \sum_{\nu'} \Re \left[
    \phi_\nu^{c \ast} \phi_{\nu'}^l
  \frac{\braket{l,\nu' |k_z| c,\nu}}{E_\nu^c - E_{\nu'}^l}
  - \phi_\nu^{c \ast} \phi_{\nu'}^s
  \frac{\braket{s,\nu' |k_z| c,\nu}}{E_\nu^c - E_{\nu'}^s}
  \right] \,\, ,
\end{align}
\end{subequations}
\end{widetext}
where we ignored the trivial $\kk_\|$-dependent part. Note that, for
a symmetric confinement $V(z)$, the sum over $\nu'$ is restricted to
terms such that the product $\nu\nu'$ is odd.  For the lowest
subband $\nu=0$, the dominant term in the sum over $\nu'$ is
$\nu'=1$.  In itinerant-electron ferromagnets with an intrinsic
imbalance in the occupation of states with opposite spins $\sigma =
\pm$, this term describes a dipolar equilibrium current.  In
itinerant-electron antiferromagnets, the quadrupolar currents in
Eq.\ (\ref{eq:AFM-EtoM:curr}) (with $\nu' = 2$) are the counterpart
of dipolar currents (\ref{eq:o-m:cur}) (with $\nu' = 1$) in
ferromagnets.
These currents are illustrated in Fig.~\ref{fig:e-vel}.


\begin{thebibliography}{134}
\bibitem [{misc:2Dmetal()}]{misc:2Dmetal}%
  \BibitemOpen
  \bibinfo {note} {These field configurations are associated with well-defined
  equilibrium-magnetoelectric effects for the metallic quasi-2D systems
  considered in our work. Complementary magnetoelectric responses, e.g., an
  in-plane electric polarization arising from the presence of a perpendicular
  magnetic field, can occur in equilibrium only for insulating 2D systems such
  as the idealized lattice model discussed in Ref.~\cite{gao14}.}\BibitemShut
  {Stop}%
\bibitem [{misc:diamond()}]{misc:diamond}%
  \BibitemOpen
  \bibinfo {note} {More precisely, the diamond structure is characterized by a
  nonsymmorphic space group ($O_h^7$, \# 227), i.e., the inversion operation
  must be combined with a nonprimitive translation in order to map the diamond
  structure onto itself. Nonetheless, the point group characterizing the
  diamond structure is the point group of a cube, $O_h$, which includes
  inversion so that electronic states are at least twofold degenerate
  throughout the Brillouin zone.}\BibitemShut {Stop}%
\bibitem [{misc:gamnas()}]{misc:gamnas}%
  \BibitemOpen
  \bibinfo {note} {In Mn-doped ferromagnetic semiconductors such as GaMnAs and
  InMnSb, the magnetic order is mediated by the holes in the valence
  band~\cite{die14}. As a result, the spontaneous magnetization
  $\vekc{M}^\mathrm{s}$ depends on the charge distribution of holes in
  heterostructures made of these materials and is therefore controllable by
  perpendicular electric fields $\Ec_z$~\cite{ohn00, saw10, mat15}. The
  mechanism for this type of magnetoelectric effect is fundamentally different
  from the new type of magnetoelectricity focused on in the present work. The
  latter effect can be cleanly accessed in experiments if the quantum well is
  equipped with both a front and a back gate that can change the electric field
  $\Ec_z$ at constant density~\cite{pap99, hab04}.}\BibitemShut {Stop}%
\bibitem [{misc:gfak()}]{misc:gfak}%
  \BibitemOpen
  \bibinfo {note} {In the relation $\vekc{Z} \equiv (g/2) \, \bohrmag \,
  \vekc{B}_\| + \vekc{X}_\|$ used in Eq.\ (\ref{eq:ham-cb:zeeman}), we assume
  for the sake of consistency between the cases with $\vekc{B}_\| \ne 0$ and
  $\vekc{X} \ne 0$ that $g > 0$ so that in Eq.\ (\ref{eq:ham-cb:rot}) the
  eigenstates of $\sigma_z$ with eigenvalue $\sigma = -1$ are predominantly
  occupied for $\mathcal{Z} \ne 0$. Indeed, for many semiconductors including
  InSb and GaAs we have $g < 0$. In such a case, when the Zeeman field
  $\mathcal{Z}$ is due to a magnetic field $\Bc_\|$, we get an additional minus
  sign in the final expressions for the magnetization [Eq.\
  (\ref{eq:EtoM:final})], the polarization [Eq.\ (\ref{eq:BtoP:final})] and the
  free energy [Eq.\ (\ref{eq:free-en:final})].}\BibitemShut {Stop}%
\bibitem [{misc:length()}]{misc:length}%
  \BibitemOpen
  \bibinfo {note} {Taking values for $d$ from Table~6.3 in Ref.\ \cite{win03},
  we find $l_d = 200\,$\AA\ for InSb, $7.24\,$\AA\ for GaAs, and $7.13\,$\AA\
  for InAs.}\BibitemShut {Stop}%
\bibitem [{misc:rashba()}]{misc:rashba}%
  \BibitemOpen
  \bibinfo {note} {The multiband Hamiltonians $\HK$ used for the numerical
  calculations presented in Fig.~\ref{fig:e-mom-150} do not explicitly include
  a Rashba term. Nonetheless, as explained, e.g., in Ref.~\cite{win03}, they
  account for Rashba spin-orbit coupling to all orders in the wave vector
  $\kk$. In this case, similar to the discussion in
  Appendix~\ref{app:orb-magnet}, Rashba spin-orbit coupling manifests itself as
  a contribution to the orbital magnetization presented in
  Fig.~\ref{fig:e-mom-150}.}\BibitemShut {Stop}%
\bibitem [{misc:toroid()}]{misc:toroid}%
  \BibitemOpen
  \bibinfo {note} {The form and interpretation of $\delta F$ in
  Eq.~(\ref{eq:free-en:ME:toroid:all}) need to be distinguished from the
  parameterization~\cite{spa08} $\delta F = - \tilde{\vek{\tau}}' \cdot \left(
  \vekc{E} \times \vekc{B} \right) - \tilde{q}_{ij}' \left( \Ec_i \Bc_j + \Ec_j
  \Bc_i \right)$ in terms of a vector $\tilde{\vek{\tau}}'$ and a symmetric
  tensor $\tilde{q}_{ij}'$. Mathematically, it is possible to determine
  $\tilde{\vek{\tau}}'$ and $\tilde{q}_{ij}'$ for any magnetoelectric tensor
  $\alpha_{ij}$. For example, starting from Eq.\ (\ref{eq:expandFmag}), we have
  $\tilde{\vek{\tau}}' = \tilde{\vek{\tau}} /2$, $\tilde{q}_{yz}' = -
  \tilde{\tau}_x'$, $\tilde{q}_{zx}' = \tilde{\tau}_y'$, and $\tilde{q}_{xy}' =
  \tilde{q}_{ii}' = 0$. Similar relations exist for Eqs.\
  (\ref{eq:free-en:qmag:toroid:all}), (\ref{eq:AFM:free-en:toroid}), and
  (\ref{eq:free-en:afm:qmag:toroid:all}). However, unless the components of
  $\tilde{\vek{\tau}}'$ and $\tilde{q}_{ij}'$ turn out to be independent (which
  they are not in, e.g., our systems of interest), they do not represent
  separate physical mechanisms for magnetoelectricity.}\BibitemShut {Stop}%
\bibitem [{misc:minus()}]{misc:minus}%
  \BibitemOpen
  \bibinfo {note} {The final results in the ferromagnetically ordered case and
  the final results in the antiferromagnetically ordered case differ by a minus
  sign that can be traced back to the fact that the Zeeman term
  (\ref{eq:ham-cb:zeeman}) favors the occupation of eigenstates with spin
  eigenvalue $\sigma = -1$.}\BibitemShut {Stop}%
\bibitem [{misc:brown:dresselhaus()}]{misc:brown:dresselhaus}%
  \BibitemOpen
  \bibinfo {note} {Our calculations assumed small coefficients $d$ and
  $\mathcal{d}$. In lowest order, $\alpha_{zi}$ is proportional to $d$ or
  $\mathcal{d}$, whereas $\sqrt{\chi^\Ec_{zz} \, \chi^{\Bc, \mathrm{p}}_{ii}}$
  is independent of these coefficients. For large values of these coefficients
  it would be necessary to include corrections proportional to higher orders of
  $d$ or~$\mathcal{d}$.}\BibitemShut {Stop}%
\bibitem [{misc:nongyrotropic()}]{misc:nongyrotropic}%
  \BibitemOpen
  \bibinfo {note} {Among 21 crystallographic point groups lacking inversion
  symmetry, 18 point groups are gyrotropic. The noncentrosymmetric,
  nongyrotropic point groups include $T_d$, $C_{3h}$, and $D_{3h}$
  \cite{ivc08}.}\BibitemShut {Stop}%
\end{thebibliography}
\end{document}